\documentclass[12pt,a4paper]{article}
\usepackage[utf8x]{inputenc}
\usepackage{jheppub}
\usepackage{rotating}

\renewcommand{\[}{\begin{equation}}
\renewcommand{\]}{\end{equation}}
\def\nn{\nonumber}

\def\bra#1{\langle #1|}
\def\ket#1{|#1 \rangle}
\def\braket#1{\langle #1 \rangle}
\def\bs{\boldsymbol}
\def\tr{\operatorname*{tr}}

\def\eg{e.g. }
\def\etc{etc. }

\def\eqn#1{eq.~\eqref{#1}}
\def\eqns#1#2{eqs.~\eqref{#1} and~\eqref{#2}}

\def\fig#1{figure~{\ref{#1}}}

\def\sec#1{section~{\ref{#1}}}
\def\secs#1#2{sections~{\ref{#1}} and~{\ref{#2}}}

\def\app#1{appendix~{\ref{#1}}}

\def\foot#1{footnote~{\ref{#1}}}
\def\rcite#1{ref.~\cite{#1}}
\def\rcites#1{refs.~\cite{#1}}

\subheader{\normalsize
	\vspace*{-3.7em}
	\begin{flushright}
		CP3-21-49\\
	\end{flushright}
}

\title{Classical observables from coherent-spin amplitudes}
\author[1]{Rafael Aoude}
\author[2]{and Alexander Ochirov}

\affiliation[1]{Centre for Cosmology, Particle Physics and Phenomenology (CP3), \\
Universit\'e catholique de Louvain, 1348 Louvain-la-Neuve, Belgium}
\affiliation[2]{Mathematical Institute, University of Oxford,
Andrew Wiles Building, Radcliffe Observatory Quarter,
Woodstock Road, Oxford, OX2 6GG, UK}

\emailAdd{rafael.aoude@uclouvain.be}
\emailAdd{ochirov@maths.ox.ac.uk}

\abstract{The quantum field-theoretic approach to classical observables due to Kosower, Maybee and O'Connell provides a rigorous pathway from on-shell scattering amplitudes to classical perturbation theory. In this paper, we promote this formalism to describe general classical spinning objects by using coherent spin states. Our approach is fully covariant with respect to the massive little group ${\rm SU}(2)$ and is therefore completely synergistic with the massive spinor-helicity formalism. We apply this approach to classical two-body scattering due gravitational interaction. Starting from the coherent-spin elastic-scattering amplitude, we derive the classical impulse and spin kick observables to first post-Minkowskian order but to all orders in the angular momenta of the massive spinning objects. From the same amplitude, we also extract an effective two-body Hamiltonian, which can be used beyond the scattering setting. As a cross-check, we rederive the classical observables in the center-of-mass frame by integrating the Hamiltonian equations of motion to the leading order in Newton's constant.
}

\begin{document}
\maketitle
\addtocontents{toc}{\protect\setcounter{tocdepth}{2}}

%%%%%%%%%%%%%%%%%%%%%%%%%%%%%%%%%%%%%%%%%%%%%%%%%%
\section{Introduction}
\label{sec:Intro}
%%%%%%%%%%%%%%%%%%%%%%%%%%%%%%%%%%%%%%%%%%%%%%%%%%

The experimental observations of gravitational waves
generated during black-hole and neutron-star mergers
\cite{Abbott:2016blz,TheLIGOScientific:2017qsa}
have accentuated the need for
an accurate and efficient theoretical description
of binary inspiral dynamics of celestial bodies.
A new perspective on this classical problem in general relativity
has recently been offered by methods rooted in quantum field theory (QFT),
which profit from a wide variety of on-shell methods
developed for the study of quantum scattering amplitudes.
In particular, many state-of-the-art post-Minkowskian (PM) computations
of the two-body dynamics of Schwarzschild and Kerr black holes
have been performed using the philosophy of effective field theory (EFT),
either in the genuinely quantum-theoretic sense
\cite{Cheung:2018wkq,Bern:2019nnu,Bern:2019crd,Bern:2020buy,Bern:2021dqo,Kosmopoulos:2021zoq}
or in the form of classical worldline effective theory
\cite{Damour:2016gwp,Damour:2017zjx,Bini:2018ywr,Kalin:2019rwq,Kalin:2019inp,Damour:2019lcq,Bini:2020flp,
Kalin:2020mvi,Kalin:2020fhe,Kalin:2020lmz,Damour:2020tta,
Mogull:2020sak,Jakobsen:2021smu,
Liu:2021zxr,Dlapa:2021npj,Jakobsen:2021lvp},
see also \rcites{Bjerrum-Bohr:2018xdl,KoemansCollado:2019ggb,Cristofoli:2020uzm,
AccettulliHuber:2020oou,Bjerrum-Bohr:2021vuf,
DiVecchia:2021bdo,Bjerrum-Bohr:2021din,Damgaard:2021ipf}
for concurrent developments using eikonal methods.

A related framework for computing classical observables
directly from expectation values of appropriately chosen QFT operators
has been developed by Kosower, Maybee and O'Connell (KMOC)
in \rcite{Kosower:2018adc}
and further extended in subsequent works
\cite{Maybee:2019jus,delaCruz:2020bbn,Cristofoli:2021vyo}.
In particular, although \rcite{Maybee:2019jus} only
considered massive particles of spin 1/2 and 1 in the KMOC formalism,
it was already mentioned
that for a well-defined classical limit of spinning-particle scattering
one should strictly speaking consider very large spin representations.
Such a limit for integer-spin representations was considered
concurrently in \rcites{Guevara:2019fsj,Arkani-Hamed:2019ymq},
albeit in a heuristic manner, as well as in the eikonal setting
in \rcites{Guevara:2018wpp,Bern:2020buy,Kosmopoulos:2021zoq},
see also \rcites{Guevara:2017csg,Chung:2018kqs,
Chung:2019duq,Chung:2019yfs,Chung:2020rrz,Aoude:2020mlg,Chen:2021huj}.

In this paper, we aim to elucidate
the more rigorous machinery behind certain unceremonious steps
that were taken in some of the mentioned works.
For this, we employ the KMOC framework
and construct the incoming massive spinning states
using the coherent-state formalism
\cite{Atkins:1971zy,Radcliffe_1971,Perelomov_1977},
which is based on the Jordan–Schwinger construction of
the general spin representations of the massive little group ${\rm SU}(2)$.
At a superficial level,
the coherent-state formalism provides a perfect ${\rm SU}(2)$ spinor
to saturate the little-group indices
that were left uncontracted in earlier approaches
to the classical limit of quantum scattering with spin
\cite{Guevara:2018wpp,Guevara:2019fsj}
--- where a heuristic notion of ``generalized expectation value''
was introduced instead.
In addition, the coherent-state formalism provides
guidance as to which quantities one should send to zero or infinity
in the classical limit and with which speed.
In particular, the normalization of these states precisely cancels with 
non-classical terms, allowing us to directly extract the classical spin vector from the amplitudes. 
This normalization also identifies the relevant three-point amplitudes which are diagonal in the spin of the massive particles.

Massless coherent states have also been very recently used
for classical radiation in \rcite{Cristofoli:2021vyo}.
In such an approach, a state corresponding
to a classically meaningful carrier-force field
(e.g. electromagnetic or gravitational)
is described as an on-shell momentum integral
over states with arbitrary numbers of massless quanta.
Similarly, here we construct massive states
with classically meaningful angular momenta
from states with arbitrarily large quantum spins.
The basic difference is the discreteness
of the set of spin states that we need to sum over
in the massive-spin setting.

Specializing to gravity in \sec{sec:ClassicalAmplitudes},
we consider general classical multipole interactions that couple
the angular momentum of a massive body to the gravitational field.
Such interactions may be included in the formalism
by considering spinning quantum particles non-minimally coupled to the graviton.
We observe an interesting feature of such non-minimal couplings:
by default they become power-suppressed in the classical limit.
In other words, we see a need to effectively rescale the quantum
coupling by inverse powers of $\hbar$ in order to reproduce
the freedom of choice in the classical multipole moments.
Interestingly, if one does not apply such a ``superclassical'' rescaling
of the non-minimal couplings, they seem to disappear in the classical limit, 
leaving the Kerr black hole as the naturally favored spinning object.

In \sec{sec:4pt} we employ our formalism
to compute the leading-order net changes in linear and angular momenta
of general spinning objects during two-body gravitational scattering.
These basic Lorentz-covariant observables,
given by \eqns{ImpulseMomLOGrav}{ImpulseSpinLOGrav},
are naturally derived from the four-point coherent-spin amplitude
\eqref{ElasticScatteringAmplitudeFinal}.
We complement our results by presenting
an effective two-body Hamiltonian in the center-of-mass (COM) frame,
which encapsulates the dynamical information contained in this amplitude.
The value of such a Hamiltonian is in its universality,
as it can easily be used for bound-state problems just as well.
We find our Hamiltonian~\eqref{ElasticScatteringHamiltonian}
to be in a different gauge as compared to the result of \rcite{Chung:2020rrz},
so we validate it by recomputing the COM versions
of the linear and angular impulse observables directly
from the corresponding Hamiltonian equations of motion.

At various points in our paper we pause to consider
the particularly interesting case of spinning black-hole scattering,
in which we find perfect agreement with the classical 1PM solution
to all orders in spin \cite{Vines:2017hyw},
as well as a new perspective on the considerations
of \rcite{Guevara:2019fsj}.

%%%%%%%%%%%%%%%%%%%%%%%%%%%%%%%%%%%%%%%%%%%%%%%%%%
\section{Classical observables via coherent spin states}
\label{sec:KMOC}
%%%%%%%%%%%%%%%%%%%%%%%%%%%%%%%%%%%%%%%%%%%%%%%%%%

In this section we review the KMOC formalism for classical observables
\cite{Kosower:2018adc,Maybee:2019jus,delaCruz:2020bbn,Cristofoli:2021vyo}
focusing on the aspects due to the spin degrees of freedom,
which are implemented using coherent states.

The starting point of the formalism is to consider
the change in the expectation value of a certain quantum operator $O$
due to scattering:\footnote{In \rcites{Kosower:2018adc,Maybee:2019jus,delaCruz:2020bbn}
the right-hand side of \eqn{KMOC0} is written as
$ i\braket{\text{in}|[O,T]|\text{in}}
+ \braket{\text{in}|T^\dagger[O,T]|\text{in}} $
due to the unitarity relation $T^\dagger = T - iT^\dagger T$.
Here we choose to deal with three terms instead of four.
At the leading perturbative order, only $\Delta_1 O$ contributes,
in which $T^\dagger$ may as well be replaced by~$T$.
}
\[
   \Delta O %\equiv \braket{\text{out}|O|\text{out}}
%    - \braket{\text{in}|O|\text{in}}
    = \braket{\text{in}|S^\dagger O S|\text{in}}
    - \braket{\text{in}|O|\text{in}}
    = \underbrace{i \braket{\text{in}|\:\![O T-T^\dagger O]\:\!|\text{in}}
                 }_{\Delta_1 O}
    + \underbrace{\braket{\text{in}|\:\!T^\dagger O T\:\!|\text{in}}
                 }_{\Delta_2 O} ,
\label{KMOC0}
\]
where we have used the scattering matrix $S=1+iT$.
As indicated, this object naturally splits into two parts,
linear and quadratic in the scattering transition operator $T$.
For this observable to have a well-defined classical interpretation,
%and assuming that the operator $O$ is hermitian,
we need the $\ket{\text{in}}$ states to behave in a predictable manner
in the classical limit.

For concreteness, we set up relativistic scattering
with an impact parameter~$b^\mu$
for two massive objects with definite classical momenta
$m_{\rm a} u_{\rm a}^\mu$ and
$m_{\rm b} u_{\rm b}^\mu$, where
$u_{\rm a, b}^2 = 1$ and $b\cdot u_{\rm a, b}=0$.
Then a convenient choice of the initial state is
\cite{AlHashimi:2009bb,Kosower:2018adc}
\[
   \ket{\text{in}} = \int_{p_1} \int_{p_2}
      \psi_{\rm a}(p_1) \psi_{\rm b}(p_2)
      e^{i b\cdot p_1/\hbar} \ket{p_1;p_2} , \qquad \quad
   \int_p \equiv \int\!\!\frac{d^4 p}{(2\pi)^3} \Theta(p^0) \delta(p^2-m^2) ,
\label{InitialStateMomentum}
\]
with the relativistic momentum-space wavefunctions of the form
\[ \psi_\xi(p) = \frac{1}{m}
      \bigg[\frac{8\pi^2}{\xi K_1(2/\xi)}\bigg]^{1/2}\!
      \exp\!\bigg({-}\frac{p \cdot u}{\xi m}\bigg) .
\label{MomentumWavefunction}
\]
Here the normalization involves
the modified Bessel function of the second kind.
These wavefunctions produce well-behaved
one-particle expectation values
\[
   \braket{p^\mu}_\xi %= m u^\mu \frac{K_2(2/\xi)}{K_1(2/\xi)}
    = m u^\mu + {\cal O}(\xi) , \qquad \quad
   \braket{p^2}_\xi = m^2 , \qquad \quad
   \braket{\bs{p}^2}_\xi \big|_{u^\mu = (1,\bs{0})}
    = \frac{3}{2} \xi m^2 + {\cal O}(\xi^2) ,
\label{MomentumExpectations}
\]
the latter of which can be interpreted
as the standard deviation (squared) in the rest frame.
The dimensionless parameter $\xi$ may therefore be thought of as the ratio
\[
   \xi \approx \frac{2\sigma_p^2}{3m^2} , \qquad \quad
   \sigma_p^2 = -\big\langle (p - \braket{p}_\xi)^2 \big\rangle_\xi ,
\]
which must naturally be sent to zero in the classical limit~\cite{Kosower:2018adc}.

The main feature of the initial states~\eqref{InitialStateMomentum}
is that they are built up from definite-momentum states,
for which we know best how to compute scattering amplitudes.
In the presence of additional degrees of freedom,
one needs to find a way to model them with quantum states in a similar manner
\cite{Maybee:2019jus,delaCruz:2020bbn}.

%%%%%%%%%%%%%%%%%%%%%%%%%%%%%%%%%%%%%%%%%%%%%%%%%%
\subsection{Quantum spin}
\label{sec:QuantumSpin}
%%%%%%%%%%%%%%%%%%%%%%%%%%%%%%%%%%%%%%%%%%%%%%%%%%

In the presence of angular degrees of freedom at the classical level,
we wish to encode them using spinning quantum states.
Let us first review the salient features of quantum angular momentum.

%%%%%%%%%%%%%%%%%%%%%%%%%%%%%%%%%%%%%%%%%%%%%%%%%%
\subsubsection{Definite-spin states}
\label{sec:DefiniteSpinStates}
%%%%%%%%%%%%%%%%%%%%%%%%%%%%%%%%%%%%%%%%%%%%%%%%%%

It will be particularly convenient
to use Schwinger's construction~\cite{Schwinger:1952dse},
in which general spin states are obtained from the zero-spin state
$\ket{s\!=\!0}$ by acting with two kinds of creation operators:
\[
   \ket{s,s_z} = \frac{ \big(a_1^\dagger\big)^{s+s_z}
                        \big(a_2^\dagger\big)^{s-s_z} }
                      { \sqrt{(s+s_z)!(s-s_z)!} }
   \ket{0} , \qquad \quad
   s_z = -s, -s+1, \ldots, s-1, s .
\label{SpinStates}
\]
Let us covariantize this construction right away ---
at this point merely with respect to the massive little group ${\rm SU}(2)$.
We then implement the angular-momentum algebra using the Pauli matrices
as follows:
\[
   [a^a, a_b^\dagger] = \delta^a_b , \qquad \quad
   \bs{S} = \frac{\hbar}{2} a_a^\dagger \bs{\sigma}^a{}_b a^b
   \qquad \Rightarrow \qquad
   [S^i, S^j] = i\hbar \epsilon^{ijk} S^k .
\label{AngularMomentumAlgebra}
\]
Note that the creation and annihilation operators are naturally equipped
with the ${\rm SU}(2)$ spinor indices that are dual to each other
under the symplectic bilinear form~$\epsilon_{ab}$.
The vectorial rotations $O\in{\rm SO}(3)$ are related to their spinorial counterparts $U\in{\rm SU}(2)$ via the standard double covering map
\[
   O^{ij} = \frac{1}{2} \tr(\sigma^i U \sigma^j U^\dagger) .
\label{DoubleCover}
\]
The operator algebra~\eqref{AngularMomentumAlgebra} is evidently
covariant with respect to such transformations,
which encode the arbitrariness of the choice
of the spin quantization axis:\footnote{The standard Pauli matrices,
in which $\sigma^3$ is diagonal, imply spin quantization along the $z$-axis.
Rotations~\eqref{SU2Rotations}, which transform
$\bs{S} \to O \bs{S} = \frac{\hbar}{2}
 a_a^\dagger (U^\dagger\bs{\sigma}U)^a{}_b a^b$,
are equivalent to choosing another spatial direction $\hat{n}^i = O^{i3}$,
for which $\hat{\bs{n}} \cdot (U^\dagger\bs{\sigma}U) = \sigma^3$ is diagonal.
}
\[
   a^a \to U^a{}_b a^b , \qquad \quad
   a_a^\dagger \to U_a{}^b a_b^\dagger = a_b^\dagger (U^\dagger)^b{}_a
   \qquad \Rightarrow \qquad S^i \to O^{ij} S^j .
\label{SU2Rotations}
\]

Starting from the scalar state,
which by definition obeys $a^a \ket{0} = 0$,
we may now construct ${\rm SU}(2)$-covariant $s$-spin states
\[
   \ket{s,\{a\}} \equiv \ket{s,\{a_1 \dots a_{2s}\}}
    = \frac{1}{\sqrt{(2s)!}} a_{a_1}^\dagger a_{a_2}^\dagger \dots
      a_{a_{2s}}^\dagger \ket{0} ~\equiv~
    \frac{\big(a_a^\dagger\big)^{\odot 2s}\!}{\sqrt{(2s)!}} \ket{0} .
\label{SpinStatesCovariant}
\]
Here the indices are automatically fully symmetrized,
and for brevity we have introduced the symbol $\odot$,
which will denote the symmetrized tensor product \cite{Guevara:2017csg}.
Note that the normalization
of the general spin states~\eqref{SpinStatesCovariant}
\[
   \braket{s,\{a_1 \dots a_{2s}\}|s'\!,\{b_1 \dots b_{2s'}\}}
    = \delta^s_{s'} \delta^{(a_1}_{(b_1} \cdots \delta^{a_{2s})}_{b_{2s})}
    = \delta^s_{s'} \big(\delta^a_b\big)^{\odot 2s}
\label{SpinStatesNormalization}
\]
involves two symmetrizations (one of which is redundant),
which include the $1/(2s)!$ denominators and imply combinatorial prefactors
when translated to the states~\eqref{SpinStates}:
\[
   \ket{s,s_z} = \sqrt{\tfrac{(2s)!}{(s+s_z)!(s-s_z)!}}
      | s, \{ \underbrace{1,\dots,1}_{s+s_z},
              \underbrace{2,\dots,2}_{s-s_z} \} \rangle , \qquad \quad
   \braket{s,s_z|s'\!,s_z'} = \delta^s_{s'} \delta^{s_z}_{s_z'} .
\]
For future reference, the spin expectation values are explicitly
\[\!\!
   \bs{S}^{\{a\}}{}_{\{b\}} \equiv
   \bra{s,\{a\}} \bs{S} \ket{s'\!,\{b\}}
    = \hbar s\,\delta^s_{s'}\,\bs{\sigma}^{(a_1}{}_{(b_1}
      \delta^{a_2}_{b_2} \cdots \delta^{a_{2s})}_{b_{2s})}
    = \hbar s\,\delta^s_{s'}\,\bs{\sigma}^a{}_b
      \odot \big(\delta^a_b\big)^{\odot (2s-1)} .\!
\label{SpinRepresentation3d}
\]
As is standard for quantum angular momentum,
this representation is diagonal in the total-spin quantum number~$s$
and not entirely diagonal in the spin-projection quantum number~$s_z$.

%%%%%%%%%%%%%%%%%%%%%%%%%%%%%%%%%%%%%%%%%%%%%%%%%%
\subsubsection{Coherent spin states}
\label{sec:CoherentSpinStates}
%%%%%%%%%%%%%%%%%%%%%%%%%%%%%%%%%%%%%%%%%%%%%%%%%%

The reason why we chose Schwinger's construction~\cite{Schwinger:1952dse}
to deal with definite-spin states is that it allows for a straightforward
implementation of coherent spin states
--- which are well-suited for setting up classical angular momentum,
see \eg \rcites{Atkins:1971zy,Radcliffe_1971,Perelomov_1977}.
These states are defined as
\[
   \ket{\alpha} = e^{-\tilde{\alpha}_a \alpha^a/2}
      e^{\alpha^a a_a^\dagger} \ket{0}
   \qquad \Rightarrow \qquad
   a^a \ket{\alpha} = \alpha^a \ket{\alpha} ,
\label{SpinStatesCoherent}
\]
starting from the same scalar state as above.
We use $\tilde{\alpha}_a$ to denote complex conjugation
of the ${\rm SU}(2)$ spinor $\alpha^a$.
The coherent spin states may of course be immediately expanded
in terms of the definite-spin states:
\[
   \ket{\alpha} = e^{-\tilde{\alpha}_a \alpha^a/2}\!\!
      \sum_{s=0,1/2}^\infty \sum_{a_1,\dots,a_{2s}\!\!}\!
      \frac{\alpha^{a_1}\!\cdots \alpha^{a_{2s}}\!}{\sqrt{(2s)!}}
      \ket{s,\{a_1\!\dots\!a_{2s}\}} %\\ &
    \equiv e^{-(\tilde{\alpha} \alpha)/2}\!
      \sum_{2s=0}^\infty \frac{(\alpha^a)^{\odot 2s}\!}{\sqrt{(2s)!}}
      \cdot \ket{s,\{a\}} ,
\label{SpinStatesCoherent2}
\]
where we have also introduced a shorthand notation for
the lengthy but straightforward contractions of little-group indices.

The crucial property of the coherent spin states is
the behavior of their one-particle expectation values
for the angular momentum operator, namely
\[
   \braket{S^i}_\alpha
    = \frac{\hbar}{2} (\tilde{\alpha}\:\!\sigma^i \alpha) , \qquad \quad
   \braket{S^i S^j}_\alpha
    = \braket{S^i}_\alpha \braket{S^j}_\alpha
    + \frac{\hbar^2}{4}
      \big[ \delta^{ij} (\tilde{\alpha} \alpha)
          + i \epsilon^{ijk} (\tilde{\alpha}\:\!\sigma^k \alpha)
      \big] .
\label{SpinExpectation}
\]
The first equation above implies that classical spin
is obtained in the limit where the spinors grow as
\[
   \|\alpha\| \equiv \sqrt{\tilde{\alpha}_a\alpha^a}
    = \sqrt{2|\bs{s}_\text{cl}|/\hbar}
    = {\cal O}(\hbar^{-1/2}) .
\]
In this limit $\braket{S^i S^j}_\alpha$ factorizes into
$\braket{S^i}_\alpha \braket{S^j}_\alpha$, and more generally we have
\[
   \braket{\alpha|\bs{S}^{\otimes n}|\alpha}
    = \big(\braket{\bs{S}}_\alpha \big)^{\otimes n} + {\cal O}(\hbar)
    ~\xrightarrow[\hbar\to0]{}~ (\bs{s}_\text{cl})^{\otimes n} .
\label{ClassicalSpinLimit}
\]

%%%%%%%%%%%%%%%%%%%%%%%%%%%%%%%%%%%%%%%%%%%%%%%%%%
\subsection{Covariant spin quantization}
\label{sec:CovariantSpinQuantization}
%%%%%%%%%%%%%%%%%%%%%%%%%%%%%%%%%%%%%%%%%%%%%%%%%%

Before we return to the computation of a classical observable
from \eqn{KMOC0}, it is worthwhile
to covariantize the above construction further
--- now with respect to the Lorentz group.
Indeed, we wish to be able to describe states with different momenta~$p^\mu$,
as opposed the rest-frame spinning states considered thus far.

%%%%%%%%%%%%%%%%%%%%%%%%%%%%%%%%%%%%%%%%%%%%%%%%%%
\subsubsection{Definite-spin wavefunctions}
\label{sec:Wavefunctions}
%%%%%%%%%%%%%%%%%%%%%%%%%%%%%%%%%%%%%%%%%%%%%%%%%%

An elegant way to promote our discussion to Minkowski space
is offered by the massive spinor-helicity formalism
\cite{Arkani-Hamed:2017jhn} (for earlier formulations see
\rcites{Kleiss:1986qc,Dittmaier:1998nn,Kosower:2004yz,Schwinn:2005pi,
Conde:2016vxs,Conde:2016izb}),
which relies on the splitting
of the four-momentum into two Weyl spinors:\footnote{An exposition
of the formalism consistent with our current conventions
may be found in \rcite{Ochirov:2018uyq}.}
\[
   p_{\alpha\dot{\beta}} = p_{\mu} \sigma^\mu_{\alpha\dot{\beta}}
    = \ket{p^a}_\alpha  [p_a|_{\dot{\beta}}
    = \epsilon_{ab} \ket{p^a}_{\;\!\!\alpha}\;\![p^b|_{\dot{\beta}} ,
   \qquad \quad
   p^2 = \det\{p_{\alpha\dot{\beta}}\} = m^2 .
\label{MassiveSpinors}
\]
Here the familiar ${\rm SU}(2)$ little-group indices~$a$ and~$b$
should not be confused with the Weyl
${\rm SL}(2,\mathbb{C})$ indices~$\alpha$ and~$\dot{\beta}$,
which represent the Lorentz group.
The latter are also raised and lowered
using the two-dimensional Levi-Civita tensors,
\eg $\bra{p^a}^\alpha = \epsilon^{\alpha\beta} \ket{p^a}_\beta$,
and the placement of the angle and square brackets helps
to differentiate between the two chiralities
and allows to keep the indices implicit.
The same notation is widely used in the massless spinor-helicity formalism
\cite{DeCausmaecker:1981jtq,Gunion:1985vca,Kleiss:1985yh,Xu:1986xb,Gastmans:1990xh},
in which the momentum splitting is more straightforward:
\[
   k_{\alpha \dot{\beta}} = k_\mu \sigma^\mu_{\alpha\dot{\beta}}
    = \ket{k}_{\alpha} [k|_{\dot{\beta}} , \qquad \quad
   k^2\!= \det\{k_{\alpha\dot{\beta}}\} = 0 .
\label{MasslessSpinors}
\]

The on-shell spinors serve as ideal building blocks
for definite-spin wavefunctions appearing in quantum field theory.
For Dirac or Majorana fermions, one may use~\cite{Arkani-Hamed:2017jhn}
\[\!\!
   u_p^a = v_{-p}^a
    = \begin{pmatrix} \ket{p^a}_\alpha\:\!\!\\ |p^a]^{\dot{\alpha}}
      \end{pmatrix} , \qquad
   \bar{u}_p^a = \bar{v}_{-p}^a
    = \begin{pmatrix} -\bra{p^a}^\alpha \\ ~~\:[p^a|_{\dot{\alpha}}
      \end{pmatrix}
   \quad \text{assuming} \quad
   \left\{
   \begin{aligned}
   \ket{{-}p} & = -\ket{p} \\
   |{-}p] & = |p] .
   \end{aligned}
   \right.\!\!
\label{DiracMassive}
\]
In the Weyl basis of the gamma matrices
$ \gamma^\mu = \left(\begin{smallmatrix} 0 & \sigma^\mu \\
  \bar{\sigma}^\mu & 0 \end{smallmatrix}\right) $,
these four-spinors obey all the standard properties,
such as the Dirac equation $(\!\not{\!p}-m) u_p = 0$.
For vector bosons,
one may adopt massive polarization vectors~\cite{Guevara:2018wpp}
\[
   \varepsilon_{p\:\!\mu}^{ab}
    = \frac{i \bra{p^{(a}}\sigma_\mu|p^{b)}]}{\sqrt{2}m} ,
\label{PolVectorMassive}
\]
which are automatically transverse and spacelike.
General higher-spin wavefunctions may then be constructed as
\cite{Guevara:2018wpp,Chung:2018kqs,Johansson:2019dnu}
\begin{subequations} \begin{align}
\label{PolTensors}
   \text{integer }s: &~~\quad
   \varepsilon_{p\:\!\mu_1\ldots\:\!\mu_s}^{\{a\}}\!
    = \varepsilon_{p\:\!\mu_1}^{(a_1 a_2}\cdots\:\!
      \varepsilon_{p\:\!\mu_s}^{a_{2s-1} a_{2s})} , \\
\label{PolSpinors}
   \text{half-integer }s: & \quad
   u_{p\:\!\mu_1\ldots\:\!\mu_{\lfloor s \rfloor}}^{\{a\}}\!
    = u_p^{(a_1} \varepsilon_{p\:\!\mu_1}^{a_2 a_3}\cdots\:\!
      \varepsilon_{p\:\!\mu_{\lfloor s \rfloor}}^{a_{2s-1} a_{2s})} .
\end{align} \label{PolTensorsSpinors}%
\end{subequations}

%%%%%%%%%%%%%%%%%%%%%%%%%%%%%%%%%%%%%%%%%%%%%%%%%%
\subsubsection{Covariant spin}
\label{sec:CovariantSpin}
%%%%%%%%%%%%%%%%%%%%%%%%%%%%%%%%%%%%%%%%%%%%%%%%%%

The spin-$s$ wavefunctions set up
a relativistic representation for angular momentum,
naturally divided into subspaces of definite total-spin quantum number.
Indeed, their spanning properties follow from their inner products
in each such subspace:
\[
   \varepsilon_{p\{a\}} \cdot \varepsilon_p^{\{b\}}
   = (-1)^s \big(\delta_a^b\big)^{\odot 2s} ,
   \qquad \quad
   \bar{u}_{p\{a\}} \cdot u_p^{\{b\}}
   = (-1)^{\lfloor s \rfloor}\;\!2m \big(\delta_a^b\big)^{\odot 2s} ,
\label{PolTensorsNormalization}
\]
where the discrepancy in the overall factors,
as compared to \eqn{SpinStatesNormalization}, is due
to the conventional properties of polarization vectors and spinors.
Moreover, appropriate spin-$s$ generalizations of the Lorentz generators
$ \Sigma^{\mu\nu,\sigma}{}_\tau = i[\eta^{\mu\sigma} \delta^\nu_\tau
                                  - \eta^{\nu\sigma} \delta^\mu_\tau] $,
namely
\begin{align}
   \text{integer }s: & \quad
   (\Sigma_s^{\mu\nu})^{\sigma_1\ldots\sigma_{s}}{}_{\tau_1\ldots\tau_{s}}
    = \Sigma^{\mu\nu,\sigma_1}{}_{\tau_1}
      \delta^{\sigma_2}_{\tau_2}\!\cdots \delta^{\sigma_s}_{\tau_s} + \ldots
    + \delta^{\sigma_1}_{\tau_1}\!\cdots \delta^{\sigma_{s-1}}_{\tau_{s-1}}
      \Sigma^{\mu\nu,\sigma_s}{}_{\tau_s} , \nn \\
   \text{half-integer }s: &\;\!\!\qquad \qquad \qquad
   \Sigma_s^{\mu\nu} = \frac{i}{4} [\gamma^\mu,\gamma^\nu]
    + \Sigma_{\lfloor s \rfloor}^{\mu\nu} ,
\label{LorentzGenerators}
\end{align}
may be combined with the on-shell momentum into
the Pauli-Lubanski spin operator
\[
   \Sigma_\lambda =
      \frac{1}{2m} \epsilon_{\lambda\mu\nu\rho} \Sigma^{\mu\nu} p^\rho .
\label{PauliLubanski}
\]
The one-particle matrix elements of this (dimensionless) operator
are explicitly
\cite{Guevara:2019fsj}
\begin{subequations} \begin{align}
   \text{integer }s: & \quad
   \frac{1}{(-1)^{s}}\,
   \varepsilon_{p\{a\}}\!\cdot \Sigma^\mu\!\cdot \varepsilon_p^{\{b\}}
    = s\,\sigma_{p\:\!\mu,(a_1}{}^{(b_1}
      \delta_{a_2}^{b_2} \cdots \delta_{a_{2s})}^{b_{2s})} , \\
   \text{half-integer }s: & \quad
   \frac{1}{(-1)^{\lfloor s \rfloor} 2m}\,
   \bar{u}_{p\{a\}}\!\cdot \Sigma^\mu\!\cdot u_p^{\{b\}}
    = s\,\sigma_{p\:\!\mu,(a_1}{}^{(b_1}
      \delta_{a_2}^{b_2} \cdots \delta_{a_{2s})}^{b_{2s})} .
\end{align} \label{PauliLubanskiMatrixElements}%
\end{subequations}
The prefactors on the left-hand side simply account
for the aforementioned difference in the normalizations
of the quantum states and their wavefunction counterparts.
The non-trivial ingredient on the right-hand side
of \eqn{PauliLubanskiMatrixElements} is the Lorentz-covariant
${\rm SU}(2)$ spin operator~\cite{Maybee:2019jus}
\[
%   \sigma_{p\:\!\mu}^{ab} = -\frac{1}{m} \bra{p^{(a}}\sigma_\mu|p^{b)}]
   \sigma_{p\:\!\mu,a}{}^b  = -\frac{1}{2m}
      \Big(\bra{p_a}\sigma_\mu|p^b] + [p_a|\bar{\sigma}_\mu\ket{p^b}]\Big) ,
\label{DiracSpinOperator}
\]
not to be confused with the ${\rm SL}(2,\mathbb{C})$ matrices
$\sigma_{\alpha\dot{\beta}}^\mu$ and $\bar{\sigma}^{\mu,\dot{\alpha}\beta}$.
It is transverse, and its properties mimic those of the Pauli matrices:
\[
\begin{aligned}
\label{DiracSpinOperatorProperties}
   p\cdot \sigma_{p\:\!\mu,a}{}^b & = 0 , \qquad \quad
   \sigma_{p\:\!\mu,a}{}^a =  0 , \qquad \quad
   \sigma_{p~a}^{\mu,}{}^b \sigma_{p\:\!\mu,c}{}^d
    = -\delta_a^d \delta_c^b - \epsilon_{ac} \epsilon^{bd} , \\
   (\sigma_{p\:\!\mu,a}{}^b)^* & = \sigma_{p\:\!\mu,b}{}^a , \qquad\;\,\qquad
   \big(\sigma_p^\mu \sigma_p^\nu \big)_a{}^b
    = -\bigg[ \eta^{\mu\nu} - \frac{p^\mu p^\nu}{m^2} \bigg] \delta_a^b
    + \frac{i}{m} \epsilon^{\mu\nu\rho\sigma} p_\rho
      \sigma_{p\:\!\sigma,a}{}^b .\!
\end{aligned}
\]
Unsurprisingly, it coincides with the Dirac spin operator
$\sigma_{p\:\!\mu,a}{}^b = \bar{u}_{p\:\!a} \gamma_\mu \gamma^5 u_p^b/2m$.
Moreover, in the rest frame $p^\mu = (m,\bs{0})$
the  operator $\sigma_{p\:\!\mu}$ reduces to $(0,U^\dagger\bs{\sigma}U)$
for some little-group rotation $U\in{\rm SU}(2)$.
Therefore,
the one-particle matrix elements~\eqref{PauliLubanskiMatrixElements}
comprise a Lorentz-covariant representation
of the rest-frame angular-momentum operator~\eqref{SpinRepresentation3d}.
Indeed, if we use the tracelessness property in \eqn{DiracSpinOperator}
to define 
$\sigma_{p\:\!\mu}\equiv \sigma_{p\:\!\mu,1}{}^1 = -\sigma_{p\:\!\mu,2}{}^2$,
such that $\sigma_p^2 = -1$,
then it can be found to serve as the spin-quantization unit four-vector
\cite{Guevara:2019fsj}:
\[
   \begin{aligned}
   \frac{ \varepsilon_{p\{a\}}\!\cdot \Sigma^\mu\!\cdot
          \varepsilon_p^{\{a\}}\!}
        { \varepsilon_{p\{a\}}\!\cdot \varepsilon_p^{\{a\}} }
    &~\:\underset{s\,\in\,\mathbb{Z}}{=} \\
   \frac{ \bar{u}_{p\{a\}}\!\cdot \Sigma^\mu\!\cdot u_p^{\{a\}}\!}
        { \bar{u}_{p\{a\}}\!\cdot u_p^{\{a\}} }
    & \underset{s\,\in\,\mathbb{Z}+\frac{1}{2}}{=}
   \end{aligned}
      \left\{\!
      \begin{aligned}
      s\:\!& \sigma_p^\mu ,~~a_1 = \ldots = a_{2s} = 1 , \\
      (s-1) & \sigma_p^\mu ,~~{\textstyle\sum_{j=1}^{2s}} a_j = 2s+1 , \\
      (s-2) & \sigma_p^\mu ,~~{\textstyle\sum_{j=1}^{2s}} a_j = 2s+2 , \\
      \dots \\
      -(s-1) & \sigma_p^\mu ,~~{\textstyle\sum_{j=1}^{2s}} a_j = 4s-1 , \\\!
      -s\:\!& \sigma_p^\mu ,~~a_1 = \ldots = a_{2s} = 2 ,
      \end{aligned}
      \right.
\label{SpinQuantization}
\]
where no summation of the little-group indices is implied.

To summarize, the one-particle angular-momentum representation
for a given momentum $p^\mu$ is (now with $\hbar$)\footnote{Raising
and lowering the ${\rm SU}(2)$ indices is performed with
$\epsilon^{ab} = -\epsilon_{ab}$
(such that $\epsilon_{ab} \epsilon^{bc} = \delta_a^b$)
and by convention either on the right,
$\alpha_a \equiv \epsilon_{ab} \alpha^b$,
or on the left,
$\tilde{\alpha}^a \equiv \tilde{\alpha}_b \epsilon^{ba}$,
so that $\tilde{\alpha}^a = (\alpha_a)^*$.
Moreover,
$ \sigma_{p\:\!\mu}{}^a{}_b \equiv
  \epsilon^{ac}\sigma_{p\:\!\mu,c}{}^d\epsilon_{db}\!
=-\sigma_{p\:\!\mu,b}{}^a $,
hence the little-group transformations
$U_a{}^b=\exp\{i\omega^\mu \sigma_{p\:\!\mu}\}_a{}^b$ and
$U^a{}_b \equiv \epsilon^{ac} U_c{}^d \epsilon_{db}
=\exp\{-i\omega^\mu \sigma_{p\:\!\mu})\}_b{}^a
$ are hermitian conjugates:
$U^a{}_b=(U^{-1})_b{}^a=(U_a{}^b)^*$.
\label{foot:SU2}
}
\[
   (S_p^\mu)_{s,\{a\}}{}^{s',\{b\}}
    = \hbar s\,\delta_s^{s'}\,\sigma_{p~(a_1}^{\mu,}{}^{(b_1}
      \delta_{a_2}^{b_2} \cdots \delta_{a_{2s})}^{b_{2s})}
    = \hbar s\,\delta_s^{s'}\,\sigma_{p~a}^{\mu,}{}^b
      \odot \big(\delta_a^b\big)^{\odot (2s-1)} ,
\label{SpinRepresentation4d}
\]
It satisfies the transverse Lie algebra
\[
  [S_p^\mu, S_p^\nu] = \frac{i\hbar}{m} \epsilon^{\mu\nu\rho\sigma}
      p_\rho S_{p\:\!\sigma} ,
\label{AngularMomentumAlgebra4d}
\]
which reduces to \eqn{AngularMomentumAlgebra} in the rest frame.
For completeness, the anticommutator is
\[
\begin{aligned}
   \big([S_p^\mu, S_p^\nu]_+\big)_{s,\{a\}}{}^{s',\{b\}} = &
    - \hbar^2 s\,\delta_s^{s'}
      \big[ \eta^{\mu\nu} - p^\mu p^\nu/m^2 \big]
      \big(\delta_a^b\big)^{\odot 2s} \\ &
    + \hbar^2 s(2s-1)\:\!\delta_s^{s'}
      \sigma_{p~a}^{\mu,}{}^b \odot \sigma_{p~a}^{\nu,}{}^b
      \odot \big(\delta_a^b\big)^{\odot (2s-2)} .
\label{AngularMomentumAnticommutation4d}
\end{aligned}
\]
A straightforward extension of the coherent-state discussion
in \sec{sec:CoherentSpinStates} allows for
a well-defined classical limit for spin
\begin{subequations} \begin{align}
\label{SpinExpectation4d}
   \braket{S_p^\mu}_\alpha &
    = \frac{\hbar}{2} (\tilde{\alpha}\:\!\sigma_p^\mu \alpha)
    ~\xrightarrow[\hbar\to0]{}~ s_\text{cl}^\mu , \qquad \qquad~\,\quad
   p \cdot s_\text{cl} = 0 , \\
\label{SpinExpectation4d2}
   \braket{S_p^\mu S_p^\nu}_\alpha &
    = \braket{S_p^\mu}_\alpha \braket{S_p^\nu}_\alpha + {\cal O}(\hbar)
    ~\xrightarrow[\hbar\to0]{}~ s_\text{cl}^\mu s_\text{cl}^\nu ,
   \qquad \quad
   \text{\etc}
\end{align} \label{SpinExpectations4d}%
\end{subequations}
%\begin{align}
%\label{SpinExpectation4d}
%   \braket{S^\mu}_\alpha &
%    = \frac{\hbar}{2} (\tilde{\alpha}\:\!\sigma_p^\mu \alpha)
%    ~\xrightarrow[\hbar\to0]{}~ s_\text{cl}^\mu , \qquad \quad
%   p \cdot s_\text{cl} = 0 , \\
%   \braket{S_p^\mu S_p^\nu}_\alpha &
%    = \braket{S_p^\mu}_\alpha \braket{S_p^\nu}_\alpha
%    + \frac{\hbar^2}{4}
%      \bigg\{ {-}\bigg[ \eta^{\mu\nu} - \frac{p^\mu p^\nu}{m^2} \bigg]
%      (\tilde{\alpha} \alpha)
%    + \frac{i}{m} \epsilon^{\mu\nu\rho\sigma} p_\rho
%      (\tilde{\alpha}\:\! \sigma_{p\:\!\sigma} \alpha)
%      \bigg\}
%    ~\xrightarrow[\hbar\to0]{}~ s_\text{cl}^\mu s_\text{cl}^\nu  . \nn
%\end{align}

%%%%%%%%%%%%%%%%%%%%%%%%%%%%%%%%%%%%%%%%%%%%%%%%%%
\subsection{Classical observables}
\label{sec:ClassicalObservables}
%%%%%%%%%%%%%%%%%%%%%%%%%%%%%%%%%%%%%%%%%%%%%%%%%%

We are now in position to write an initial state for two massive objects
``${\rm a}$'' and ``${\rm b}$'' with definite classical linear and angular momenta:
\begin{align}
\label{InitialStateSpin}
 & \ket{\text{in}} = \int_{p_1} \int_{p_2}
      \psi_{\rm a}(p_1) \psi_{\rm b}(p_2)
      e^{i b\cdot p_1/\hbar} \ket{p_1,\alpha;p_2,\beta} \\ &
    = e^{-(\|\alpha\|^2 + \|\beta\|^2)/2}
      \sum_{s_1,s_2} \int_{p_1,p_2}\!\!\!\!
      e^{i b\cdot p_1/\hbar} \psi_{\rm a}(p_1) \psi_{\rm b}(p_2)
      \frac{ (\alpha^a)^{\odot 2s_1} (\beta^b)^{\odot 2s_2}\!}
           { \sqrt{(2s_1)!(2s_2)!} }
      \cdot \ket{p_1,s_1,\{a\};p_2,s_2,\{b\}} . \nn
\end{align}
Here the ${\rm SU}(2)$ spinors $\alpha$ and $\beta$
transform in the little groups of $p_1$ and $p_2$, respectively,
which are integrated over.
However, let us recall that their main purpose is to define
classical spin vectors $s_{\rm a}^\mu$ and $s_{\rm b}^\mu$
in the sense of \eqn{SpinExpectation4d},
which in presence of the momentum-wavefunction integration should be upgraded to
\[
   \braket{S^\mu}_{\xi,\alpha}
    = \frac{\hbar}{2} \int_p\!|\psi_{\xi}(p)|^2
      \tilde{\alpha}(p) \sigma_p^\mu \alpha(p)
      ~\xrightarrow[\hbar\to0]{}~ s_\text{cl}^\mu
      \equiv \lim_{\hbar\to0}
   \frac{\hbar}{2} (\tilde{\alpha}(u) \sigma_u^\mu \alpha(u)) ,
\label{SpinClassicalLimit}
\]
such that $u \cdot s_\text{cl} = 0$.
Therefore, in the context of computing classical observables,
we only need to consider the spinors $\alpha$ and $\beta$
which depend on the momenta in a unitary fashion ---
exclusively to account for the misalignment
between the little-group representations of $p^\mu$ and~$u^\mu$, namely
\[
   \alpha^a(p_1) = U^a{}_c(p_1/m_{\rm a},u_{\rm a})\:\!
   \alpha^c(u_{\rm a}) ,
\label{SU2SpinorTransform}
\]
and likewise for $\beta(p_2)$.
This is why in \eqn{InitialStateSpin} we allowed their ${\rm SU}(2)$-invariant norms to be pulled outside of the momentum integration.
We leave the residual momentum dependence of $\alpha$ and $\beta$ implicit, until we need to specify it further.

Let us now return to the observable $\Delta O$,
which we split into two parts:
\begin{align}\!\!\!
   \Delta_1 O &
   =\;\!\!\!\int_{p_1',p_2',p_1,p_2}\!\!\!\!\!e^{-i k \cdot b/\hbar}
      \psi_{\rm a}^*(p_1') \psi_{\rm b}^*(p_2')
      \psi_{\rm a}(p_1) \psi_{\rm b}(p_2)
      i\bra{p_1',\alpha;p_2',\beta} [O T\!-\!T^\dagger O]
      \ket{p_1,\alpha;p_2,\beta} , \nn \\\!\!\!
   \Delta_2 O &
   =\;\!\!\!\int_{p_1',p_2',p_1,p_2}\!\!\!\!\!e^{-i k \cdot b/\hbar}
      \psi_{\rm a}^*(p_1') \psi_{\rm b}^*(p_2')
      \psi_{\rm a}(p_1) \psi_{\rm b}(p_2)
      \bra{p_1',\alpha;p_2',\beta} T^\dagger O T
      \ket{p_1,\alpha;p_2,\beta} ,\!
\label{KMOC1}
\end{align}
where $k^\mu = p_1'^\mu - p_1^\mu$ is a momentum mismatch.
In the present context of relativistic spinning objects,
$O$ should be thought of as either a momentum or spin operator.
Depending on whether it is the latter or the former (or a function thereof),
we may need or not need to further expand the coherent spin states
in terms of definite spins.

%%%%%%%%%%%%%%%%%%%%%%%%%%%%%%%%%%%%%%%%%%%%%%%%%%
\subsubsection{Impulse formulae}
\label{sec:Impulse}
%%%%%%%%%%%%%%%%%%%%%%%%%%%%%%%%%%%%%%%%%%%%%%%%%%

Let us consider the more involved case of the angular impulse observable.
The contribution linear in the scattering transition operator $T$ is
\[
\begin{aligned}\!\!\!\!
   \Delta_1 S_{\rm a}^\mu
    = e^{-\|\alpha\|^2} \sum_{s_1,s_1'}
      \int_{p_1',p_2',p_1,p_2}\!\!e^{-i k \cdot b/\hbar}
      \psi_{\rm a}^*(p_1') \psi_{\rm b}^*(p_2')
      \psi_{\rm a}(p_1) \psi_{\rm b}(p_2)
      \frac{ (\tilde{\alpha}_{a'}\!)^{\odot 2s_1'} (\alpha^a)^{\odot 2s_1}\!}
           { \sqrt{(2s_1')!(2s_1)!} } & \\ \cdot\,
      i\bra{p_1',s_1',\{a'\};p_2',\beta}
      [S_{\rm a}^\mu T - T^\dagger S_{\rm a}^\mu]
      \ket{p_1,s_1,\{a\};p_2,\beta} & .\!
\label{ImpulseSpin1}
\end{aligned}
\]
Here and below, we will make use of
the completeness relation in the Hilbert subspace
involving at least the two massive particles ${\rm a}$ and ${\rm b}$,\footnote{The basic one-particle completeness relation
for coherent spin states is
\begin{equation*}
   \int\!\frac{d^4 \alpha}{\pi^2} \ket{\alpha} \bra{\alpha}
    = \sum_{2s=0}^\infty \ket{s,\{a\}} \cdot \bra{s,\{a\}}
    = \sum_{2s=0}^\infty\,\sum_{s_z=-s}^s \ket{s,s_z} \bra{s,s_z} = 1 .
%    = \int_0^\infty\!\!\!\|\alpha\|^3 d\|\alpha\|
%      \int\!\frac{d^3 U}{\pi^2} \ket{\alpha} \bra{\alpha}
%      \bigg|_{\alpha^a = U^a{}_b (\|\alpha\|,0)^b } .
%\label{SpinStatesCompleteness}
\end{equation*}
\label{foot:SpinStatesCompleteness}
}
\[
   1_{n\geq2} =\!\int_{p_1,p_2} \sum_{s_1} 
      \int\!\frac{d^4\beta}{\pi^2} \sum_X
      \ket{p_1,s_1,\{a\}; p_2,\beta; X} \cdot
      \bra{p_1,s_1,\{a\}; p_2,\beta; X} .
\label{HilbertCompletenessRelation}
\]
Here the dot is again used as a shorthand for the contraction
of the little-group indices $\{a_1,\ldots,a_{s_1}\}$.
Of course, the spin operator $S_{\rm a}^\mu$
does not change the number of particles
and is diagonal in $p_1$ and $s_1$.
Therefore, in the case of \eqn{ImpulseSpin1}
the matrix elements of the quantum operators in the second line become
\begin{align}
\label{ImpulseSpin1MatrixElement}
 & \bra{p_1',s_1',\{a'\}; p_2',\beta}
      [S_{\rm a}^\mu T - T^\dagger S_{\rm a}^\mu]
      \ket{p_1,s_1,\{a\};p_2,\beta} \\ &
%   =\!\int_{r_1,r_2} \sum_{s_1''}
%      \int\!\frac{d^4\beta'}{\pi^2} \nn \\ & \quad \times
%      \Big[
%      \bra{p_1',s_1',\{a'\};p_2',\beta} S_{\rm a}^\mu
%      \ket{r_1,s_1'',\{c\};r_2,\beta'} \cdot
%      \bra{r_1,s_1'',\{c\};r_2,\beta'} T
%      \ket{p_1,s_1,\{a\};p_2,\beta} \nn \\ &~\:\quad
%    - \bra{p_1',s_1',\{a'\};p_2',\beta} T^\dagger
%      \ket{r_1,s_1'',\{c\};r_2,\beta'} \cdot
%      \bra{r_1,s_1'',\{c\};r_2,\beta'} S_{\rm a}^\mu
%      \ket{p_1,s_1,\{a\};p_2,\beta}
%      \Big] \nn \\ &
   =\!\int_{r_1,r_2} \sum_{s_1''}
      \bigg[ \delta_{p_1'-r_1} \delta_{p_2'-r_2}
      \big(S_{p_1'}^\mu\big)^{s_1',\{a'\}}{}_{s_1'',\{c\}} \cdot
      \bra{r_1,s_1'',\{c\};r_2,\beta} T
      \ket{p_1,s_1,\{a\};p_2,\beta} \nn \\ & \qquad\:\:\qquad
    - \bra{p_1',s_1',\{a'\};p_2',\beta} T^\dagger
      \ket{r_1,s_1'',\{c\};r_2,\beta} \cdot
      \big(S_{p_1}^\mu\big)^{s_1'',\{c\}}{}_{s_1,\{a\}}
      \delta_{r_1-p_1} \delta_{r_2-p_2}
      \bigg] \nn \\ &
    = (2\pi)^4 \delta^{(4)}(p_1+p_2-p_1'-p_2') \bigg[
      \sum_{c_1,\dots,c_{s_1'}\!}\!\!
      \big(S_{p_1'}^\mu\big)^{\{a'\}}{}_{\{c\}}\,
      {\cal A}^{\{c\}}{}_{\{a\}}(p_1',s_1';p_2',\beta|p_1,s_1;p_2,\beta)
      \nn \\ & \qquad \qquad \qquad \qquad \qquad~~\:\qquad
   -\!\sum_{c_1,\dots,c_{s_1}\!}\!\!
%      {\cal A}^{\{a'\}}{}_{\{c\}}(p_1',s_1';p_2',\beta|p_1,s_1;p_2,\beta)\,
      \big({\cal A}^*(p_1,s_1;p_2,\beta|p_1',s_1';p_2',\beta)
      \big)_{\{c\}}{}^{\{a'\}}
      \big(S_{p_1}^\mu\big)^{\{c\}}{}_{\{a\}}
      \bigg] , \nn
\end{align}
where $\delta_{p-p'}$ is a shorthand for the on-shell delta function
$2p^0 (2\pi)^3\delta^{(3)}(\bs{p}-\bs{p}')$.
We have thus converted the transition matrix elements
to scattering amplitudes, in which we write the outgoing particles first
and the incoming particles after the vertical line ---
so as to mimic the structure of the matrix elements
and preserve consistency with the placement of the ${\rm SU}(2)$ indices.
The indices of the complex-conjugated amplitude in the last line
are dualized but still written in the ``out-in'' order.
The spin operators
in the definite-spin representation~\eqref{SpinRepresentation4d}
have also been reduced with respect to the total-spin quantum number
in a natural way:
$(S_p^\mu)^{\{a\}}{}_{\{b\}} \equiv (S_p^\mu)^{s,\{a\}}{}_{s'=s,\{b\}}$.
Note that in the explicit summations
in the last two lines of \eqn{ImpulseSpin1MatrixElement}
the number of the contracted ${\rm SU}(2)$ indices are different,
and they correspond to distinct little groups.

After integrating the momentum-conservation delta function
\[
\begin{aligned}
 & \int_{p_1',p_2'}\!\!(2\pi)^4 \delta^{(4)}(p_1+p_2-p_1'-p_2') \\ & \qquad
    = \int\!\!\frac{d^4 k}{(2\pi)^2} \Theta(p_1^0+k^0) \Theta(p_2^0-k^0) 
      \delta(2p_1\!\cdot k +k^2) \delta(2p_2\!\cdot k - k^2) ,
\label{Fourier2d}
\end{aligned}
\]
and replacing $p_1'=p_1+k$ and $p_2'=p_2-k$ in the rest of the integrand,
we obtain
\begin{align}\!\!\!
   \Delta_1 S_{\rm a}^\mu &
    = e^{-\|\alpha\|^2} \sum_{s_1,s_1'}
      \int_{p_1,p_2} \int_k\!e^{-i k \cdot b/\hbar}
      \psi_{\rm a}^*(p_1+k) \psi_{\rm b}^*(p_2-k)
      \psi_{\rm a}(p_1) \psi_{\rm b}(p_2)
      \frac{ (\tilde{\alpha}_{a'}\!)^{\odot 2s_1'} (\alpha^a)^{\odot 2s_1}\!}
           { \sqrt{(2s_1')!(2s_1)!} }\!\!\nn \\ & \qquad \qquad \cdot
      \Big[
      \big(S_{p_1+k}^\mu\big)^{\{a'\}}{}_{\{c'\}} \cdot
      i{\cal A}^{\{c'\}}{}_{\{a\}}(p_1\!+\!k,s_1';p_2\!-\!k,\beta
      |p_1,s_1;p_2,\beta) \nn \\ & \qquad \qquad \quad
    - i\big({\cal A}^*(p_1,s_1;p_2,\beta|p_1\!+\!k,s_1';p_2\!-\!k,\beta)
       \big)_{\{c\}}{}^{\{a'\}} \cdot
      \big(S_{p_1}^\mu\big)^{\{c\}}{}_{\{a\}}
      \Big] ,\!
\label{ImpulseSpin1Quantum}
\end{align}
where $\int_k$ is a shorthand for
the two-dimensional integration measure in \eqn{Fourier2d}.

After similar manipulations involving the completeness relation
\eqref{HilbertCompletenessRelation},
the second angular-impulse contribution
\[
\begin{aligned}\!\!\!
   \Delta_2 S_{\rm a}^\mu
    = e^{-\|\alpha\|^2} \sum_{s_1,s_1'}
      \int_{p_1',p_2',p_1,p_2}\!\!e^{-i k \cdot b/\hbar}
      \psi_{\rm a}^*(p_1') \psi_{\rm b}^*(p_2')
      \psi_{\rm a}(p_1) \psi_{\rm b}(p_2)
      \frac{ (\tilde{\alpha}_a')^{\odot 2s_1'} (\alpha^a)^{\odot 2s_1}\!}
           { \sqrt{(2s_1')!(2s_1)!} } & \\ \cdot\,
      \bra{p_1',s_1',\{a'\};p_2',\beta} T^\dagger S_{\rm a}^\mu T
      \ket{p_1,s_1,\{a\};p_2,\beta} & ,\!
\label{ImpulseSpin2}
\end{aligned}
\]
which is quadratic in the transition operator $T$, may be rewritten as
\begin{align}\!\!\!
 & \Delta_2 S_{\rm a}^\mu
    = e^{-\|\alpha\|^2} \sum_{s_1,s_1'}
      \int_{p_1,p_2} \int_k\!e^{-i k \cdot b/\hbar}
      \psi_{\rm a}^*(p_1+k) \psi_{\rm b}^*(p_2-k)
      \psi_{\rm a}(p_1) \psi_{\rm b}(p_2)
      \frac{ (\tilde{\alpha}_{a'}\!)^{\odot 2s_1'} (\alpha^a)^{\odot 2s_1}\!}
           { \sqrt{(2s_1')!(2s_1)!} }\!\!\nn \\ &\!\cdot\!
      \sum_{s_1'',s_2''} \int\!\frac{d^4 w_1 d^4 w_2}{(2\pi)^2}
      \Theta(p_1^0+w_1^0) \Theta(p_2^0+w_2^0)
      \delta(2p_1\!\cdot w_1 + w_1^2)
      \delta(2p_2\!\cdot w_2 + w_2^2) \nn \\ &~\times
      \sum_X \delta^{(4)}(w_1+w_2+p_X)
      \big({\cal A}^*(p_1\!+\!w_1,s_1'';p_2\!+\!w_2,s_2'',\{b\};X
                     |p_1\!+\!k,s_1';p_2\!-\!k,\beta)
      \big)_{\{c\}}{}^{\{a'\}} \nn \\ & \qquad\:\,\,\cdot
      \big(S_{p_1+w_1}^\mu\big)^{\{c\}}{}_{\{e\}} \cdot
      {\cal A}^{\{e\}}{}_{\{a\}}{}(p_1\!+\!w_1,s_1'';
         p_2\!+\!w_2,s_2'',\{b\};X|p_1,s_1;p_2,\beta) .
\label{ImpulseSpin2Quantum}
\end{align}

Analogously, the momentum-change contributions
$\Delta_1 P_{\rm a}^\mu$ and $\Delta_2 P_{\rm a}^\mu$
together give
\begin{align}
\label{ImpulseMomQuantum}
 & \Delta P_{\rm a}^\mu
    = \int_{p_1,p_2} \int_k\!e^{-i k \cdot b/\hbar}
      \psi_{\rm a}^*(p_1+k) \psi_{\rm b}^*(p_2-k)
      \psi_{\rm a}(p_1) \psi_{\rm b}(p_2) \\ &\!\!\times\!\bigg\{
      (p_1\!+\!k)^\mu
      i{\cal A}(p_1\!+\!k,\alpha;p_2\!-\!k,\beta|p_1,\alpha;p_2,\beta)
    - p_1^\mu
      i{\cal A}^*(p_1,\alpha;p_2,\beta|p_1\!+\!k,\alpha;p_2\!-\!k,\beta)
      \nn \\ &\:\,
   +\!\sum_{s_1'',s_2''} \int\!\frac{d^4 w_1 d^4 w_2}{(2\pi)^2}
      \Theta(p_1^0+w_1^0) \Theta(p_2^0+w_2^0)
      \delta(2p_1\!\cdot w_1 + w_1^2)
      \delta(2p_2\!\cdot w_2 + w_2^2) \nn \\ &\,\,\quad \times\!
      \sum_X\,\delta^{(4)}(w_1\!+\!w_2\!+\!p_X)\,
      {\cal A}^*(p_1\!+\!w_1,s_1'',\{a\};p_2\!+\!w_2,s_2'',\{b\};X
                     |p_1\!+\!k,\alpha;p_2\!-\!k,\beta)
       \nn \\ & \qquad \qquad \qquad \qquad \qquad\:\cdot
      (p_1\!+\!w_1)^\mu
      {\cal A}(p_1\!+\!w_1,s_1'',\{a\};
               p_2\!+\!w_2,s_2'',\{b\};X|p_1,\alpha;p_2,\beta) \bigg\} , \nn
\end{align}
where the spin degrees of freedom merely play a spectator role,
as they are not affected by the momentum operator.

%%%%%%%%%%%%%%%%%%%%%%%%%%%%%%%%%%%%%%%%%%%%%%%%%%
\subsubsection{$\hbar$ power-counting}
\label{sec:hbarCounting}
%%%%%%%%%%%%%%%%%%%%%%%%%%%%%%%%%%%%%%%%%%%%%%%%%%

The impulse formulae above hold in fully quantum field theory.
Let us now analyze the powers of $\hbar$ to determine
the simplifications due to the classical limit.

Formally, the classical limit should be defined
in terms of dimensionless quantities;
namely, for each particle we take
\begin{subequations}
\[
   \xi \approx \frac{2\sigma_p^2}{3m^2} ~\to~ 0 , \qquad \quad
   \|\alpha\|^2 = \frac{2}{\hbar} \sqrt{-s_\text{cl}^2} ~\to~ \infty .
\label{ClassicalLimitParticles}
\]
These conditions on the incoming objects, however, are not sufficient
to guarantee that the scattering outcome will be classically calculable.
To rule out the possibility of head-on or deeply inelastic collisions,
which tend to heavily depend on the internal structure of the projectiles,
we must additionally impose that the impact parameter be much larger
than the wavepacket spreads
(which we estimate via Heisenberg's uncertainty principle):
\[
   |b| \equiv \sqrt{-b^2} ~\gg~ (\sigma_x)_{\rm a,b}
       \geq \frac{\hbar}{2(\sigma_p)_{\rm a,b}}
       \propto \frac{\hbar}{\sqrt{\xi}\:\!m_{\rm a,b}} .
\label{ClassicalLimitImpact}
\]
In other words, we require that
$|b| \gg \sigma_x \gg (\lambda_\text{Compton})_{\rm a,b}$
at the same time,
where of course $\lambda_\text{Compton} \equiv 2\pi\hbar/m$.
Due to the Fourier transformation to momentum space
via $e^{-i k \cdot b/\hbar}$, this naturally translates to
\[
   |k| \equiv \sqrt{-k^2} ~\ll~ (\sigma_p)_{\rm a,b}
%       \equiv \sqrt{3\xi/2}\:\!m_{\rm a,b} 
       \propto \sqrt{\xi}\:\!m_{\rm a,b}  .
\label{ClassicalLimitMomentum}
\]
\label{ClassicalLimit}%
\end{subequations}
Indeed, outside of this classically relevant region of small $|k|$
the Fourier integral becomes highly oscillatory at least for some $b$.
The condition \eqref{ClassicalLimitMomentum} is further improved
in the longitudinal directions, as both copies of the momentum wavefunction,
$\psi_{\rm a}(p_1)$ and $\psi_{\rm a}^*(p_1+k)$, become sharply peaked
around the same classical momentum~$m_{\rm a} u_{\rm a}^\mu$
(and likewise for particle ${\rm b}$),
thus constraining $k \cdot u_{\rm a,b} \ll \xi m_{\rm a,b} \ll (\sigma_p)_{\rm a,b} $.

Perhaps a simpler way to keep track of the above limits is offered
by the heuristic classical limit $\hbar \to 0$,
for which we adopt the following rules:
\[
   \sigma_x, \sigma_p \propto \hbar^{1/2} , \qquad
   \xi \propto \hbar , \qquad
   \|\alpha\| \propto \hbar^{-1/2} , \qquad
   |k| \propto \hbar , \qquad
   k \cdot u_{\rm a,b} \propto \hbar^{3/2} .
\label{ClassicalLimitSummary}
\]
The remaining quantities,
such as $m_{\rm a,b}$, $u_{\rm a,b}^\mu$ and $s_{\rm a,b}^\mu$,
remain classically meaningful.
In fact, all force-carrier momenta inside scattering amplitudes
should be thought of
in terms of their wavenumbers $\bar{k}^\mu \equiv k^\mu/\hbar$,
as argued on unitarity grounds in \rcite{Kosower:2018adc}.

%%%%%%%%%%%%%%%%%%%%%%%%%%%%%%%%%%%%%%%%%%%%%%%%%%
\subsubsection{Leading classical impulse}
\label{sec:LOImpulse}
%%%%%%%%%%%%%%%%%%%%%%%%%%%%%%%%%%%%%%%%%%%%%%%%%%

Here we wish to focus on the classical observables
at leading order in the coupling constant.
Both in gauge theory and gravity, it is convenient
to absorb factors of $1/\sqrt{\hbar}$
into the coupling constants \cite{Kosower:2018adc},
namely
\[
   \sqrt{\alpha_e} = e/\sqrt{4\pi\hbar} , \qquad \quad
   \kappa = \sqrt{32\pi G/\hbar} .
\label{CouplingConstants}
\]
Indeed, since we choose to work with momentum-space amplitudes
of mass dimension $M^{4-n}$,
the natural expansion parameters for them must have a pure mass dimension,
such as $0$ for the fine-structure constant~$\alpha_e$
and $-1$ for the gravitational coupling~$\kappa$ defined above.
The need for these additional powers of $\hbar$ arises from the mismatch
between the dimensions of momenta and coordinate derivatives.
(We still keep $c=1$.)

At leading order, unitarity restricts the amplitudes
to be hermitian via the identity $T^\dagger = T - iT^\dagger T \approx T$.
Hence we have tree-level conjugation rules such as
\[
   {\cal A}^{(0)\{a'\}}{}_{\{a\}}(p_1',s_1';p_2',\beta'
      |p_1,s_1;p_2,\beta)
    = \big({\cal A}^{(0)*}(p_1,s_1;p_2,\beta|p_1',s_1';p_2',\beta')
       \big)_{\{a\}}{}^{\{a'\}} .
\]
Therefore, the linear- and angular-impulse formulae~\eqref{ImpulseMomQuantum}
and~\eqref{ImpulseSpin1Quantum} simplify to
\begin{subequations} \begin{align}\!\!
\label{ImpulseMomLO}
   \Delta P_{\rm a}^\mu &
    = i\!\int_{p_1,p_2} \int_k\!e^{-i \bar{k} \cdot b}
      |\psi_{\rm a}(p_1)|^2 |\psi_{\rm b}(p_2)|^2\,k^\mu
      {\cal A}^{(0)}(p_1\!+\!k,\alpha;p_2\!-\!k,\beta
                    |p_1,\alpha;p_2,\beta) ,\!\\\!\!
\label{ImpulseSpinLO}
   \Delta S_{\rm a}^\mu &
    = ie^{-\|\alpha\|^2} \sum_{s_1,s_1'}
      \int_{p_1,p_2} \int_k\!e^{-i \bar{k} \cdot b}
      |\psi_{\rm a}(p_1)|^2 |\psi_{\rm b}(p_2)|^2
      \frac{ (\tilde{\alpha}_{a'}\!)^{\odot 2s_1'} (\alpha^a)^{\odot 2s_1}\!}
           { \sqrt{(2s_1')!(2s_1)!} }\!\!\nn \\ & \qquad~\qquad \cdot
      \Big[
      \big(S_{p_1+k}^\mu\big)^{\{a'\}}{}_{\{c'\}} \cdot
      {\cal A}^{(0)\{c'\}}{}_{\{a\}}(p_1\!+\!k,s_1';p_2\!-\!k,\beta
      |p_1,s_1;p_2,\beta) \\ & \qquad \qquad~\quad
    - {\cal A}^{(0)\{a'\}}{}_{\{c\}}(p_1\!+\!k,s_1';p_2\!-\!k,\beta
      |p_1,s_1;p_2,\beta) \cdot
      \big(S_{p_1}^\mu\big)^{\{c\}}{}_{\{a\}}
      \Big] , \nn
\end{align} \label{ImpulsesLO}%
\end{subequations}
where we have also neglected the shifts by $\pm k = {\cal O}(\hbar)$
in the momentum wavepackets.

In fact, in these leading-order equations
$\hbar$ may be set to zero everywhere but in the denominators,
for instance, in
$\delta(2p_1\!\cdot k + k^2) = \delta(2p_1\!\cdot \bar{k})/\hbar + {\cal O}(h^0)$.
This is because the leading contribution in $\hbar$
will end up ${\cal O}(\hbar^0)$ by itself
--- which is not the case at higher orders,
where the leading contributions develop poles in $\hbar$
that cancel only after summing multiple contributions,
and it is the subleading terms ${\cal O}(\hbar^0)$
that give the classical observables
\cite{Kosower:2018adc}.
In view of this, and keeping in mind that
in the KMOC formalism both $p_1^\mu$ and $p_1'^\mu=p_1^\mu+k^\mu$
correspond to the initial-state momentum $m_{\rm a} u_{\rm a}^\mu$,
we propose to treat them democratically 
\[
   p_{\rm a}^\mu = (p_1^\mu + p_1'^\mu)/2 = p_1^\mu + k^\mu/2 , \qquad \quad
   p_{\rm b}^\mu = (p_2^\mu + p_2'^\mu)/2 = p_2^\mu - k^\mu/2 ,
\]
such that the wavefunctions~\eqref{MomentumWavefunction}
satisfy the following exact identities
\[
   \psi_{\rm a}^*(p_1+k) \psi_{\rm a}(p_1)
    = |\psi_{\rm a}(p_{\rm a})|^2 , \qquad \quad
   \psi_{\rm b}^*(p_2-k) \psi_{\rm b}(p_2)
    = |\psi_{\rm b}(p_{\rm b})|^2 .
\]
The overall integration measure can then be expressed as
\begin{align}
 & \int_{p_1,p_2} \int_k
%    = \int\!\frac{d^4 p_1 d^4 p_2 d^4 k}{(2\pi)^8}
%      \Theta(p_1^0) \Theta(p_2^0) \nn
%      \Theta(p_1^0+k^0) \Theta(p_2^0-k^0) \\ & \qquad \qquad \quad \times 
%      \delta(p_1^2-m_{\rm a}^2) \delta(p_2^2-m_{\rm b}^2)
%      \delta(2p_1\!\cdot k +k^2) \delta(2p_2\!\cdot k - k^2) \nn \\ &
    = \int\!\frac{d^4 p_{\rm a} d^4 p_{\rm b} d^4 k}{(2\pi)^8}
      \Theta(p_{\rm a}^0-k^0\!/2) \Theta(p_{\rm a}^0+k^0\!/2)
      \Theta(p_{\rm b}^0-k^0\!/2) \Theta(p_{\rm b}^0+k^0\!/2)
      \nn \\ & \qquad \qquad \quad~~\times 
      \delta(p_{\rm a}^2-m_{\rm a}^2+k^2\!/4)
      \delta(p_{\rm b}^2-m_{\rm b}^2+k^2\!/4)
      \delta(2p_{\rm a}\!\cdot k) \delta(2p_{\rm b}\!\cdot k) ,
\end{align}
where in the classically relevant region $|k| = {\cal O}(\hbar)$
the four theta functions amount to simply
$\Theta(p_{\rm a}^0) \Theta(p_{\rm b}^0)$.
We also see that the masses of $p_{\rm a}$ and $p_{\rm b}$
are both shifted by $-k^2/4={\cal O}(\hbar^2)$.
Since these shifts may be safely ignored in the classical limit,
the only important $k$ dependence in the measure
remains in the transversality delta functions.
Therefore, we are allowed to make the integration-measure replacement
\[
   \int_{p_1,p_2} \int_k ~\to~
   \int_{p_{\rm a},p_{\rm b}} \int_k , \qquad \quad
   \int_k \equiv \int\!\frac{d^4 k}{(2\pi)^2}
      \delta(2p_{\rm a}\!\cdot k) \delta(2p_{\rm b}\!\cdot k) ,
\label{EikonalMeasure}
\]
where $\int_k$ is now the standard eikonal measure,
while $\int_{p_{\rm a}}$ and $\int_{p_{\rm b}}$ are defined
just as in \eqn{InitialStateMomentum}
without any additional reference to $k$.

Furthermore, the LO impulse formula~\eqref{ImpulseMomLO} may be rewritten
in terms of a partial derivative in the impact parameter:
\[
\begin{aligned}
   \Delta P_{\rm a}^\mu
    = -\hbar \frac{\partial~}{\partial b_\mu\!} &
      \int_{p_{\rm a},p_{\rm b}}\!\!\!
      |\psi_{\rm a}(p_{\rm a})|^2 |\psi_{\rm b}(p_{\rm b})|^2 \\ \times\! &
      \int_k e^{-i \bar{k} \cdot b}
      {\cal A}^{(0)}(p_{\rm a}\!+\!k/2,\alpha;p_{\rm b}\!-\!k/2,\beta
                    |p_{\rm a}\!-\!k/2,\alpha;p_{\rm b}\!+\!k/2,\beta) .
\label{ImpulseMomLOFinal}
\end{aligned}
\]
This seemingly four-dimensional derivative should be understood
in the two-dimensional sense within the transverse subspace
\[
   {\rm E}^\perp_{p_{\rm a},p_{\rm b}} \equiv
   \{ x \in \mathbb{R}^4: x \cdot p_{\rm a} = x \cdot p_{\rm b} = 0 \} .
\label{TransverseSubspace}
\]
A simple way to enforce this is to additionally contract
$\partial/\partial b_\mu$ with the transverse projector
\[
   \Pi^\mu{}_\nu(p_{\rm a},p_{\rm b})
    = \frac{\epsilon^{\mu\rho\alpha\beta} \epsilon_{\nu\rho\gamma\delta}}
           {(p_{\rm a} \cdot p_{\rm b})^2-m_{\rm a}^2 m_{\rm b}^2}
           p_{{\rm a}\:\!\alpha} p_{{\rm b}\:\!\beta}
           p_{\rm a}^\gamma p_{\rm b}^\delta ,
\label{TransverseProjector}
\]
which we leave implicit for the time being.
Note that at this order there is no need to distinguish
${\rm E}^\perp_{p_{\rm a},p_{\rm b}} \ni k$ and
${\rm E}^\perp_{u_{\rm a},u_{\rm b}} \ni b$.

%%%%%%%%%%%%%%%%%%%%%%%%%%%%%%%%%%%%%%%%%%%%%%%%%%
\subsubsection{Leading classical spin kick}
\label{sec:LOSpinKick}
%%%%%%%%%%%%%%%%%%%%%%%%%%%%%%%%%%%%%%%%%%%%%%%%%%

In order to understand the classical limit
of the angular impulse \eqref{ImpulseSpinLO} more closely,
we need to simplify its ${\rm SU}(2)$-index structure
(which currently refers to multiple little groups)
by boosting the spin operators to the same reference momentum.
In view of the tensor-product structure~\eqref{SpinRepresentation4d}
of these operators,
it is only the spin-1/2 operator~\eqref{DiracSpinOperator}
that is boosted non-trivially:
\[
%   \sigma_{p_1+k}^\mu{}^a{}_b = U^a{}_c(p_1\!+\!k,p_1)
%      \bigg[ \sigma_{p_1}^\mu
%           - \frac{\hbar}{m_{\rm a}^2} p_1^\mu \bar{k}^\nu
%             \sigma_{p_1\:\!\mu} + {\cal O}(\hbar)
%      \bigg]^c_{~d} U^d{}_b(p_1,p_1\!+\!k)
   \sigma_{p_{\rm a} \pm k/2}^\mu{}^a{}_b
    = U^a{}_c(p_{\rm a}\!\pm\!k/2,p_{\rm a})
      \bigg[ \sigma_{p_{\rm a}}^\mu
           \mp \frac{\hbar}{2m_{\rm a}^2} p_{\rm a}^\mu \bar{k}^\nu
             \sigma_{p_{\rm a}\:\!\mu} + {\cal O}(\hbar)
      \bigg]^c_{~d} U^d{}_b(p_{\rm a},p_{\rm a}\!\pm\!k/2) ,
\label{SpinBoostLinear}
\]
see \rcites{Maybee:2019jus,Guevara:2019fsj} or \sec{sec:MinimalCoupling} below
for more details.
The ${\rm SU}(2)$ transformations here satisfy
\[
   U^a{}_c(p_{\rm a}\!\pm\!k/2,p_{\rm a})
   U^c{}_b(p_{\rm a},p_{\rm a}\!\pm\!k/2) = \delta^a_b
\]
and have the same nature as those introduced in \eqn{SU2SpinorTransform}
for the coherent-state spinors.
We may therefore identify
\[
\begin{aligned}
   \tilde{\alpha}_{a'}(p_{\rm a}\!+\!k/2)
   U^a{}_c(p_{\rm a}\!+\!k/2,p_{\rm a}) &
    = \tilde{\alpha}_c(p_{\rm a}) , \\
   U^c{}_a(p_{\rm a},p_{\rm a}\!-\!k/2) \alpha^a(p_{\rm a}\!-\!k/2) &
    = \alpha^c(p_{\rm a}) .
\end{aligned}
\]
This allows us to rewrite the leading angular impulse~\eqref{ImpulseSpinLO}
with all ${\rm SU}(2)$ indices associated with the little group
of the same momentum $p_{\rm a}$:
\begin{align}
\label{ImpulseSpinLO2}
 & \Delta S_{\rm a}^\mu
    = ie^{-\|\alpha\|^2} \sum_{s_1,s_1'}
      \int_{p_{\rm a},p_{\rm b}}\!\!\!
      |\psi_{\rm a}(p_{\rm a})|^2 |\psi_{\rm b}(p_{\rm b})|^2
      \int_k
      \frac{ (\tilde{\alpha}_{a'}(p_{\rm a}))^{\odot 2s_1'}
             (\alpha^a(p_{\rm a}))^{\odot 2s_1}\!}
           { \sqrt{(2s_1')!(2s_1)!} } \\ &\!\cdot\!
      \Bigg\{
      \bigg[ S_{p_{\rm a}}^\mu\!
           - \frac{p_{\rm a}^\mu k_\nu}{2m_{\rm a}^2} S_{p_{\rm a}}^\nu
      \bigg]^{\{a'\}}_{~~~\,\{c'\}}\!\cdot
      \tilde{\cal A}^{(0)\{c'\}}{}_{\{a\}}(p_{\rm a}\!+\!k/2,s_1';
      p_{\rm b}\!-\!k/2,\beta|p_{\rm a}\!-\!k/2,s_1;p_{\rm b}\!+\!k/2,\beta)
      \nn \\ &\;\!
    - \tilde{\cal A}^{(0)\{a'\}}{}_{\{c\}}(p_{\rm a}\!+\!k/2,s_1';
      p_{\rm b}\!-\!k/2,\beta|p_{\rm a}\!-\!k/2,s_1;
      p_{\rm b}\!+\!k/2,\beta) \cdot
      \bigg[ S_{p_{\rm a}}^\mu\!
           + \frac{p_{\rm a}^\mu k_\nu}{2m_{\rm a}^2} S_{p_{\rm a}}^\nu
      \bigg]^{\{c\}}_{~~\;\{a\}}
      \Bigg\} , \nn
\end{align}
where we have introduced little-group reduced amplitudes
\[
\begin{aligned}
   \tilde{\cal A}^{\{a'\}}{}_{\{b\}}(p_{\rm a}\!+\!k/2,s_1';
   p_{\rm b}\!-\!k/2,\beta|p_{\rm a}\!-\!k/2,s_1;p_{\rm b}\!+\!k/2,\beta)
   \equiv U^{\{a'\}}{}_{\{c'\}}(p_{\rm a},p_{\rm a}\!+\!k/2) & \\ \cdot\,
      {\cal A}^{\{c'\}}{}_{\{d\}}(p_{\rm a}\!+\!k/2,s_1';
      p_{\rm b}\!-\!k/2,\beta|p_{\rm a}\!-\!k/2,s_1;
      p_{\rm b}\!+\!k/2,\beta) \cdot
      U^{\{d\}}{}_{\{b\}}(p_{\rm a}\!-\!k/2,p_{\rm a}) & ,
\end{aligned}
\]

These amplitudes may be expressed as functions of momenta
and spin operators.
The external spin operators in the formula~\eqref{ImpulseSpinLO2}
occur in the form of a commutator and an anticommutator
with the tree amplitude.
Due to the crucial property~\eqref{SpinExpectation}
of the coherent spin states, the leading classical contribution
comes from replacing the spin operator with its expectation value
$\braket{S_{p_{\rm a}}^\mu}_\alpha$.
Hence the anticommutator amounts to a factor of two:
\[
   \braket{ [S_{p_{\rm a}}^\nu , {\cal A}]_+ }_\alpha
    = 2 \braket{S_{p_{\rm a}}^\nu}_\alpha \braket{{\cal A} }_\alpha
      + {\cal O}(\hbar) .
\]
It may seem that the commutator term should be negligible
with respect to the anticommutator.
However, since the latter is multiplied by $k_\nu = \hbar \bar{k}_\nu$,
the leading classical contribution of the former is just as important.
Now the only operators inside the amplitude
that $S_{p_{\rm a}}^\mu$ does not commute with are its own components,
as encoded by the transverse spin algebra~\eqref{AngularMomentumAlgebra4d}.
In view of the contraction with the coherent spin states,
we may ignore the order of multiplication
in the leading non-vanishing commutator contribution,
which becomes \cite{Guevara:2019fsj}
\begin{align}
   \braket{ [S_{p_{\rm a}}^\mu , {\cal A}] }_\alpha &
    = \Big\langle [S_{p_{\rm a}}^\mu , S_{p_{\rm a}}^\sigma]
      \frac{\partial {\cal A}}{\partial S_{p_{\rm a}}^\sigma\!}
      \Big\rangle_{\!\alpha} + {\cal O}(\hbar)
%    = \frac{i\hbar}{m_{\rm a}}
%      \epsilon^{\mu\nu\rho\sigma} p_{{\rm a}\:\!\nu}
%      \Big\langle S_{p_{\rm a}\rho}
%      \frac{\partial {\cal A}}{\partial S_{p_{\rm a}}^\sigma\!}
%      \Big\rangle_{\!\alpha} + {\cal O}(\hbar) \\ &
    = \frac{i\hbar}{m_{\rm a}} \epsilon^{\mu\nu\rho\sigma} p_{{\rm a}\:\!\nu}
      \braket{S_{p_{\rm a}\rho}}_\alpha
      \frac{\partial \braket{\cal A}_\alpha}
           {\partial \braket{S_{p_{\rm a}}^\sigma}_\alpha\!}
    + {\cal O}(\hbar) .
\end{align}
Here the coherent-spin amplitude $\braket{\cal A}_\alpha$
is understood to be a function of the classical spin
$\braket{S_{p_{\rm a}}^\mu}_\alpha$ at momentum $p_{\rm a}$,
and the partial derivative should be understood in the
three-dimensional sense within the subspace transverse to $p_{\rm a}$.

It is convenient to have a shorthand notation
for spin-length expectation values
\[
   a_{\rm a}^\mu \equiv \frac{1}{m_{\rm a}\!}
      \braket{S_{p_{\rm a}}^\mu}_\alpha , \qquad \quad
   a_{\rm b}^\mu \equiv \frac{1}{m_{\rm b}\!}
      \braket{S_{p_{\rm b}}^\mu}_\beta .
\label{SpinLength}
\]
They are in correspondence with classical spins
$s_{\rm a}^\mu = \braket{S_{u_{\rm a}}^\mu}_\alpha$ and
$s_{\rm b}^\mu = \braket{S_{u_{\rm b}}^\mu}_\beta$,
to which they become proportional, but strictly speaking only after integration
over $p_{\rm a}$ and $p_{\rm b}$, as in \eqn{SpinClassicalLimit}.
In terms of these intermediate spin lengths,
the leading angular impulse may therefore be expressed as \cite{Guevara:2019fsj}
\[
\begin{aligned}
   \Delta S_{\rm a}^\mu
    = \frac{\hbar}{m_{\rm a}\!} &
      \int_{p_{\rm a},p_{\rm b}}\!\!\!
      |\psi_{\rm a}(p_{\rm a})|^2 |\psi_{\rm b}(p_{\rm b})|^2
      \bigg[
      p_{\rm a}^\mu a_{\rm a}^\nu \frac{\partial~}{\partial b^\nu\!}
    - \epsilon^{\mu\nu\rho\sigma} p_{{\rm a}\:\!\nu} a_{{\rm a}\rho}
      \frac{\partial~~}{\partial a_{\rm a}^\sigma}
      \bigg] \\ \times\! &
      \int_k e^{-i \bar{k} \cdot b}
      {\cal A}^{(0)}(p_{\rm a}\!+\!k/2,\alpha;p_{\rm b}\!-\!k/2,\beta
                    |p_{\rm a}\!-\!k/2,\alpha;p_{\rm b}\!+\!k/2,\beta) .
\label{ImpulseSpinLOFinal}
\end{aligned}
\]

%\footnote{For brevity, we also adopt
%the hatted notation for momentum measures and delta functions
%to absorb appropriate powers of $2\pi$; in particular,
%$\dd^4k \equiv d^4k/(2\pi)^4$ and $\del(p^2-m^2) \equiv 2\pi\del(p^2-m^2)$.}

%%%%%%%%%%%%%%%%%%%%%%%%%%%%%%%%%%%%%%%%%%%%%%%%%%
\section{Classical spinning amplitudes}
\label{sec:ClassicalAmplitudes}
%%%%%%%%%%%%%%%%%%%%%%%%%%%%%%%%%%%%%%%%%%%%%%%%%%

In this section we analyze the coherent-spin amplitudes,
in terms of which we have already written
the leading-order impulse formulae~\eqref{ImpulseMomLOFinal}
and~\eqref{ImpulseSpinLOFinal}.
These amplitudes naturally appear
in the expectation value of the scattering transition operator
\begin{align}
 & \braket{\text{in}|T|\text{in}}
    = \int_{p_1',p_2',p_1,p_2}\!\!\!\!\!e^{-i k \cdot b/\hbar}
      \psi_{\rm a}^*(p_1') \psi_{\rm b}^*(p_2')
      \psi_{\rm a}(p_1) \psi_{\rm b}(p_2)
      \bra{p_1',\alpha;p_2',\beta} T
      \ket{p_1,\alpha;p_2,\beta} \\ &
%    = \int_{p_1,p_2} \int_k\!e^{-i k \cdot b/\hbar}
%      \psi_{\rm a}^*(p_1\!+\!k) \psi_{\rm b}^*(p_2\!-\!k)
%      \psi_{\rm a}(p_1) \psi_{\rm b}(p_2)
%      {\cal A}(p_1\!+\!k,\alpha;p_2\!-\!k,\beta
%      |p_1,\alpha;p_2,\beta) \nn \\ &
    = \int_{p_{\rm a},p_{\rm b}}\!\!\!
      |\psi_{\rm a}(p_{\rm a})|^2 |\psi_{\rm b}(p_{\rm b})|^2
      \int_k\!e^{-i \bar{k} \cdot b}
      {\cal A}(p_{\rm a}\!+\!k/2,\alpha;p_{\rm b}\!-\!k/2,\beta
              |p_{\rm a}\!-\!k/2,\alpha;p_{\rm b}\!+\!k/2,\beta) , \nn
\end{align}
which may be called ``the scattering function''
in the impact-parameter space.
It is, roughly speaking, an eikonal Fourier transform
of ``the classical scattering amplitude''
in momentum space.

%%%%%%%%%%%%%%%%%%%%%%%%%%%%%%%%%%%%%%%%%%%%%%%%%%
\begin{figure}[t]
\centering
\includegraphics[height=0.25\textwidth]{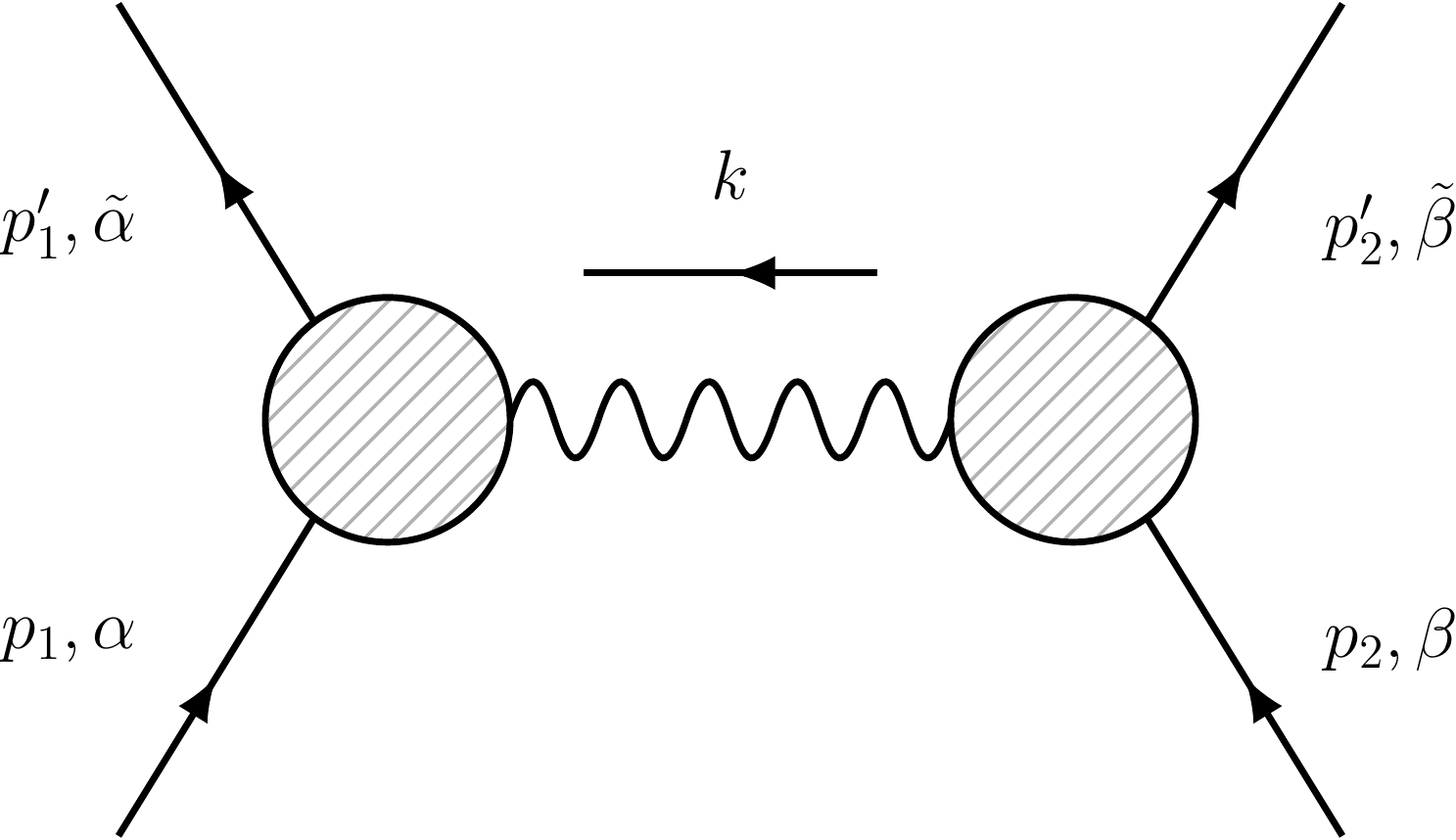}
\vspace{-5pt}
\caption{\label{fig:4pt} Residue of
tree-level four-point scattering amplitude in $t$ channel}
\end{figure}
%%%%%%%%%%%%%%%%%%%%%%%%%%%%%%%%%%%%%%%%%%%%%%%%%%

The classical limit is dominated by small momentum transfer
\[
   t = k^2 = \hbar^2 \bar{k}^2 .
\]
where the tree-level four-point amplitude
is factorized into two three-point ones:
\[
\begin{aligned}
\label{ElasticScatteringAmplitude}
  {\cal A}^{(0)}(p_1',\alpha;p_2',\beta|p_1,\alpha;p_2,\beta)
    =-\frac{1}{\hbar^2 \bar{k}^2}\!\sum_\pm &
      {\cal A}^{(0)}(p_1',\alpha|p_1,\alpha;k,\pm) \\ \times\,&
      {\cal A}^{(0)}(p_2',\beta;k,\mp|p_2,\beta) + {\cal O}(1/\hbar) ,
\end{aligned}
\]
as illustrated in \fig{fig:4pt}.
For real external momenta on the mass shell,
$k$ is always spacelike.
However, we may extract the $t$-channel residue
by first considering complex on-shell momenta consistent with $k^2=0$
and then analytically continuing the result to real spacelike $k$,
along the lines of the Holomorphic Classical Limit of
\rcite{Guevara:2017csg}.

%%%%%%%%%%%%%%%%%%%%%%%%%%%%%%%%%%%%%%%%%%%%%%%%%%
\subsection{Three-point amplitudes}
\label{sec:3pt}
%%%%%%%%%%%%%%%%%%%%%%%%%%%%%%%%%%%%%%%%%%%%%%%%%%

\begin{figure}[t]
\centering
\includegraphics[height=0.25\textwidth]{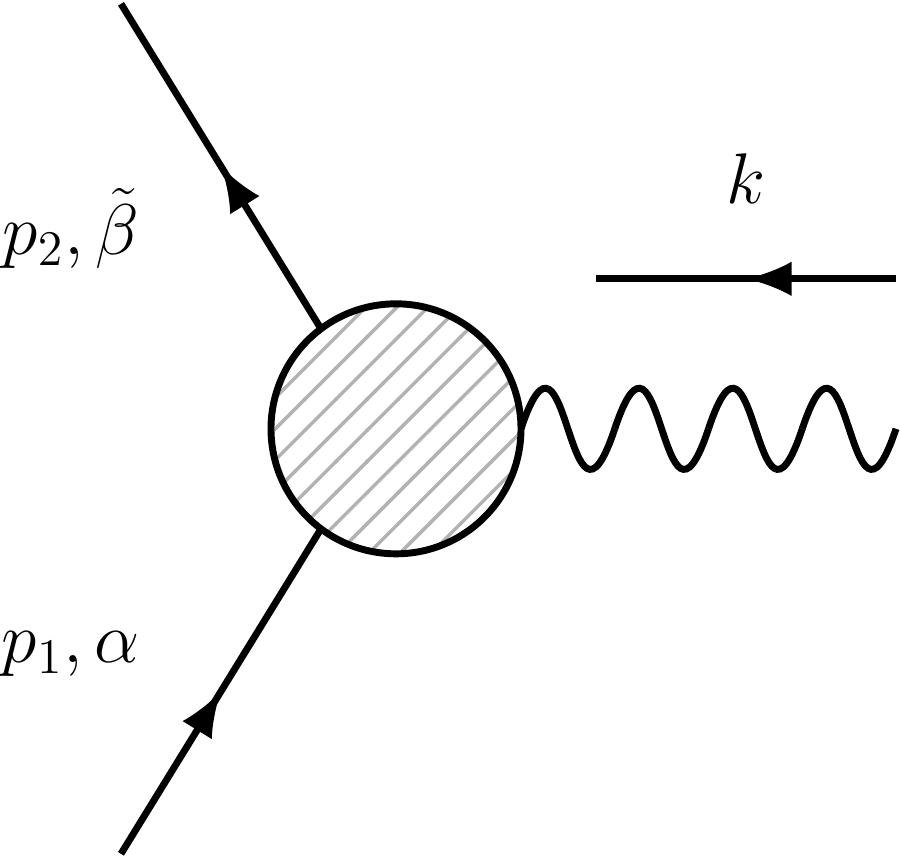}
\vspace{-5pt}
\caption{\label{fig:3pt} Tree-level three-point scattering amplitude}
\end{figure}

Let us now focus on the classical limit of the three-point amplitude
shown in \fig{fig:3pt}
\[
\begin{aligned}
\label{3pt}
   {\cal A}_3^h & \equiv {\cal A}^{(0)}(p_2,\beta|p_1,\alpha;k,h) \\ &
    = e^{-(\|\alpha\|^2 + \|\beta\|^2)/2} \sum_{s_1,s_2}
      \frac{(\tilde{\beta}_b)^{\odot 2s_2} (\alpha^a)^{\odot 2s_1}}
           {\sqrt{(2s_1)!(2s_2)!}} \cdot
      {\cal A}^{(0)\{b\}}{}_{\{a\}}(p_2,s_2|p_1,s_1;k,h) ,
\end{aligned}
\]
which we have relabeled with respect to the amplitudes
appearing in \eqn{ElasticScatteringAmplitude},
so as to unclutter the notation within this subsection.
Although we have allowed the angular-momentum spinors
to be different for the incoming and outgoing massive states,
we will keep in mind that, in view of the classical impulse formulae,
we are particularly interested in the case where $\alpha = \beta$.
Note, however, that the state expansion above
involves a double summation over amplitudes
with all possible combinations of incoming and outgoing massive spins.
In fact, one can think of coherent-state amplitudes
as generating functions for various definite-spin amplitudes,
for instance
\[
\label{FiniteFromCoherent}
   {\cal A}^{(0)\{b\}}{}_{\{a\}}(p_2,s_2|p_1,s_1;k,h)
    = \frac{1}{\sqrt{(2s_1)!(2s_2)!}}
   \bigg[ \frac{\partial~}{\partial\tilde{\beta}_b} \bigg]^{\odot 2s_2} 
   \bigg[ \frac{\partial~}{\partial\alpha^a} \bigg]^{\odot 2s_1} 
   {\cal A}_3^h\,
%   \frac{ \partial^{2(s_1+s_2)} {\cal A}^{(0)}(p_2,\beta|p_1,\alpha;k,h) }
%        { \partial\tilde{\beta}_{(b_1}\!\ldots
%          \partial\tilde{\beta}_{b_{2s_2})}
%          \partial\tilde{\alpha}^{(a_1}\!\ldots
%          \partial\tilde{\alpha}^{a_{2s_1})} }
   \bigg|_{\alpha=\beta=0} .
\]

%%%%%%%%%%%%%%%%%%%%%%%%%%%%%%%%%%%%%%%%%%%%%%%%%%
%%%%%%%%%%%%%%%%%%%%%%%%%%%%%%%%%%%%%%%%%%%%%%%%%%

For concreteness, we work with the case of gravitational interaction.
We start with the minimal-coupling amplitudes\footnote{Stricty speaking,
the amplitudes \eqref{GravityMatter3pt}
should appear with an additional prefactor
of $(-1)^{\lfloor s \rfloor}$, as in \rcite{Johansson:2019dnu}.
However, this prefactor is merely due to the conventional normalization
\eqref{PolTensorsNormalization} of the external wavefunctions,
by which we have taken care to divide in \eqn{GravityMatter3pt}
and the following.
}
\begin{subequations} \begin{align}
\label{GravityMatter3ptPlus}
   {\cal A}_\text{min}^{(0)\{b\}}{}_{\{a\}}(p_2,s|p_1,s;k,+) &
    =-\frac{\kappa}{2}
      \frac{\braket{2^b 1_a}^{\odot 2s}\!}{m^{2s-2}} x^2 , \\*
\label{GravityMatter3ptMinus}
   {\cal A}_\text{min}^{(0)\{b\}}{}_{\{a\}}(p_2,s|p_1,s;k,-) &
    = (-1)^{2s+1} \frac{\kappa}{2}
      \frac{[2^b 1_a]^{\odot 2s}\!}{m^{2s-2}} x^{-2} ,
\end{align} \label{GravityMatter3pt}%
\end{subequations}
where $x$ is the unit-helicity factor
that has multiple equivalent representations
in terms of massless helicity spinors or polarization vectors:
\[
   x = \frac{[k|p_1\ket{r}}{m \braket{k\;\!r}}
     = \frac{m [k\;\!r]}{\bra{k}p_1|r]}
     = -\frac{\sqrt{2}}{m} (p_1\cdot\varepsilon^+)
     = \bigg[ \frac{\sqrt{2}}{m} (p_1\cdot\varepsilon^-) \bigg]^{-1} .
\label{HelicityFactor}
\]
These simplest general-spin amplitudes
were proposed by Arkani-Hamed, Huang and Huang
\cite{Arkani-Hamed:2017jhn}
based on their tame behavior in the massless limit,
but were soon realized \cite{Guevara:2018wpp,Chung:2018kqs}
to correspond to scattering of Kerr's rotating black hole (also obtained by means of on-shell heavy particle theories~\cite{Aoude:2020onz}).
Let us derive this result in our current formalism.

%%%%%%%%%%%%%%%%%%%%%%%%%%%%%%%%%%%%%%%%%%%%%%%%%%
\subsection{From minimal coupling to Kerr}
\label{sec:MinimalCoupling}
%%%%%%%%%%%%%%%%%%%%%%%%%%%%%%%%%%%%%%%%%%%%%%%%%%

The minimal-coupling amplitudes correspond to the diagonal slice
$s_1=s_2=s$ of the summation in the total-spin quantum numbers,
so their contribution to the coherent-spin amplitude~\eqref{3pt} is simply
\[
\begin{aligned}
   {\cal A}_{3,\text{min}}^+ &
    = -\frac{\kappa}{2} x^2
      e^{-(\|\alpha\|^2 + \|\beta\|^2)/2} \sum_{2s=0}^\infty
      \frac{1}{(2s)!}
      (\tilde{\beta}_b)^{\odot 2s} \cdot
      \frac{\braket{2^b 1_a}^{\odot 2s}\!}{m^{2s-2}} \cdot
      (\alpha^a)^{\odot 2s} \\ &
    = -\frac{\kappa}{2} m^2 x^2 e^{-(\|\alpha\|^2 + \|\beta\|^2)/2}
      \exp\bigg\{ \frac{1}{m} \tilde{\beta}_b \braket{2^b 1_a} \alpha^a
          \bigg\} .
\label{A3ptPlus}
\end{aligned}
\]
Recall that in \sec{sec:LOSpinKick} we boosted
the initial- and final-state spin operators
to the same intermediate momentum $p_{\rm a}$.
In the same spirit,
we are now going to determine the dependence of the above exponent
on the angular-momentum operator~$S_{p_{\rm a}}^\mu$,
which generates the little group of the average momentum
$p_{\rm a} = (p_1+p_2)/2 = p_1 + k/2$.
The three-point on-shell kinematics implies
that this momentum is also on-shell due to
$p_1 \cdot k = p_2 \cdot k = p_{\rm a} \cdot k = 0$,
and it is related to $p_1$ and $p_2$ via boosts
\[
   p_1^\rho = \exp\!\bigg[ {-}\frac{i p_{\rm a}^\mu k^\nu\!}{2m^2}
              \Sigma_{\mu\nu} \bigg]^\rho_{~\sigma} p_{\rm a}^\sigma ,
   \qquad \quad
   p_2^\rho = \exp\!\bigg[ \frac{i p_{\rm a}^\mu k^\nu\!}{2m^2}
              \Sigma_{\mu\nu} \bigg]^\rho_{~\sigma} p_{\rm a}^\sigma .
\label{Boost12}
\]
The initial- and final-state spinors may be similarly boosted
using the chiral Lorentz generators
$\sigma^{\mu\nu} = i \sigma^{[\mu} \bar{\sigma}^{\nu]}/2$:
\begin{align}
\label{Boost12a}
   \ket{1_a} & = U_a{}^b(p_1,p_{\rm a})
      \exp\!\bigg[ {-}\frac{i p_{\rm a}^\mu k^\nu\!}{2m^2} 
                   \sigma_{\mu\nu} \bigg] \ket{{\rm a}_b}
    = U_a{}^b(p_1,p_{\rm a})
      \bigg( \ket{{\rm a}_b} - \frac{1}{4m} |k|{\rm a}_b] \bigg) ,  \\
   \ket{2^a} & = U^a{}_b(p_2,p_{\rm a})
      \exp\!\bigg[ \frac{i p_{\rm a}^\mu k^\nu\!}{2m^2}
                   \sigma_{\mu\nu} \bigg] \ket{{\rm a}^b}
   ~~\Rightarrow~~
   \bra{2^a} = U^a{}_b(p_2,p_{\rm a})
%      \bra{{\rm a}^b}
%      \exp\!\bigg[ {-}\frac{i p_{\rm a}^\mu k^\nu\!}{2m^2}
%                   \sigma_{\mu\nu} \bigg]
      \bigg( \bra{{\rm a}^b} - \frac{1}{4m} [{\rm a}^b|k| \bigg) , \nn
\end{align}
where we have used the nilpotency of $p_{\rm a}^\mu k^\nu \sigma_{\mu\nu}$
for $k^2=0$.\footnote{While here we consider the boosts from $p_{1,2}$ to $p_{\rm a}$, similar relations between spinors, whose momenta differ by quantum fluctuations, can be obtained by means of on-shell heavy-particle EFT variables, as discussed in \rcite{Damgaard:2019lfh,Aoude:2020onz}.}
Recalling the mild momentum dependence~\eqref{SU2SpinorTransform}
of the ${\rm SU}(2)$ spinors,
as well as the properties of ${\rm SU}(2)$ transformations
pointed out in \foot{foot:SU2},
we may rewrite the exponent in \eqref{A3ptPlus} as
\[
\begin{aligned} %&
   \tilde{\beta}_b(p_2) \braket{2^b 1_a} \alpha^a(p_1) & %\\ &
%    = \tilde{\beta}_b(p_2) U^b{}_c(p_2,p_{\rm a})
%      \bigg( \bra{{\rm a}^c} - \frac{1}{4m} [{\rm a}^c|k| \bigg)
%      \bigg( \ket{{\rm a}_d} - \frac{1}{4m} |k|{\rm a}_d] \bigg)
%      U_e{}^d(p_1,p_{\rm a}) \alpha^e(p_1) \nn \\ &
%    = \tilde{\beta}_c(p_{\rm a})
%      \bigg( \bra{{\rm a}^c} - \frac{1}{4m} [{\rm a}^c|k| \bigg)
%      \bigg( \ket{{\rm a}_d} - \frac{1}{4m} |k|{\rm a}_d] \bigg)
%      \alpha^d(p_{\rm a}) \\ &
    = \tilde{\beta}_b(p_{\rm a})
      \bigg( \braket{{\rm a}^b {\rm a}_a}
           - \frac{1}{4m} \Big( [{\rm a}^b|k\ket{{\rm a}_a}
                              + \bra{{\rm a}^b}k|{\rm a}_a] \Big)\!\bigg)
      \alpha^a(p_{\rm a}) . %\\ &
%    = \tilde{\beta}_b(p_{\rm a})
%      \bigg( m \delta^b_a
%           - \frac{1}{2} \hbar \bar{k}_\mu \sigma_{p_{\rm a}}^{\mu,b}{}_a
%      \bigg) \alpha^a(p_{\rm a}) .
\label{SpinorProduct2Spin}
\end{aligned}
\]
Here $\braket{{\rm a}^b {\rm a}_a} = m \delta_a^b$, and we may recognize the ${\rm SU}(2)$ spin generator\footnote{The
apparent sign difference between \eqns{DiracSpinOperator}{DiracSpinOperator2}
is due to all indices being always raised
``on the left'' in the spinor-helicity formalism, whereas in ${\rm SU}(2)$
we have
$\sigma_p^{\mu,a}{}_b = \epsilon^{ac} \sigma_{p~c}^{\mu,}{}^d \epsilon_{db}$.}
\[
   \sigma_{p_{\rm a}}^{\mu,a}{}_b  = \frac{1}{2m}
      \Big(\bra{{\rm a}^a}\sigma^\mu|{\rm a}_b] + [{\rm a}^a|\bar{\sigma}^\mu\ket{{\rm a}_b}\Big) ,
\label{DiracSpinOperator2}
\]
showing the separation between spinless and spin effects.

Having thus evaluated the positive-helicity amplitude~\eqref{A3ptPlus},
for completeness
we display the coherent-spin amplitudes for both graviton helicities:
\[
   {\cal A}_{3,\text{min}}^\pm
    = -\frac{\kappa}{2} m^2 x^{\pm 2}
      e^{-(\|\alpha\|^2 + \|\beta\|^2)/2 + \tilde{\beta} \alpha}
      \exp\bigg\{ {\mp}\frac{\hbar}{2m} \bar{k}_\mu
      (\tilde{\beta}\:\!\sigma_{p_{\rm a}}^\mu\;\!\!\alpha) \bigg\} ,
\label{A3ptMin}
\]
where all spinors are understood to correspond
to the little group of $p_{\rm a}$.
A remarkable property here is that we have factored out
the standard coherent-state overlap function
\[
   \braket{\beta|\alpha}
    = e^{-(\|\alpha\|^2 + \|\beta\|^2)/2 + \tilde{\beta} \alpha} .
\label{CoherentOverlap}
\]
Note that in the classical limit
of the four-point amplitude~\eqref{ElasticScatteringAmplitude}
we take $\tilde{\beta}_a = (\alpha^a)^*$ and can therefore
identify the spin expectation value~\eqref{SpinExpectation4d}:
\[
   {\cal A}_{3,\text{min}}^\pm \big|_{\beta=\alpha}
    =-\frac{\kappa}{2} m^2 x^{\pm 2}
      \exp\bigg\{ {\mp}\frac{1}{m} \bar{k}_\mu
      \braket{S_{p_{\rm a}}^\mu}_\alpha \bigg\}
    =-\frac{\kappa}{2} m^2 x^{\pm 2}
      e^{\mp \bar{k} \cdot a_{\rm a}} .
\label{A3ptMinClassical}
\]
Importantly, this result would have been classically vanishing
if the exponential suppression by the original prefactor
$e^{-(\|\alpha\|^2 + \|\beta\|^2)/2} = e^{-\|\alpha\|^2}$,
as $\|\alpha\|^2$ grows as $1/\hbar$,
had not been canceled by the exponential growth of
$e^{\tilde{\beta} \alpha} = e^{\|\alpha\|^2}$.

%%%%%%%%%%%%%%%%%%%%%%%%%%%%%%%%%%%%%%%%%%%%%%%%%%
\subsubsection{Connection to Kerr black hole}
\label{sec:KerrStressTensor}
%%%%%%%%%%%%%%%%%%%%%%%%%%%%%%%%%%%%%%%%%%%%%%%%%%

The exponential spin-multipole structure
that we have obtained is a firm basis for the identification
of the minimal-coupling amplitudes
with gravitational scattering of black holes.
The argument for this is given in \rcites{Guevara:2018wpp,Guevara:2019fsj}
and is based on the classical stress-energy tensor \cite{Vines:2017hyw}
\[
   T^{\mu\nu}_\text{Kerr}(x) = m\!\int\!d\tau\;\!
      u^{(\mu} \exp(a*\partial)^{\nu)}{}_\rho u^\rho
      \delta^{4}(x-u \tau) ,
\label{KerrEnergyTensor}
\]
which in linearized gravity serves as an effective source for a Kerr black hole with mass $m$, velocity $u^\mu$ and spin length $a^\mu$. 
The exponential $e^{a*\partial}$, involving the Levi-Civita contraction
\[  
   (a*b)^{\mu\nu} = \epsilon^{\mu\nu\alpha\beta} a_\alpha b_\beta ,
\label{LeviCivitaStar}
\]
may be shown to yield precisely $e^{\mp \bar{k} \cdot a_{\rm a}}$
when coupled to an on-shell graviton
(see \app{app:Nonminimal} for a matching calculation encompassing
the Kerr case).
In fact, formulae identical to the right-hand side of \eqn{A3ptMinClassical}
appeared in \rcite{Guevara:2019fsj}
but featured chiral-spinor versions
of the Pauli-Lubanski operator~\eqref{PauliLubanski}
instead of the spin expectation value $a_{\rm a}^\mu$.
An obvious advantage of our current approach
is that it enables us to directly identify
the classical spin vector right away,
instead of heuristically replacing quantum-mechanical operators
with the corresponding classical quantities.

%%%%%%%%%%%%%%%%%%%%%%%%%%%%%%%%%%%%%%%%%%%%%%%%%%
\subsubsection{Multipoles from lower spins}
\label{sec:LowerSpins}
%%%%%%%%%%%%%%%%%%%%%%%%%%%%%%%%%%%%%%%%%%%%%%%%%%

As first noticed in \rcite{Vaidya:2014kza},
lower spin-induced black-hole multipoles may be extracted
from considering particles with finite quantum spin.
Namely, the $2^{2s}$-pole interaction appears
for spin-$s$ particles.
To see how this occurs in our formalism,
let us truncate the exponential~\eqref{A3ptPlus}
in the argument above:
\begin{align}
\label{A3ptPlusTruncated1}
 & {\cal A}_{3,\text{min}}^{+,\text{trunc}} \big|_{\beta=\alpha}
    = -\frac{\kappa}{2} x^2 e^{-\|\alpha\|^2}
      \sum_{2s=0}^r \frac{1}{(2s)!}
      (\tilde{\alpha}_b(p_2))^{\odot 2s} \cdot
      \frac{\braket{2^b 1_a}^{\odot 2s}\!}{m^{2s-2}} \cdot
      (\alpha^a(p_1))^{\odot 2s} \nn \\ &
    = -\frac{\kappa}{2} m^2 x^2 e^{-\|\alpha\|^2}
      \sum_{2s=0}^r \frac{1}{(2s)!}
      (\tilde{\alpha}_b(p_{\rm a}))^{\odot 2s}\!\cdot
      \bigg( \delta^b_a - \frac{\hbar}{2m} \bar{k}_\mu
             \sigma_{p_{\rm a}}^{\mu,b}{}_a\!\bigg)^{\!\odot 2s}\!\!\cdot
      (\alpha^a(p_{\rm a}))^{\odot 2s} \\ &
    = -\frac{\kappa}{2} m^2 x^2 e^{-\|\alpha\|^2}
      \sum_{2s=0}^r \sum_{n=0}^{2s}
      \frac{(\tilde{\alpha}_b(p_{\rm a}))^{\odot 2s}\!}{n!(2s-n)!} \cdot
      \bigg( {-}\frac{\hbar}{2m} \bar{k}\cdot\sigma_{p_{\rm a}}^{~\,b}{}_a\!
      \bigg)^{\!\odot n}\!\!\odot
      (\delta^b_a)^{\odot (2s-n)}\!\cdot
      (\alpha^a(p_{\rm a}))^{\odot 2s} . \nn
\end{align}
To deal with this truncated sum,
we may use the following property
of the quantum angular-momentum representation~\eqref{SpinRepresentation4d},
true for any lightlike~$k$ satisfying $p \cdot k = 0$:
\[
   \big[(k \cdot S_p)^n\big]^{s,\{a\}}{}_{s',\{b\}}
    = \frac{\hbar^n (2s)!}{2^n (2s-n)!}\;\!\delta^s_{s'}
      \big(k\cdot\sigma_p{}^a{}_b\big)^{\odot n} \odot
      \big(\delta^a_b\big)^{\odot (2s-n)} .
\label{AngularMomentumPower3pt}
\]
Whenever $n>2s$, the gamma function $\Gamma(2s-n+1)=(2s-n)!$
in the denominator develops poles,
implying that the right-hand side then vanishes,
so we can still obtain an exponential:
\begin{align}
   {\cal A}_{3,\text{min}}^{+,\text{trunc}} \big|_{\beta=\alpha} &
    = -\frac{\kappa}{2} m^2 x^2 e^{-\|\alpha\|^2}
      \sum_{2s=0}^r \frac{1}{(2s)!} \sum_{n=0}^{2s} \frac{1}{n!}
      (\tilde{\alpha}_b)^{\odot 2s}\!\cdot\!
      \bigg[ \bigg({-}\frac{1}{m} \bar{k}\cdot S_{p_{\rm a}} \bigg)^{\!n}
      \bigg]^{\{b\}}_{~\;\;\;\{a\}}\!\!\cdot
      (\alpha^a)^{\odot 2s} \nn \\ &
    = -\frac{\kappa}{2} m^2 x^2 e^{-\|\alpha\|^2}
      \sum_{2s=0}^r \frac{1}{(2s)!}
      (\tilde{\alpha}_b)^{\odot 2s}\!\cdot
      \exp\!\bigg[{-}\frac{1}{m} \bar{k}\cdot S_{p_{\rm a}}
      \bigg]^{\{b\}}_{~\;\;\;\{a\}}\!\!\cdot
      (\alpha^a)^{\odot 2s} \nn \\ &
    = -\frac{\kappa}{2} m^2 x^2 e^{-\|\alpha\|^2}
      \sum_{2s=0}^r \frac{\|\alpha\|^{4s}}{(2s)!}
      \bigg\langle\!\exp\!\bigg[{-}\frac{1}{m} \bar{k}\cdot S_{p_{\rm a}}
                          \bigg] \bigg\rangle_{\!2s} ,
\label{A3ptPlusTruncated2}
\end{align}
where we have used a finite-spin expectation value
defined in an obvious way as
$\braket{O}_{2s} = (\tilde{\alpha}_a)^{\odot 2s} \cdot
 O^{s,\{a\}}{}_{s'=s,\{b\}} \cdot (\alpha^b)^{\odot 2s}/\|\alpha\|^{4s}$.
For any finite truncation $r$, taking $\|\alpha\|~\to~\infty$ according to
the classical limit~\eqref{ClassicalLimit} would nullify
the coherent-spin amplitude.
Moreover, for any finite $s=0,1/2,\dots,r/2$,
the spin exponential in \eqn{A3ptPlusTruncated2} is naturally truncated
at the $2^{2s}$ multipole,
exactly as in \rcites{Guevara:2018wpp,Guevara:2019fsj,Aoude:2020onz}.
However, if we presume a classical-limit property
of the type~\eqref{SpinExpectation4d2}, we may recognize that
the multipoles which are present in the finite-spin contributions above
very well correspond to those in our full result~\eqref{A3ptMinClassical}
--- except for the summation over~$s$
and the normalization prefactors
$e^{-\|\alpha\|^2} \|\alpha\|^{4s}/(2s)!$,
which are important in the coherent-spin formalism
but should rather be ignored in a finite-spin approach.

%%%%%%%%%%%%%%%%%%%%%%%%%%%%%%%%%%%%%%%%%%%%%%%%%%
\subsection{Non-minimal coupling}
\label{sec:Nonminimal}
%%%%%%%%%%%%%%%%%%%%%%%%%%%%%%%%%%%%%%%%%%%%%%%%%%

Now that we have explored how the multipole structure
of a Kerr black hole arises from the minimal-coupling amplitudes,
we may as well consider more general massive particles
which couple to gravity in a non-minimal way.
We write the corresponding three-point amplitudes still for equal spins
but otherwise in full generality as \cite{Arkani-Hamed:2017jhn}
\begin{align}
\label{GravityMatter3ptNonMin}
%\label{GravityMatter3ptNonMinPlus}
   {\cal A}_\text{gen}^{(0)\{b\}}{}_{\{a\}}(p_2,s|p_1,s;k,+) &
    =-\frac{\kappa}{2} \sum_{n=0}^{2s} g_n^+
      \frac{x^{n+2} \braket{2^b 1_a}^{\odot (2s-n)}\!}{m^{2s+n-2}} \odot
      \big( \braket{2^b k} \braket{k\:\!1_a} \big)^{\odot n} , \\*
%\label{GravityMatter3ptNonMinMinus}
   {\cal A}_\text{gen}^{(0)\{b\}}{}_{\{a\}}(p_2,s|p_1,s;k,-) &
    = (-1)^{2s+1} \frac{\kappa}{2} \sum_{n=0}^{2s} g_n^-
      \frac{x^{-n-2} [2^b 1_a]^{\odot (2s-n)}\!}{m^{2s+n-2}} \odot
      \big( [2^b k] [k\:\!1_a] \big)^{\odot n} . \nn
\end{align}
The dimensionless coupling constants $g_n^+$ and $g_n^-$
are understood to be related by complex conjugation
due to parity conservation.
Moreover, as we have seen above and will be able to further confirm below,
the minimal couplings $g_0^\pm$
determine the gravitational interaction at zero spin
and are therefore pegged to unity by the equivalence principle.
Importantly, we assume that the coupling constants $g_{n>0}^\pm$
depend only on their ``non-minimalness''~$n$
and not on the spin quantum number~$s$.

Let us construct the non-minimal coherent-spin
amplitude for positive helicity:
\begin{align}
   {\cal A}_{3,\text{gen}}^+ &
    = -\frac{\kappa}{2} e^{-(\|\alpha\|^2 + \|\beta\|^2)/2}
      \sum_{2s=0}^\infty \sum_{n=0}^{2s}
      \frac{g_n^+ x^{n+2}}{(2s)! m^{2s+n-2}}
      \big(\tilde{\beta}_b \braket{2^b 1_a} \alpha^a\big)^{2s-n}
      \big(\tilde{\beta}_b \braket{2^b k} \braket{k\:\!1_a} \alpha^a
      \big)^n \nn \\ &
    = -\frac{\kappa}{2} m^2 x^2 e^{-(\|\alpha\|^2 + \|\beta\|^2)/2}
      \sum_{n=0}^\infty g_n^+ x^n
      \bigg(\frac{\tilde{\beta}_b \braket{2^b k} \braket{k\:\!1_a} \alpha^a}
                 {m\:\!\tilde{\beta}_b \braket{2^b 1_a} \alpha^a}
      \bigg)^{\!n}
      \sum_{2s=n}^\infty
      \frac{\big(\tilde{\beta}_b \braket{2^b 1_a} \alpha^a\big)^{2s}\!}
           {(2s)! m^{2s}} \nn \\ &
    = -\frac{\kappa}{2} m^2 x^2 e^{-(\|\alpha\|^2 + \|\beta\|^2)/2}
      \exp\bigg\{ \frac{1}{m} \tilde{\beta}_b \braket{2^b 1_a} \alpha^a
          \bigg\}
\label{A3ptGenPlus} \\ & \qquad\:\!\!\times
      \sum_{n=0}^\infty g_n^+ x^n
      \bigg(\frac{\tilde{\beta}_b \braket{2^b k} \braket{k\:\!1_a} \alpha^a}
                 {m\:\!\tilde{\beta}_b \braket{2^b 1_a} \alpha^a}
      \bigg)^{\!n}
      \bigg[ 1 - \frac{1}{(n-1)!}
      \Gamma\bigg(n, \frac{1}{m} \tilde{\beta}_b \braket{2^b 1_a} \alpha^a
            \bigg)
      \bigg] , \nn
\end{align}
where changing the order of summation allowed us to
evaluate the sum in the total-spin quantum number~$s$.
Setting $\tilde{\beta}_a = (\alpha^a)^*$, we can again identify
\[
   \frac{1}{m} \tilde{\alpha}_b(p_2) \braket{2^b 1_a} \alpha^a(p_1)
    = \|\alpha\|^2 - \bar{k} \cdot a_{\rm a} ,
\label{MinimalSpinIdentity1}
\]
as in \sec{sec:MinimalCoupling}.
In addition, we will now also need the equalities
\[
   x\;\!\tilde{\alpha}_b(p_2)
      \braket{2^b k} \braket{k\:\!1_a} \alpha^a(p_1)
    = x^{-1} \tilde{\alpha}_b(p_2)
      [2^b k] [k\:\!1_a] \alpha^a(p_1)
    = 2 m^2 (\bar{k} \cdot a_{\rm a}) .
\label{NonMinimalSpinIdentity}
\]
They may be proven by using the three-point spinorial identities
\[
   [2^b 1_a]
    = -\braket{2^b 1_a} - \frac{x}{m} \braket{2^b k}\braket{k\:\!1_a} ,
%   \braket{2^b 1_a} = -[2^b 1_a] - \frac{x^{-1}\!}{m} [2^b k] [k\:\!1_a] ,
   \qquad \quad
   [2^b k] = x \braket{2^b k} , \qquad \quad
   [k\:\!1_a] = x \braket{k\:\!1_a} ,
\label{Square2Angle3pt}
\]
together with the parity-conjugated version of \eqn{MinimalSpinIdentity1}
\[
  -\frac{1}{m} \tilde{\alpha}_b(p_2) [2^b 1_a] \alpha^a(p_1)
    = \|\alpha\|^2 + \bar{k} \cdot a_{\rm a} ,
\label{MinimalSpinIdentity2}
\]
which was implicitly used earlier to arrive at
the negative-helicity version of \eqn{A3ptMin}.
%\begin{align}
%   {\cal A}_{3,\text{gen}}^- &
%    = -\frac{\kappa}{2} e^{-(\|\alpha\|^2 + \|\beta\|^2)/2}
%      \sum_{2s=0}^\infty \sum_{n=0}^{2s}
%      \frac{(-1)^{2s} g_n^- x^{-n-2}}{(2s)! m^{2s+n-2}}
%      \big(\tilde{\beta}_b [2^b 1_a] \alpha^a\big)^{2s-n}
%      \big(\tilde{\beta}_b [2^b k] [k\:\!1_a] \alpha^a
%      \big)^n \nn \\ &
%    = -\frac{\kappa}{2} m^2 x^{-2}
%      e^{-(\|\alpha\|^2 + \|\beta\|^2)/2}
%      \sum_{n=0}^\infty g_n^- x^{-n}
%      \bigg(\frac{\tilde{\beta}_b [2^b k] [k\:\!1_a] \alpha^a}
%                 {m\:\!\tilde{\beta}_b [2^b 1_a] \alpha^a}
%      \bigg)^{\!n}
%      \sum_{2s=n}^\infty
%      \frac{\big({-}\tilde{\beta}_b [2^b 1_a] \alpha^a\big)^{2s}\!}
%           {(2s)! m^{2s}} \nn \\ &
%    = -\frac{\kappa}{2} m^2 x^{-2}
%      e^{-(\|\alpha\|^2 + \|\beta\|^2)/2}
%      \exp\bigg\{{-}\frac{1}{m} \tilde{\beta}_b [2^b 1_a] \alpha^a \bigg\}
%\label{A3ptGenMinus} \\ & \qquad\:\!\!\times
%      \sum_{n=0}^\infty g_n^- x^{-n}
%      \bigg(\frac{\tilde{\beta}_b [2^b k] [k\:\!1_a] \alpha^a}
%                 {m\:\!\tilde{\beta}_b [2^b 1_a] \alpha^a}
%      \bigg)^{\!n}
%      \bigg[ 1 - \frac{1}{(n-1)!}
%      \Gamma\bigg(n,{-}\frac{1}{m} \tilde{\beta}_b [2^b 1_a] \alpha^a
%            \bigg)
%      \bigg] , \nn
%\end{align}
In this way, we obtain
\[\!\!
   {\cal A}_{3,\text{gen}}^\pm \big|_{\beta=\alpha}\!
    = -\frac{\kappa}{2} m^2 x^{\pm 2}
      e^{\mp \bar{k} \cdot a_{\rm a}}\!
      \sum_{n=0}^\infty g_n^\pm
      \bigg[\frac{\pm 2 \bar{k} \cdot a_{\rm a}}
                 {\|\alpha\|^2 \mp \bar{k} \cdot a_{\rm a}}
      \bigg]^{\!n}
      \Bigg[ 1 - \frac{\Gamma\big(n, \|\alpha\|^2
                       \mp \bar{k} \cdot a_{\rm a} \big)}{(n-1)!}
      \Bigg] ,\!
\label{A3ptGen}
\]
where we have again displayed both helicity amplitudes for completeness.
%\begin{align}
%\label{Boosts12b}
%   |1_a] & = U_a{}^b(p_1,p_{\rm a})
%      \exp\!\bigg[ {-}\frac{i p_{\rm a}^\mu k^\nu\!}{2m^2} 
%                   \bar{\sigma}_{\mu\nu} \bigg] |{\rm a}_b]
%    = U_a{}^b(p_1,p_{\rm a})
%      \bigg( |{\rm a}_b] - \frac{1}{4m} |k\ket{{\rm a}_b} \bigg) ,  \\
%   |2^a] & = U^a{}_b(p_2,p_{\rm a})
%      \exp\!\bigg[ \frac{i p_{\rm a}^\mu k^\nu\!}{2m^2}
%                   \bar{\sigma}_{\mu\nu} \bigg] |{\rm a}^b]
%   ~~\Rightarrow~~
%   [2^a| = U^a{}_b(p_2,p_{\rm a})
%      [{\rm a}^b|
%      \bigg( [{\rm a}^b| - \frac{1}{4m} \bra{{\rm a}^b}k| \bigg) . \nn
%\end{align}
In the classical limit~\eqref{ClassicalLimit},
the gamma-function term is exponentially suppressed,
since $\Gamma(n,\|\alpha\|^2) \sim \|\alpha\|^{2(n-1)} e^{-\|\alpha\|^2}$
as $\|\alpha\| \to \infty$.
Even still, the non-minimal couplings $g_{n>0}^\pm$ seem to be
polynomially suppressed by $\|\alpha\|^{-2n}$.

This means that, in order to be able to model
a classically spinning massive object
with generic spin-induced multipoles using a quantum particle,
one needs to introduce non-minimal coupling constants $g_{n>0}^\pm$
that scale as ${\cal O}(\hbar^{-n})$.\footnote{Alternatively, one could
rescale the non-minimal coupling constants $g_{n>0}^\pm$
already in \eqn{GravityMatter3ptNonMin} ---
by switching to massless spinors $\ket{\bar{k}}$ and~$|\bar{k}]$
in place of $\ket{k}$ and~$|k]$, which are both ${\cal O}(\hbar^{1/2})$.}
In general relativity, the dynamics of such classical objects
is conveniently described by the worldline effective theory,
in which the spin-induced multipole contributions
linear in the curvature tensor enter via the interaction Lagrangian
\cite{Porto:2006bt,Porto:2008tb,Levi:2015msa}
\begin{align}
\label{LSInteractions}
   S_R = m\!\int\!d\tau \bigg[
        & \sum_{n=1}^\infty \frac{(-1)^n}{(2n)!} C_{{\rm ES}^{2n}}
          (a \cdot \nabla)^{2n-2}
          R_{\lambda\mu\,\nu\rho} u^\lambda a^\mu u^\nu a^\rho \\ &
        + \sum_{n=1}^\infty \frac{(-1)^n}{(2n+1)!} C_{{\rm BS}^{2n+1}}
          (a \cdot \nabla)^{2n-1}\,
          {}^*\!R_{\lambda\mu\,\nu\rho} u^\lambda a^\mu u^\nu a^\rho
   \bigg]_{x = r(\tau)}  , \nn
\end{align}
where
${}^*\!R_{\lambda\mu\,\nu\rho}
= \sqrt{-g}\;\!\epsilon_{\lambda\mu\sigma\tau} R^{\sigma\tau}{}_{\nu\rho}/2$,
and $r^\mu(\tau)$, $u^\mu(\tau)=dr^\mu/d\tau$ and $a^\mu(\tau)$
are the coordinate, velocity and spin functions of proper time, respectively.
In fact, it is possible to establish a correspondence between
our non-minimal couplings $g_{n>0}^\pm$
and the dimensionless worldline Wilson coefficients
$C_{{\rm ES}^{2n}}$ and $C_{{\rm BS}^{2n+1}}$.
To do this, here we rely on the scattering amplitude
\[
   {\cal A}^\pm_\text{gen}(p,k)
    = -\frac{\kappa}{2} m^2 x^{\pm 2}
      \bigg[ \sum_{n=0}^\infty \frac{C_{{\rm ES}^{2n}}}{(2n)!}
             (\bar{k} \cdot a)^{2n}
             \pm \sum_{n=0}^\infty \frac{C_{{\rm BS}^{2n+1}}}{(2n+1)!}
             (\bar{k} \cdot a)^{2n+1}
      \bigg] ,
\label{LSAmplitude}
\]
which follows from the worldline action above
and is derived in \app{app:Nonminimal}.
The classical limit of the amplitude~\eqref{A3ptGen}
may be reorganized in a similar fashion:
\[
\begin{aligned}
\label{A3ptGen2}
   {\cal A}_{3,\text{gen}}^\pm \big|_{\beta=\alpha} &
    = -\frac{\kappa}{2} m^2 x^{\pm 2}
      e^{\mp \bar{k} \cdot a_{\rm a}}\!
      \sum_{n=0}^\infty g_n^\pm
      \bigg[\frac{\pm 2 \bar{k} \cdot a_{\rm a}}
                 {\|\alpha\|^2}
      \bigg]^{\!n} \\ &
%    = -\frac{\kappa}{2} m^2 x^{\pm 2}
%      \sum_{t=0}^\infty
%      \frac{1}{t!} \big({\mp}\bar{k} \cdot a_{\rm a}\big)^t
%      \sum_{r=0}^\infty g_r^\pm
%      \bigg[\frac{\pm 2 \bar{k} \cdot a_{\rm a}}
%                 {\|\alpha\|^2}
%      \bigg]^{\!r} \\ &
   = -\frac{\kappa}{2} m^2 x^{\pm 2}
      \sum_{n=0}^\infty \big({\mp}\bar{k} \cdot a_{\rm a}\big)^n
      \sum_{r=0}^n \frac{(-2)^r g_r^\pm}{(n-r)!\|\alpha\|^{2r}} .
\end{aligned}
\]
We can therefore read off the worldline Wilson coefficients
implied by the non-minimal amplitudes~\eqref{GravityMatter3ptNonMin}:
\[
   C_{{\rm ES}^{2n}} = \sum_{r=0}^{2n}
      \frac{(2n)! (-2)^r g_r^\pm}{(2n-r)! \|\alpha\|^{2r}} , \qquad \quad
   C_{{\rm BS}^{2n+1}} = -\sum_{r=0}^{2n+1}
      \frac{(2n+1)! (-2)^r g_r^\pm}{(2n-r+1)! \|\alpha\|^{2r}} .
\label{WilsonCoefficientMap}
\]
This matching clearly requires that $g_n^+$ and $g_n^-$
be real and equal to each other.
Moreover, let us point out the fact that,
as explained in \app{app:Nonminimal},
in passing from the action~\eqref{LSInteractions}
to the amplitude~\eqref{LSAmplitude}
we had to introduce the $n=0$ terms
with $C_{{\rm ES}^0} = -C_{{\rm BS}^1} = 1$,
which unequivocally follow from the worldline kinetic terms.
Hence
\[
   g_0^\pm = C_{{\rm ES}^0} = 1 , \qquad \quad
   g_1^\pm = \|\alpha\|^2(C_{{\rm BS}^1}+g_0^\pm)/2 = 0 .
\]

The crucial feature of
the Wilson-coefficient map~\eqref{WilsonCoefficientMap} is that,
in order to describe a generic classical particle
using the three-point amplitudes~\eqref{GravityMatter3ptNonMin},
one must consider the non-minimal coupling constants $g_{n>0}$
that depend on the classical spin length~$a$ of the particle via
\[
   \|\alpha\|^2 = \frac{2m}{\hbar} \sqrt{-a^2} .
\]
The only gravitational objects escaping this rule
seem to be black holes, for which $g_0^\pm = 1$ and $g_{n>0}^\pm=0$.

In this section, we have naturally landed on equal coefficients
$g_n^+=g_n^-$ for positive- and negative- helicity amplitudes.
Non-equal coefficients can be motivated by the electric-magnetic duality,
which in electromagnetism mixes the electric and magnetic charges.
The gravitational electric-magnetic duality
relates the mass and NUT charge parameter
in a similar manner~\cite{Henneaux:2004jw}.
Together with double copy and the Newman-Janis shift~\cite{Newman:1965tw},
the duality generates a whole web of theories described in
\rcites{Huang:2019cja,Emond:2020lwi},
in which the three-point coupling coefficients $g_n^\pm$
develop complex phases while still being related by complex conjugation.

%%%%%%%%%%%%%%%%%%%%%%%%%%%%%%%%%%%%%%%%%%%%%%%%%%
\subsection{Unequal spin amplitudes}
\label{sec:UnequalSpin}
%%%%%%%%%%%%%%%%%%%%%%%%%%%%%%%%%%%%%%%%%%%%%%%%%%

So far we have been focusing on the diagonal slice $s_1=s_2$
of the double sum in the three-point coherent-spin amplitude~\eqref{3pt}.
This is because, as we will show in this section,
the off-diagonal contributions
vanish in the classical limit,
or at least may be considered to vanish
unless certain artificial assumptions are made beforehand.

For concreteness,
let us consider the positive-helicity amplitudes with $s_1>s_2$,
which may be written in full generality as \cite{Arkani-Hamed:2017jhn}
\begin{align}
\label{GravityMatter3ptUnequal}
   {\cal A}_{s_1>s_2}^{(0)\{b\}}{}_{\{a\}}(p_2,s_2|p_1,s_1;k,+) &
%    =-\frac{\kappa}{2} \sum_{n=s_1-s_2}^{s_1+s_2}\!g_{n,s_1,s_2}
%      \frac{x^{n+2} \braket{2^b 1_a}^{\odot (s_1+s_2-n)}\!}
%           {m^{s_1+s_2+n-2}} \\ & \qquad
%      \odot \braket{2^b k}^{\odot (n-s_1+s_2)}
%      \odot \braket{k\:\!1_a}^{\odot (n+s_1-s_2)} . \nn \\ &
    =-\frac{\kappa}{2} \sum_{n=0}^{2s_2} g_{n,s_1,s_2}
      \frac{x^{s_1-s_2+n+2}}{m^{2s_1+n-2}} \\ & \qquad\!\times
      \braket{2^b 1_a}^{\odot (2s_2-n)}
      \odot \braket{2^b k}^{\odot n}
      \odot \braket{k\:\!1_a}^{\odot (2s_1-2s_2+n)} . \nn
\end{align}
Dressing it with coherent states gives
\[
\begin{aligned}
   {\cal A}_{3,s_1>s_2}^+ &
    = -\frac{\kappa}{2} m^2 x^2 e^{-(\|\alpha\|^2 + \|\beta\|^2)/2}\!
      \sum_{2s_2=0}^\infty\,\sum_{2s_1=2s_2+1}^\infty\,
      \sum_{n=0}^{2s_2}
      \frac{g_{n,s_1,s_2} x^{s_1-s_2+n}}
           {\sqrt{(2s_1)!(2s_2)!} m^{2s_1+n}} \\ & \qquad\,\qquad \times
      \big(\tilde{\beta}_b \braket{2^b 1_a} \alpha^a\big)^{2s_2-n}
      \big(\tilde{\beta}_b \braket{2^b k} \braket{k\:\!1_a} \alpha^a\big)^n
      \big(\braket{k\:\!1_a} \alpha^a\big)^{2(s_1-s_2)} .
\end{aligned}
\]
We are now interested in the behavior of the above triple sum
in the classical limit, in which
\[
   \|\alpha\| = {\cal O}(\hbar^{-1/2}) = \|\beta\| , \qquad \quad
   \ket{k} = {\cal O}(\hbar^{1/2})  .
\]
So the only factor that has a chance to counteract
the vanishing of the exponential prefactor is the term
\[
   \frac{1}{m}\tilde{\beta}_b(p_2) \braket{2^b 1_a} \alpha^a(p_1)
    = \tilde{\beta}_a(p_{\rm a}) \alpha^a(p_{\rm a})
    - \frac{1}{m} \bar{k}_\mu \braket{\beta|S_{p_{\rm a}}^\mu|\alpha} ,
\label{SpinorProduct2Spin2}
\]
in which the second contribution is classically finite
but the first one grows as ${\cal O}(1/\hbar)$.
Indeed, we have seen in \sec{sec:MinimalCoupling}
how this contribution can in principle combine with the exponential prefactor
to produce the coherent-state overlap $\braket{\beta|\alpha}$,
which is finite unless $\|\alpha-\beta\|\to\infty$,
as follows from $|\braket{\beta|\alpha}|^2 = e^{-\|\alpha-\beta\|^2}$.
Let us therefore denote the finite dimensionless contributions
${\cal O}(\hbar^0)$ as
\[
   \frac{\tilde{\beta}_b(p_2) \braket{2^b 1_a} \alpha^a(p_1)}
        {m \tilde{\beta}_a(p_{\rm a}) \alpha^a(p_{\rm a})} = A , \quad
   \frac{x}{m^2} \tilde{\beta}_b(p_2) \braket{2^b k}
      \braket{k\:\!1_a} \alpha^a(p_1) = B , \quad
   \frac{\sqrt{x}}{m} \braket{k\:\!1_a} \alpha^a(p_1) = C ,
\]
in terms of which the amplitude contribution writes
\[\!\!
   {\cal A}_{3,s_1>s_2}^+\!
    = -\frac{\kappa}{2} m^2 x^2 e^{-(\|\alpha\|^2 + \|\beta\|^2)/2}\!
      \sum_{2s_2=0}^\infty \sum_{r=1}^\infty \sum_{n=0}^{2s_2}
      \frac{g_{n,s_2+r/2,s_2} B^n C^r}{\sqrt{(2s_2+r)!(2s_2)!}}
      \big(A \tilde{\beta}_a \alpha^a \big)^{2s_2-n}\!.\!
\]

Moreover, although we have allowed the coupling constants $g_{n,s_1,s_2}$
to depend on the spin quantum numbers,
in the following estimations let us replace them with
\[
   y_n \equiv \sup_{s_1>s_2} |g_{n,s_1,s_2}| .
\]
This is going to allow us to change the order of summation:
\begin{align}
   \bigg| \frac{2 {\cal A}_{3,s_1>s_2}^+}{\kappa\;\!m^2 x^2} \bigg| & \leq
%      e^{-(\|\alpha\|^2 + \|\beta\|^2)/2}\!
%      \sum_{2s_2=0}^\infty \sum_{r=1}^\infty \sum_{n=0}^{2s_2}
%      \frac{y_n |A|^{2s_2-n}\,|B|^n\,|C|^r}{\sqrt{(2s_2+r)!(2s_2)!}}
%      |\tilde{\beta}_a \alpha^a|^{2s_2-n} \\ & =
      e^{-(\|\alpha\|^2 + \|\beta\|^2)/2}
      \sum_{n=0}^\infty \frac{y_n |B|^n}{|A \tilde{\beta}_a \alpha^a|^n}
      \sum_{r=1}^\infty |C|^r\!
      \sum_{2s_2=n}^\infty\!
      \frac{|A \tilde{\beta}_a \alpha^a|^{2s_2}}{\sqrt{(2s_2+r)!(2s_2)!}} .
\end{align}
In order to deal with the square roots of the factorials,
we use the following inequality
\[
   (t+r)!\,t! = \Gamma(t+r+1)\,\Gamma(t+1)
    ~>~ \Gamma^2(t+3/2)\,\Gamma(r) , \qquad \quad
   t , r = 1, 2, \ldots
\]
The resulting sum may be evaluated using
\[
   \sum_{t=n}^\infty \frac{a^t}{\Gamma(t+3/2)}
    = \frac{e^a}{\sqrt{a}}
      \bigg[ 1 - \frac{\Gamma(n+1/2,a)}{\Gamma(n+1/2)} \bigg]
    < \frac{e^a}{\sqrt{a}} , \qquad \quad a>0 .
\]
Hence our estimation becomes
\[
   \bigg| \frac{2 {\cal A}_{3,s_1>s_2}^+}{\kappa\;\!m^2 x^2} \bigg| <
      e^{-(\|\alpha\|^2 + \|\beta\|^2)/2}
      \sum_{n=0}^\infty \frac{y_n |B|^n}{|A \tilde{\beta}_a \alpha^a|^n}
      \sum_{r=1}^\infty \frac{|C|^r}{\sqrt{(r-1)!}}
      \frac{\exp|A \tilde{\beta}_a \alpha^a|}
           {\sqrt{|A \tilde{\beta}_a \alpha^a|}} .
\]
Here we may recall that $A \tilde{\beta}_a \alpha^a$
is precisely given by \eqn{SpinorProduct2Spin2},
so we finally obtain\footnote{In the discussion
following \eqn{NonequalSpinEstimation},
we assume that the sums in the second line of the equation converge
regardless of the $\hbar$ scaling.
In fact, the series in $r$ is absolutely convergent
by the d'Alembert ratio test.
The convergence of the series in $n$ seems to depend
on the values of $A$, $B$ and $\tilde{\beta}_a \alpha^a$,
but it is natural to assume that the coupling constants~$g_{n,s_1,s_2}$
(and their suprema~$y_n$) contain a factorial dependence on $n$,
such as that implied by
the Wilson-coefficent map~\eqref{WilsonCoefficientMap} for equal spins.}
\[
\begin{aligned}
   \big| {\cal A}_{3,s_1>s_2}^+ \big|
    < \frac{\kappa}{2} m^2 x^2
      \frac{e^{-(\|\alpha\| - \|\beta\|)^2/2}\!}
           {|\tilde{\beta}_a \alpha^a|^{1/2}}
      \exp\Big| \frac{1}{m} \bar{k}_\mu
                 \braket{\beta|S_{p_{\rm a}}^\mu|\alpha} \Big| & \\ \times
      \sum_{n=0}^\infty
      \frac{y_n |B|^n}{|\tilde{\beta}_a \alpha^a|^n\,|A|^{n+1/2}}
      \sum_{r=1}^\infty \frac{|C|^r}{\sqrt{(r-1)!}} & .
\label{NonequalSpinEstimation}
\end{aligned}
\]

In the classical limit, we are interested in $\beta \approx \alpha$,
including the case where $\tilde{\beta}_a = (\alpha^a)^*$,
which means that the relative ``cosine''
inside the inner product $\tilde{\beta}_a \alpha^a$
should be regarded as at least finite.
Therefore, the product $\tilde{\beta}_a \alpha^a$
grows as ${\cal O}(1/\hbar)$.
Similarly to situation with the equal-spin non-minimal couplings
in \sec{sec:Nonminimal},
the coupling-constants~$y_n$ are accompanied by factors of
$1/(\tilde{\beta}_a \alpha^a)^n$ and are thus power-suppressed
--- unless they or their counterparts $g_{n,s_1,s_2}$ are rescaled
at the level of the scattering amplitude~\eqref{GravityMatter3ptUnequal}.
For equal spins, however, such a rescaling procedure was motivated
by the knowledge of classical multipole interactions,
which could be modeled by proportionally amplified coupling constants.
In addition, the crucial difference between equal and unequal spins
is the presence of an overall factor
$1/|\tilde{\beta}_a \alpha^a|^{1/2}={\cal O}(\hbar^{1/2})$
in the estimation~\eqref{NonequalSpinEstimation},
which means that even the ``almost minimal'' couplings $g_{0,s_1,s_2}$
would need to scale as ${\cal O}(\hbar^{-1/2})$.
Since we are not aware of classical interactions
that would benefit from such an elaborate rescaling procedure ---
which in this case could not be consistently implemented
by a $\ket{k} \to \ket{\bar{k}}$ switch,
and moreover the classical interpretation of the expression
$C = \sqrt{x} \braket{k\:\!1_a} \alpha^a/m$ seems rather obscure,
we see no interest in trying to retain
the non-equal spin couplings in the classical limit.
We are thus vindicated for ignoring the off-diagonal contributions
in the coherent-spin amplitude~\eqref{3pt}.

%%%%%%%%%%%%%%%%%%%%%%%%%%%%%%%%%%%%%%%%%%%%%%%%%%
\section{Elastic gravitational scattering}
\label{sec:4pt}
%%%%%%%%%%%%%%%%%%%%%%%%%%%%%%%%%%%%%%%%%%%%%%%%%%

In this section we return to the four-point amplitude
${\cal A}^{(0)}(p_1',\alpha;p_2',\beta|p_1,\alpha;p_2,\beta)$,
which determines the leading-order impulse and spin-kick observables
via the formulae~\eqref{ImpulseMomLOFinal} and~\eqref{ImpulseSpinLOFinal}.
We have already explained that the classical limit of this amplitude
is naturally factorized into a product of two three-point amplitudes,
as depicted in \fig{fig:4pt}.
Now that we have extensively dissected these lower-point ingredients,
we may proceed to constructing the four-point amplitude.

For the sake of generality,
we will use the ``classical'' amplitudes~\eqref{LSAmplitude},
which in \app{app:Nonminimal} are derived
from the worldline effective action with free Wilson coefficients
\[
   C_{2n} \equiv C_{{\rm ES}^{2n}} , \qquad \quad
   C_{2n+1} \equiv C_{{\rm BS}^{2n+1}} .
\]
In view of the matching~\eqref{WilsonCoefficientMap},
we may easily reinterpret these amplitudes as the classical limit
of the coherent-state amplitudes~\eqref{A3ptGen}.
After the above Wilson-coefficient relabeling,
we can write them simply as
\[
   {\cal A}^{(0)}(p_1',\alpha|p_1,\alpha;k,\pm)
    = -\frac{\kappa}{2} m_{\rm a}^2 x_{\rm a}^{\pm 2}
      \sum_{n=0}^\infty
      \frac{C_{{\rm a}\:\!n}}{n!} \big({\pm}\bar{k} \cdot a_{\rm a}\big)^n
    + {\cal O}(\hbar^0) ,
\label{LSAmplitude2}
\]
and similarly for particle ${\rm b}$, except a sign switch for $\bar{k}$.
If we plug these amplitudes into \eqn{ElasticScatteringAmplitude},
we run into the helicity-factor ratios,
which on the $t$-channel pole kinematics evaluate to
\[
   x_{\rm a}/x_{\rm b} = \gamma(1-v) , \qquad \quad
   x_{\rm b}/x_{\rm a} = \gamma(1+v) .
\label{xFactorRatios}
\]
Here we have used the relative Lorentz factor $\gamma$
and the corresponding velocity $v$ defined by
\[
   \gamma = \frac{1}{\sqrt{1-v^2}}
    = \frac{p_{\rm a}\!\cdot p_{\rm b}}{m_{\rm a} m_{\rm b}} .
\label{gammaFactor}
\]
The physical meaning of these quantities is that
in the rest frame of one of the incoming particles
the other one moves with speed $v$, as in
\[
   \left\{
   \begin{aligned}
      p_{\rm a}^\mu & = \gamma m_{\rm a} (1,0,0,v) , \\
      p_{\rm b}^\mu & = (m_{\rm b},0,0,0) ,
   \end{aligned}
   \right. \qquad
   \overset{\substack{\text{Lorentz boost}\\~}}{\Leftrightarrow} \qquad
   \left\{
   \begin{aligned}
      p_{\rm a}^\mu & = (m_{\rm a},0,0,0) , \\
      p_{\rm b}^\mu & = \gamma m_{\rm b} (1,0,0,-v) .
   \end{aligned}
   \right.
\]

Therefore, the classical limit of the elastic scattering amplitude
\eqref{ElasticScatteringAmplitude} becomes
\[
\begin{aligned}
\label{ElasticScatteringAmplitude2}
   {\cal A}^{(0)}(p_1',\alpha;p_2',\beta|p_1,\alpha;p_2,\beta)
    =-\frac{\kappa^2 m_{\rm a}^2 m_{\rm b}^2}{4\hbar^2 \bar{k}^2}
      \sum_\pm\;\!&\:\!\! \gamma^2 (1 \mp v)^2 \\ \times
      \sum_{n_1=0}^\infty
      \frac{C_{{\rm a}\:\!n_1}}{n_1!}
      \big({\pm}\bar{k} \cdot a_{\rm a}\big)^{n_1}
      \sum_{n_2=0}^\infty &
      \frac{C_{{\rm b}\:\!n_2}}{n_2!}
      \big({\pm}\bar{k} \cdot a_{\rm b}\big)^{n_2}
    + {\cal O}(\hbar^{-5/2}) .
\end{aligned}
\]
We are still not entirely ready to analytically continue this expression
away from the $t$-channel pole kinematics,
since it contains parity-odd products of the type~$\bar{k} \cdot a$
--- which make sense on the three-point kinematics
and are naturally accompanied by the helicity-dependent signs,
but are alien to real-valued classical physics.
Note, for example, that in order to make the transition between
a clearly parity-even action~\eqref{LSInteractions}
and the amplitude expression~\eqref{LSAmplitude}
we need the three-point identity~\eqref{Even2Odd3pt}
involving the massless polarization vector.
Now that we wish to go away from the on-shell kinematics for the graviton,
using \eqn{Even2Odd3pt} is not an option.
There are, however, four-point identities
which involve the Levi-Civita tensor
and are valid on the $t$-channel pole kinematics,
namely \cite{Guevara:2019fsj}
\[
   i\epsilon_{\mu\nu\rho\sigma}
      p_{\rm a}^\mu p_{\rm b}^\nu \bar{k}^\rho a_{\rm a}^\sigma
    = m_{\rm a} m_{\rm b} \gamma v (\bar{k} \cdot a_{\rm a}) , \qquad \quad
   i\epsilon_{\mu\nu\rho\sigma}
      p_{\rm a}^\mu p_{\rm b}^\nu \bar{k}^\rho a_{\rm b}^\sigma
    = m_{\rm a} m_{\rm b} \gamma v (\bar{k} \cdot a_{\rm b}) .
\label{Even2Odd4pt}
\]
For brevity, we follow \rcite{Vines:2017hyw} in introducing the notation
\[
   w^{\mu\nu} = \frac{2p_{\rm a}^{[\mu} p_{\rm b}^{\nu]}}
                     {m_{\rm a} m_{\rm b} \gamma v} , \qquad \quad
   [w*a]_\lambda = (*w)_{\lambda\mu} a^\mu
    = \frac{\epsilon_{\lambda\mu\nu\rho} p_{\rm a}^\mu p_{\rm b}^\nu a^\rho}
           {m_{\rm a} m_{\rm b} \gamma v} ,
\label{wInForm}
\]
which allows us to write
a new expression for the elastic scattering amplitude
that may be understood beyond the $t$-channel pole kinematics:
\begin{align}
\label{ElasticScatteringAmplitudeFinal}
   {\cal A}^{(0)}(p_{\rm a}\!+\!k/2,\alpha;p_{\rm b}\!-\!k/2,\beta
                 |p_{\rm a}\!-\!k/2,\alpha;p_{\rm b}\!+\!k/2,\beta)
    =-\frac{8\pi G m_{\rm a}^2 m_{\rm b}^2 \gamma^2}{\hbar^3 \bar{k}^2}
      ~~\: & \\ \times
      \sum_\pm (1 \mp v)^2\!\!
      \sum_{n_1,n_2=0}^\infty\!\!
      \frac{C_{{\rm a}\:\!n_1} C_{{\rm b}\:\!n_2}\!}{n_1!n_2!}
      \big({\pm} i\bar{k} \cdot [w*a_{\rm a}]\big)^{n_1}
      \big({\pm} i\bar{k} \cdot [w*a_{\rm b}]\big)^{n_2} &
    + {\cal O}(\hbar^{-5/2}) . \nn
\end{align}

%%%%%%%%%%%%%%%%%%%%%%%%%%%%%%%%%%%%%%%%%%%%%%%%%%
\subsection{Eikonal phase}
\label{sec:EikonalIntegration}
%%%%%%%%%%%%%%%%%%%%%%%%%%%%%%%%%%%%%%%%%%%%%%%%%%

Our formulae~\eqref{ImpulseMomLOFinal} and~\eqref{ImpulseSpinLOFinal}
for both of the leading-order impulse observables
involve the eikonal Fourier transform of the above amplitude
\[
   {\cal A}_4^{(0)}(b)
    = \int_k e^{-i \bar{k} \cdot b}
      {\cal A}^{(0)}(p_{\rm a}\!+\!k/2,\alpha;p_{\rm b}\!-\!k/2,\beta
                    |p_{\rm a}\!-\!k/2,\alpha;p_{\rm b}\!+\!k/2,\beta) ,
\label{EikonalAmplitude}
\]
where $\int_k$ is the measure \eqref{EikonalMeasure}.
This object is widely referred to as
the eikonal phase \cite{Bjerrum-Bohr:2018xdl,
Bern:2019nnu,Bern:2019crd,Bern:2020buy,Aoude:2020ygw,Kosmopoulos:2021zoq,
Bern:2021dqo,Bjerrum-Bohr:2021vuf,Bjerrum-Bohr:2021din}.
In view of the form of the integrand~\eqref{ElasticScatteringAmplitudeFinal},
let us proceed to computing the integrals
\[
   I^{(n)}_{\mu_1\dots\mu_n}(b_\perp)
    = \int_k \frac{e^{-i \bar{k} \cdot b}}{\bar{k}^2}
      \bar{k}_{\mu_1}\!\cdots \bar{k}_{\mu_n}
    = \frac{\hbar^2}{4}\!\int\!\frac{d^4 \bar{k}}{(2\pi)^2}
      \delta(p_{\rm a}\!\cdot \bar{k}) \delta(p_{\rm b}\!\cdot \bar{k})
      \frac{e^{-i \bar{k} \cdot b_\perp}}{\bar{k}^2}
      \bar{k}_{\mu_1}\!\cdots \bar{k}_{\mu_n} ,
\label{EikonalIntegrals}
\]
where we have extracted the Planck constants inherent to the eikonal measure.
We have also taken care to indicate
that these integrals only depend on
$b_\perp^\mu = \Pi^\mu{}_\nu(p_{\rm a},p_{\rm b}) b^\nu$,
which lies in ${\rm E}^\perp_{p_{\rm a},p_{\rm b}}$,
as opposed to the original impact parameter
$b \in {\rm E}^\perp_{u_{\rm a},u_{\rm b}}$.

The simplest of these integrals is a scalar
that may be computed in the center-of-mass (COM) frame, in which
$\bar{k}$ is constrained to be
$(0,\bar{k}^1,\bar{k}^2,0) \in {\rm E}^\perp_{p_{\rm a},p_{\rm b}}$.
The result is
\[
   I^{(0)}(b_\perp) = \frac{\hbar^2}{8\pi m_{\rm a} m_{\rm b} \gamma v}
      \Big( \log\sqrt{-b_\perp^2} + C_\epsilon \Big) , \qquad \quad
   C_\epsilon
    = \log\frac{\epsilon}{2} + \gamma_\text{Euler} + {\cal O}(\epsilon^2) .
\label{EikonalIntegral0}
\]
where $C_\epsilon$ is a logarithmically divergent constant,
which we wrote in terms of the infrared regulator
$\epsilon < |\bar{\bs{k}}_\perp|$.
Being interested in the observables, the formulae for which only depend
on the derivatives of ${\cal A}^{(0)}(b)$,
we may safely omit this infinite constant later.
All of the tensor integrals~\eqref{EikonalIntegrals}
may now be obtained as partial derivatives
\[
   I_{(n)}^{\mu_1\dots\mu_n}(b_\perp)
    = i^n \Pi^{\mu_1\nu_1} \cdots \Pi^{\mu_n\nu_n}
      \frac{\partial~}{\partial b_\perp^{\nu_1}} \cdots
      \frac{\partial~}{\partial b_\perp^{\nu_n}} I^{(0)}(b_\perp) .
\]
The projectors~\eqref{TransverseProjector} are important
to ensure that that the resulting expression stays transverse
in all of its indices.
In this way, one can obtain explicit answers like
\[
   I_{(1)}^\mu(b_\perp) = \frac{i\hbar^2}{8\pi m_{\rm a} m_{\rm b} \gamma v}
      \frac{b_\perp^\mu}{b_\perp^2} , \qquad \quad
   I_{(2)}^{\mu\nu}(b_\perp)
    = \frac{-\hbar^2}{8\pi m_{\rm a} m_{\rm b} \gamma v}
      \frac{b_\perp^2 \Pi^{\mu\nu} - 2b_\perp^\mu b_\perp^\nu}{b_\perp^4} ,
\]
which match those in \eg \rcite{Maybee:2019jus}.
However, we choose to exploit the fact
that the numerator of our amplitude~\eqref{ElasticScatteringAmplitudeFinal}
depends on $\bar{k}$ exclusively via $\bar{k} \cdot [w*a_{\rm a}]$,
which already does the job of projecting the non-transverse components
of any such integral for us.
%(In fact, $\Pi^\mu{}_\nu = (*w)^{\mu\rho} (*w)_{\nu\rho}$.)
We may therefore write the eikonal phase~\eqref{EikonalAmplitude}
directly in terms of derivatives:
\begin{align}
\label{EikonalAmplitudeFinal}
   {\cal A}_4^{(0)}(b) = & 
     -\frac{G m_{\rm a} m_{\rm b} \gamma}{\hbar\:\!v}
      \sum_\pm (1 \pm v)^2\!\!
      \sum_{n_1,n_2=0}^\infty\!\!
      \frac{(\pm 1)^{n_1+n_2}\!}{n_1!n_2!}
      C_{{\rm a}\:\!n_1} C_{{\rm b}\:\!n_2} \\ & ~~\;\times
      \big([w*a_{\rm a}] \cdot \partial_{b_\perp} \big)^{n_1}
      \big([w*a_{\rm b}] \cdot \partial_{b_\perp} \big)^{n_2}
      \log\sqrt{-b_\perp^2}
%      \Big( \log\sqrt{-b_\perp^2} + C_\epsilon \Big)
    + {\cal O}(\hbar^{-1/2}) . \nn
\end{align}

As a cross-check, we can switch to Kerr scattering by setting
$C_{{\rm a}\:\!n} = C_{{\rm b}\:\!n} = (-1)^n$.
Then the infinite sums organize themselves into translation operators,
and \eqn{EikonalAmplitudeFinal} becomes
\begin{align}
   {\cal A}_4^{(0)}(b) &
%   =-\frac{G m_{\rm a} m_{\rm b} \gamma}{\hbar\:\!v}
%      \sum_\pm (1 \pm v)^2
%      \exp\!\big({\mp}[w*(a_{\rm a}+a_{\rm b})]\cdot\partial_{b_\perp}\big)
%      \Big( \log\sqrt{-b_\perp^2} + C_\epsilon \Big)
%    + {\cal O}(\hbar^{-1/2}) \nn \\ &
   =-\frac{G m_{\rm a} m_{\rm b} \gamma}{\hbar\:\!v} \sum_\pm (1 \pm v)^2
      \log\sqrt{-\big(b_\perp \mp w*(a_{\rm a}\!+a_{\rm b})\big)^2}
%      \Big( \log\sqrt{-\big(b_\perp \mp w*(a_{\rm a}\!+a_{\rm b})\big)^2}
%          + C_\epsilon \Big)
    + {\cal O}(\hbar^{-1/2}) ,
\label{EikonalAmplitudeKerr}
\end{align}
which matches the eikonal phase in \rcite{Guevara:2019fsj}.

%%%%%%%%%%%%%%%%%%%%%%%%%%%%%%%%%%%%%%%%%%%%%%%%%%
\subsection{Impulse observables}
\label{sec:ImpulseObservables}
%%%%%%%%%%%%%%%%%%%%%%%%%%%%%%%%%%%%%%%%%%%%%%%%%%

We are now ready to compute the leading-order linear and angular impulses.
First, we apply the formula~\eqref{ImpulseMomLOFinal}
to the eikonal phase~\eqref{EikonalAmplitudeFinal} and obtain
\begin{align}
\label{ImpulseMomLOGrav}
 & \Delta P_{\rm a}^\mu
    = -\hbar \frac{\partial~}{\partial b_\mu\!}
      \int_{p_{\rm a},p_{\rm b}}\!\!\!
      |\psi_{\rm a}(p_{\rm a})|^2 |\psi_{\rm b}(p_{\rm b})|^2
      {\cal A}_4^{(0)}(b) \\ &\!
    = G m_{\rm a} m_{\rm b} \frac{\gamma}{v} \sum_\pm (1 \pm v)^2\!\!
      \sum_{n_1,n_2=0}^\infty\!\!
      \frac{(\pm 1)^{n_1+n_2}\!}{n_1!n_2!}
      C_{{\rm a}\:\!n_1} C_{{\rm b}\:\!n_2}
      \big([w*a_{\rm a}] \cdot \partial_b \big)^{n_1}
      \big([w*a_{\rm b}] \cdot \partial_b \big)^{n_2}
      \frac{b^\mu}{b^2} \bigg|_\text{cl}\!. \nn
\end{align}
Here ``cl'' identification indicates
that the wave-function integration has localized
the initial momenta~$p_{\rm a,b}^\mu$ on their classical values
$m_{\rm a,b} u_{\rm a,b}^\mu$.
This has allowed us to replace the transverse projection
of the impact parameter $b_\perp \in {\rm E}^\perp_{p_{\rm a},p_{\rm b}}$
by the original quantity $b \in {\rm E}^\perp_{u_{\rm a},u_{\rm b}}$.
Other quantities are also naturally understood in terms
of the initial four-velocities:
\[
   \gamma_\text{cl} = \frac{1}{\sqrt{1-v_\text{cl}^2}}
    = u_{\rm a}\!\cdot u_{\rm b} , \qquad \quad
   [w_\text{cl}*a]_\lambda = \frac{1}{\sqrt{\gamma_\text{cl}^2-1}}
      \epsilon_{\lambda\mu\nu\rho} u_{\rm a}^\mu u_{\rm b}^\nu a^\rho .
\label{ClassicalAnalogues}
\]
In particular, the spin-length expectation values,
defined at momenta $p_{\rm a,b}$ in \eqn{SpinLength},
are now identified with the initial classical angular momenta
$s_{\rm a,b} = m_{\rm a,b} a_{\rm a,b}^\text{cl}$.

Similarly, the spin kick formula~\eqref{ImpulseSpinLOFinal}
yields the following answer:
\begin{align}
   \Delta S_{\rm a}^\mu &
    = \frac{\hbar}{m_{\rm a}\!}
      \int_{p_{\rm a},p_{\rm b}}\!\!\!
      |\psi_{\rm a}(p_{\rm a})|^2 |\psi_{\rm b}(p_{\rm b})|^2
      \bigg[
      p_{\rm a}^\mu a_{\rm a}^\nu \frac{\partial~}{\partial b^\nu\!}
    - \epsilon^{\mu\nu\rho\sigma} p_{{\rm a}\:\!\nu} a_{{\rm a}\rho}
      \frac{\partial~~}{\partial a_{\rm a}^\sigma}
      \bigg] {\cal A}_4^{(0)}(b) \nn \\ &
    = -G m_{\rm a} m_{\rm b} \frac{\gamma}{v} \sum_\pm (1 \pm v)^2\!\!
      \sum_{n_1,n_2=0}^\infty\!\!
      \frac{(\pm 1)^{n_1+n_2}\!}{n_1!n_2!}
      C_{{\rm a}\:\!n_1} C_{{\rm b}\:\!n_2}
      \big([w*a_{\rm b}] \cdot \partial_{b} \big)^{n_2}
\label{ImpulseSpinLOGrav} \\ & \qquad \qquad \qquad \qquad\:\,\,\times\!
      \bigg[
      u_{\rm a}^\mu a_{\rm a}^\nu \frac{\partial~}{\partial b^\nu\!}
    - \epsilon^{\mu\nu\rho\sigma} u_{{\rm a}\:\!\nu} a_{{\rm a}\rho}
      \frac{\partial~~}{\partial a_{\rm a}^\sigma}
      \bigg]
      \big([w*a_{\rm a}] \cdot \partial_{b} \big)^{n_1}
      \log\sqrt{-b^2} \bigg|_\text{cl}\!, \nn %\\ &\!
%    = -G m_{\rm a} m_{\rm b} \frac{\gamma}{v} \sum_\pm (1 \pm v)^2\!\!
%      \sum_{n_1,n_2=0}^\infty\!\!
%      \frac{(\pm 1)^{n_1+n_2}\!}{n_1!n_2!}
%      C_{{\rm a}\:\!n_1} C_{{\rm b}\:\!n_2}
%      \big([w*a_{\rm b}] \cdot \partial_{b} \big)^{n_2}
%      \nn \\ & \quad~~\times\!
%      \bigg[
%      \big([w*a_{\rm a}] \cdot \partial_{b} \big)^{n_1}
%      u_{\rm a}^\mu \frac{b \cdot a_{\rm a}}{b^2}
%    + \frac{n_1}{\gamma v}
%      \big([w*a_{\rm a}] \cdot \partial_{b} \big)^{n_1-1}
%      \frac{1}{b^2}
%      \big[ (u_{\rm b} \cdot a_{\rm a}) b^\mu
%          - (u_{\rm b}^\mu - \gamma u_{\rm a}^\mu) (b \cdot a_{\rm a}) \big]
%      \bigg] \nn
\end{align}
where the derivatives in the square brackets could be easily
evaluated further for the price of enlarging the final expression.
Note there is always at least one derivative acting on the logarithm,
thus ensuring that the answer stays independent
of the implicit infrared singularity.
More structure expectedly arises in the Kerr black-hole scattering case,
in which we derive
\begin{subequations} \begin{align}
\label{ImpulsesMomLOKerr}
 & \Delta P_{\rm a}^\mu
    = G m_{\rm a} m_{\rm b} \frac{\gamma}{v} \sum_\pm (1 \mp v)^2
      \frac{[b \pm w*(a_{\rm a}\!+a_{\rm b})]^\mu}
           {[b \pm w*(a_{\rm a}\!+a_{\rm b})]^2} \bigg|_\text{cl}\!, \\
 & \Delta S_{\rm a}^\mu
%    = -G m_{\rm a} m_{\rm b} \frac{\gamma}{2v} \sum_\pm (1 \mp v)^2
%      \bigg[
%      u_{\rm a}^\mu a_{\rm a}^\nu \frac{\partial~}{\partial b^\nu\!}
%    - \epsilon^{\mu\nu\rho\sigma} u_{{\rm a}\:\!\nu} a_{{\rm a}\rho}
%      \frac{\partial~~}{\partial a_{\rm a}^\sigma}
%      \bigg]
%      \log\big({-}[b \pm w*(a_{\rm a}\!+a_{\rm b})]^2\big) \\ &
    = -G m_{\rm a} m_{\rm b} \frac{\gamma}{v}
      \sum_\pm \frac{(1 \mp v)^2}{[b \pm w*(a_{\rm a}\!+a_{\rm b})]^2\!}
      \bigg[
      (a_{\rm a}\!\cdot [b \pm w*a_{\rm b}])\;\!u_{\rm a}^\mu \\ &
\label{ImpulsesSpinLOKerr} \qquad~\,\,\quad
  \pm \frac{1}{\gamma v}
      \Big( (u_{\rm b}\!\cdot a_{\rm a})\;\!
            [b \pm w*(a_{\rm a}\!+a_{\rm b})]^\mu            
          - (a_{\rm a}\!\cdot [b \pm w*a_{\rm b}])\;\!
            [u_{\rm b}\!- \gamma u_{\rm a}]^\mu
      \Big)
      \bigg] \bigg|_\text{cl}\!. \nn
\end{align} \label{ImpulsesLOKerr}%
\end{subequations}
We have verified that these expressions
are entirely equivalent to
the leading-order black-hole results
first obtained by Vines \cite{Vines:2017hyw}.

%%%%%%%%%%%%%%%%%%%%%%%%%%%%%%%%%%%%%%%%%%%%%%%%%%
\subsection{Effective Hamiltonian}
\label{sec:Hamiltonian}
%%%%%%%%%%%%%%%%%%%%%%%%%%%%%%%%%%%%%%%%%%%%%%%%%%

An alternative route to classical mechanics due to gravity,
that is notably also usable for bound-state problems
such as binary compact-object inspirals,
is to pass via an effective two-body Hamiltonian.
It is convenient to set it up in the center-of-mass frame, in which
\[\!
   p_{\rm a} = (E_{\rm a}, \bs{p}) ,\!\!\qquad
   p_{\rm b} = (E_{\rm b}, -\bs{p}) ,\!\!\qquad
   E_j =  \sqrt{\bs{p}^2 + m_j^2} ,\!\!\qquad
   \gamma v = \frac{(E_{\rm a}+E_{\rm b}) |\bs{p}|}{m_{\rm a} m_{\rm b}} .
\label{COMFrame}
\]
Then the conservative Hamiltonian is composed of
the kinetic energy $(E_{\rm a}+E_{\rm b})$
and an effective potential $V$
\cite{Damour:2001tu,Porto:2016pyg,Bern:2020buy}:
\[
   H(\bs{r},\bs{p},\bs{S}_{\rm a},\bs{S}_{\rm b})
    = \sqrt{\bs{p}^2 + m_{\rm a}^2} + \sqrt{\bs{p}^2 + m_{\rm b}^2}
    + V(\bs{r},\bs{p},\bs{S}_{\rm a},\bs{S}_{\rm b}) .
\]
The leading-order effective potential may be extracted directly
from the tree-level classical scattering amplitude simply as
\[
   V^{(1)}(\bs{r},\bs{p},\bs{S}_{\rm a},\bs{S}_{\rm b})
    = -\frac{\hbar^3}{4 E_{\rm a} E_{\rm b}}\!
      \int\!\frac{d^3 \bar{\bs{k}}}{(2\pi)^3} e^{i \bar{\bs{k}}\cdot\bs{r}}
      {\cal A}^{(0)}(\bar{\bs{k}},\bs{p},\bs{S}_{\rm a},\bs{S}_{\rm b}) ,
\label{Amplitude2Potential}
\]
whereas a more intricate EFT matching is needed
\cite{Cheung:2018wkq,Cristofoli:2019neg} at higher orders.
Another important difference of this approach from the KMOC formalism
is that it does not involve any additional momentum-wavefunction integration,
so the momenta $p_{\rm a,b}^\mu$ are identified with the classical
incoming momenta $m_{\rm a,b} u_{\rm a,b}^\mu$ from the start.
In fact, the amplitude input is then taken to be
${\cal A}(p_{\rm a}+k,p_{\rm b}-k|p_{\rm a},p_{\rm b})$ --- as opposed to
${\cal A}(p_{\rm a}+k/2,p_{\rm b}-k/2|p_{\rm a}-k/2,p_{\rm b}+k/2)$.
This difference is, however, irrelevant at leading order,
since the classical limit sends $k \to 0$ in the numerator anyway.
Therefore, the COM amplitude may be read off directly from
our result in \eqn{ElasticScatteringAmplitudeFinal}:
\begin{align}
\label{ElasticScatteringAmplitudeCOM}
   {\cal A}^{(0)}(\bar{\bs{k}},\bs{p},\bs{S}_{\rm a},\bs{S}_{\rm b}) = &\;
      \frac{8\pi G m_{\rm a}^2 m_{\rm b}^2 \gamma^2}{\hbar^3 \bar{\bs{k}}^2}
      \sum_\pm (1 \mp v)^2\!\! \\ & \times\!\!
      \sum_{n_1,n_2=0}^\infty\!\!
      \frac{C_{{\rm a}\:\!n_1} C_{{\rm b}\:\!n_2}\!}{n_1!n_2!}
      \bigg( {\pm} \frac{i}{m_{\rm a}\!} \bar{\bs{k}} \cdot
            [\hat{\bs{p}} \times \bs{S}_{\rm a}]
      \bigg)^{\!n_1}
      \bigg( {\pm} \frac{i}{m_{\rm b}\!} \bar{\bs{k}} \cdot
            [\hat{\bs{p}} \times \bs{S}_{\rm b}]
      \bigg)^{\!n_2} , \nn
\end{align}
where the Levi-Civita contractions have naturally been converted
to cross products:
\[
   [w*a]^0 = 0 , \qquad \quad
   [w*a]^i = -[\hat{\bs{p}} \times \bs{a}]^i , \qquad \quad
   \hat{\bs{p}} = \bs{p}/|\bs{p}| .
\]

Let us further justify how we have just traded the spin vectors
$\bs{a}_{\rm a,b}$ for $\bs{S}_{\rm a,b}/m_{\rm a,b}$
in translating between
\eqns{ElasticScatteringAmplitudeFinal}{ElasticScatteringAmplitudeCOM}.
We wish to follow the EFT approach of \rcite{Bern:2020buy},
in which the integer-spin amplitudes are constructed
directly in terms of the rest-frame spin operators $\hat{\bs{S}}_{\rm a,b}$
acting on external polarization tensors.
By invoking a coherent spin-state construction~\cite{Klauder_1985},
these operators themselves were identified with the classical spin vectors
obeying the equations of motion
\[
   \dot{\bs{r}} = \frac{\partial H}{\partial \bs{p}} , \qquad \quad
   \dot{\bs{p}} = -\frac{\partial H}{\partial \bs{r}} , \qquad \quad
   \dot{\bs{S}}_j = -\bs{S}_j \times
      \frac{\partial H}{\partial \bs{S}_j} , \qquad
   j = {\rm a}, {\rm b} .
\label{EoM3d}
\]
Note that the minimal boost from the rest frame of particle ${\rm a}$
to the COM frame is
\begin{align}
   \bs{a}_{\rm a}
    = \frac{\bs{S}_{\rm a}}{m_{\rm a}}
    + \frac{(\bs{p} \cdot \bs{S}_{\rm a}) \bs{p}}
           {m_{\rm a}^2 (E_{\rm a}+m_{\rm a})} , \qquad \quad
   a_{\rm a}^0 = \frac{\bs{p} \cdot \bs{S}_{\rm a}}{m_{\rm a}^2} .
\label{SpinBoost}
\end{align}
(Here and below, the corresponding expressions for particle ${\rm b}$
may be easily obtained by flipping the sign in front of $\bs{p}$.)
Clearly, the triple product
$\bar{\bs{k}} \cdot [\hat{\bs{p}} \times \bs{a}_{\rm a}]$
is insensitive to the spin-length contribution proportional to $\bs{p}$,
so the COM amplitude is indeed given
simply by \eqn{ElasticScatteringAmplitudeCOM}.

Taking its three-dimensional Fourier transform~\eqref{Amplitude2Potential},
we immediately obtain the 1PM effective conservative potential
at first PM order:
\begin{align}
\label{ElasticScatteringHamiltonian}
   V^{(1)}(\bs{r},\bs{p},\bs{S}_{\rm a},\bs{S}_{\rm b}) & =
   {-}\frac{G m_{\rm a}^2 m_{\rm b}^2 \gamma^2}{2 E_{\rm a} E_{\rm b}}
      \sum_\pm (1 \mp v)^2\!\! \\ &\:\!\times\!\!
      \sum_{n_1,n_2=0}^\infty\!\!
      \frac{C_{{\rm a}\:\!n_1} C_{{\rm b}\:\!n_2}\!}{n_1!n_2!}
      \bigg( {\pm} \frac{1}{m_{\rm a}\!}
            [\hat{\bs{p}} \times \bs{S}_{\rm a}] \cdot \nabla_{\bs{r}}
      \bigg)^{\!n_1}
      \bigg( {\pm} \frac{1}{m_{\rm b}\!}
            [\hat{\bs{p}} \times \bs{S}_{\rm b}] \cdot \nabla_{\bs{r}}
      \bigg)^{\!n_2}\!
      \frac{1}{|\bs{r}|} . \nn
\end{align}
This closed-form expression is spin-exact but remarkably simple
and may be easily expanded to any required order in the angular momenta
by repeated differentiation.
In the particularly interesting case of two Kerr black holes,
the effective potential becomes
\[
\label{ElasticScatteringHamiltonianBH}
   V^{(1)}(\bs{r},\bs{p},\bs{S}_{\rm a},\bs{S}_{\rm b}) =
   {-}\frac{G m_{\rm a}^2 m_{\rm b}^2 \gamma^2}{2 E_{\rm a} E_{\rm b}}
      \sum_\pm 
      \frac{ (1 \pm v)^2 }
           { |\bs{r} \pm \hat{\bs{p}} \times
             (\bs{a}_{\rm a} + \bs{a}_{\rm b})| } ,
\]
where we are still free to use either $\bs{a}_{\rm a,b}$
or $\bs{S}_{\rm a,b}/m_{\rm a,b}$.

An expression very similar to \eqn{ElasticScatteringHamiltonian}
in the form of a Fourier integral has been written
by Chung, Huang, Kim and Lee \cite{Chung:2020rrz},
who then expanded it to the first four orders in spin and found agreement
with the literature \cite{Levi:2014sba,Levi:2014gsa,Levi:2015msa,Levi:2015uxa,Levi:2016ofk,Levi:2019kgk}.
Our results, however, are different from those of \rcite{Chung:2020rrz}.
This is perhaps most evident
in our black-hole potential \eqref{ElasticScatteringHamiltonianBH},
which has a simpler denominator than in \rcite{Chung:2020rrz}.
We claim, however, that our Hamiltonian~\eqref{ElasticScatteringHamiltonian}
is physically equivalent to that in \rcite{Chung:2020rrz}
and differs from it by something that would constitute a gauge choice
in a more traditional derivation of an effective two-body Hamiltonian
from general relativity.
Indeed, the absence of terms involving $(\bs{r}\cdot\bs{p})$
in the potential of \rcite{Chung:2020rrz}
corresponds to the so-called isotropic gauge,
whereas we start to run into such terms already at quadratic order in spin:
\[
   ([\hat{\bs{p}} \times \bs{S}_{\rm a}] \cdot \nabla_r)^2 \frac{1}{|\bs{r}|}
    = \frac{1}{|\bs{r}|^3}
      \begin{aligned}
      \big\{ 2 \bs{S}_{\rm a}^2     
           - 2 (\hat{\bs{p}}\cdot\bs{S}_{\rm a})^2
           + 6 (\hat{\bs{r}}\cdot\hat{\bs{p}})
               (\hat{\bs{r}}\cdot\bs{S}_{\rm a})
               (\hat{\bs{p}}\cdot\bs{S}_{\rm a}) & \\
           -\,3(\hat{\bs{r}}\cdot\bs{S}_{\rm a})^2
           - 3(\hat{\bs{r}}\cdot\hat{\bs{p}})^2 \bs{S}_{\rm a}^2 &\:\!
      \big\} .
      \end{aligned}
\]

Since finding canonical transformations between Hamiltonians
is a non-trivial task,
here we choose to follow a more instructive path
and rederive (the COM version of) the impulse observables
\eqref{ImpulseMomLOGrav} and \eqref{ImpulseSpinLOGrav}
directly from our potential~\eqref{ElasticScatteringHamiltonianBH}.

%%%%%%%%%%%%%%%%%%%%%%%%%%%%%%%%%%%%%%%%%%%%%%%%%%
\subsection{Observables from motion}
\label{sec:Motion}
%%%%%%%%%%%%%%%%%%%%%%%%%%%%%%%%%%%%%%%%%%%%%%%%%%

The starting point for solving the equations of motion~\eqref{EoM3d}
perturbatively is to acknowledge the fact that
the kinetic part of the Hamiltonian depends exclusively on momenta.
This means that in absence of interaction
only the relative trajectory $\bs{r}(t)$ has a non-trivial evolution.
Therefore, we can set up the initial conditions
by assuming the momenta and spins to be constant at zeroth order in $G$:
\begin{subequations}
\[
   \bs{p}(t) = \bs{p}_\text{in} + {\cal O}(G) , \qquad \quad
   \bs{S}_j(t) = \bs{S}_j^\text{in} + {\cal O}(G) ,
\label{ZeroSolutionMom}
\]
whereas the relative trajectory becomes
\[
   \dot{\bs{r}} = \frac{\partial H}{\partial \bs{p}}
    = \frac{\bs{p}}{E_{\rm a}} + \frac{\bs{p}}{E_{\rm b}} + {\cal O}(G)
   \qquad \Rightarrow \qquad
   \bs{r}(t) = \bs{b}
    + \bigg[\frac{E_{\rm a} + E_{\rm b}}{E_{\rm a} E_{\rm b}}
      \bs{p} \bigg]_\text{in} t + {\cal O}(G) .
\label{ZeroSolutionTraj}
\]
\label{ZeroSolution}%
\end{subequations}
Here we have set $\bs{r}(0) = \bs{b}$,
so the impact parameter,
such that $\bs{b}\cdot \bs{p}_\text{in} = 0$,
naturally coincides
with the minimal distance between the two massive objects.
Here and below, we use the subscript ``$\text{in}$''
to freeze the affected dynamical variables at their initial values;
\eg
$E_{\rm a}^\text{in} = \sqrt{m_{\rm a}^2 + \bs{p}_\text{in}^2}$,
$ [\hat{\bs{p}} \times \bs{S}_{\rm b}]_\text{in}
= \hat{\bs{p}}_\text{in} \times \bs{S}_{\rm b}^\text{in} $,
\etc

%%%%%%%%%%%%%%%%%%%%%%%%%%%%%%%%%%%%%%%%%%%%%%%%%%
\subsubsection{Impulse from motion}
\label{sec:ImpulseMom3d}
%%%%%%%%%%%%%%%%%%%%%%%%%%%%%%%%%%%%%%%%%%%%%%%%%%

In order to handle the two-body motion of general spinning objects
with more ease, it is convenient to introduce a shorthand
for the following differential operators
\[
   {\cal C}_j^\pm(\nabla_{\bs{r}}) \equiv
%   {\cal C}_\pm(\hat{\bs{p}},\bs{S}_j;\nabla_{\bs{r}}) \equiv
      \sum_{n=0}^\infty
      \frac{C_{j\:\!n}\!}{n!}
      \bigg( {\pm} \frac{1}{m_j\!}
            [\hat{\bs{p}} \times \bs{S}_j] \cdot \nabla_{\bs{r}}
      \bigg)^{\!n} , \qquad \quad
      j = {\rm a}, {\rm b} ,
\label{CoeffDiffOperator}
\]
which are ubiquitous to the 1PM effective potential
\eqref{ElasticScatteringHamiltonian}.
Its coordinate derivative
\[
   \frac{\partial H}{\partial \bs{r}} =
      \frac{G m_{\rm a}^2 m_{\rm b}^2 \gamma^2}{2 E_{\rm a} E_{\rm b}}
      \sum_\pm (1 \mp v)^2
      {\cal C}_{\rm a}^\pm(\nabla_{\bs{r}})
      {\cal C}_{\rm b}^\pm(\nabla_{\bs{r}})
      \frac{\bs{r}}{|\bs{r}|^3\!} + {\cal O}(G^2)
    = -\dot{\bs{p}}
\]
governs the evolution of the momentum variable.
In the non-spinning case ${\cal C}_j^\pm = 1$,
the leading-order solution is obtained by straightforward time integration
\[
   \int\!dt \frac{\bs{r}}{|\bs{r}|^3\!}
      \bigg|_{ \bs{r}\,=\,\bs{b}\,+
               \big[\frac{E_{\rm a} + E_{\rm b}}{E_{\rm a} E_{\rm b}} \bs{p}
               \big]_\text{in}\!t}\!\!
    =\!\int\!dt
      \frac{ \bs{b} + \big[\frac{E_{\rm a} + E_{\rm b}}{E_{\rm a} E_{\rm b}}
             \bs{p} \big]_\text{in} t }
           {\!\big( \bs{b}^2\!
           + \big[\frac{m_{\rm a} m_{\rm b} \gamma v}{E_{\rm a} E_{\rm b}}
             \big]^2_\text{in} t^2\big)^{3/2}\!}
    = \frac{ \frac{\bs{b}}{\bs{b}^2} t
           - \big[ \frac{ E_{\rm a} E_{\rm b} \bs{p} }
                        { (E_{\rm a} + E_{\rm b}) \bs{p}^2 }
             \big]_\text{in} }
           {\!\big( \bs{b}^2\!
           + \big[\frac{m_{\rm a} m_{\rm b} \gamma v}{E_{\rm a} E_{\rm b}}
             \big]^2_\text{in} t^2\big)^{1/2}\!}
    + \bs{C} ,
\]
where in the denominator
we have used the kinematic relation~\eqref{COMFrame}.
To do the same in the general spinning case,
we need to convince ourselves in the following:
\[\!\!
   {\cal C}_{{\rm a},\text{in}}^\pm(\nabla_{\bs{r}})
   {\cal C}_{{\rm b},\text{in}}^\pm(\nabla_{\bs{r}})
      \frac{\bs{r}}{|\bs{r}|^3\!}
      \bigg|_{ \bs{r}\,=\,\bs{b}\,+
               \big[\frac{E_{\rm a} + E_{\rm b}}{E_{\rm a} E_{\rm b}} \bs{p}
               \big]_\text{in}\!t}\!\!
    = {\cal C}_{{\rm a},\text{in}}^\pm(\nabla_{\bs{b}})
      {\cal C}_{{\rm b},\text{in}}^\pm(\nabla_{\bs{b}})
      \frac{ \bs{b} + \big[\frac{E_{\rm a} + E_{\rm b}}{E_{\rm a} E_{\rm b}}
             \bs{p} \big]_\text{in} t }
           {\!\big( \bs{b}^2\!
           + \big[\frac{m_{\rm a} m_{\rm b} \gamma v}{E_{\rm a} E_{\rm b}}
             \big]^2_\text{in} t^2\big)^{3/2}\!}\,.
\label{r2b}
\]
This is indeed true due to the structure of
${\cal C}_{j,\text{in}}^\pm(\nabla_{\bs{r}})$,
which differentiate exclusively along directions
orthogonal to $\bs{p}_\text{in}$.
In other words, these differential operators only depend
on the transverse gradient\footnote{Note that we use
subscripts in the gradients $\nabla_{\bs{r}}$ and $\nabla_{\bs{b}}$
simply to specify the differentiation variables.
Directional derivatives are then constructed via an explicit
scalar product, as in $(\hat{\bs{p}}_\text{in}\!\cdot \nabla_{\bs{r}})$.}
\[
   \nabla_{\bs{r}}^\perp = \Pi_\text{in} \nabla_{\bs{r}} , \qquad
   \Pi_\text{in}^{ij} = \delta^{ij}
    - p^i_\text{in} p^j_\text{in} / p^2_\text{in}
    \qquad \Rightarrow \qquad
   \nabla_{\bs{r}} = \hat{\bs{p}}_\text{in}
                    (\hat{\bs{p}}_\text{in}\!\cdot \nabla_{\bs{r}})
    + \nabla_{\bs{r}}^\perp ,
\]
where we have used the COM-frame version
of the transverse projector~\eqref{TransverseProjector}.
Therefore, we can formulate a more general statement:
\[
   (\nabla_{\bs{r}}^\perp)^{\otimes n} f(\bs{r})
      \big|_{\bs{r} = \bs{b}+\bs{c}}
    = (\nabla_{\bs{b}}^\perp)^{\otimes n} f(\bs{b}+\bs{c})
   \qquad \text{provided that} \qquad
   \bs{c}^\perp = \Pi_\text{in} \bs{c} = 0 ,
\]
from which \eqn{r2b} follows directly.
We also remind the reader that the impact-parameter differentiation
is always assumed to be transverse,
$\nabla_{\bs{b}}\equiv\nabla_{\bs{b}}^\perp$.
%In other words, \eqn{r2b} can be viewed as a special case of
%the identity for directional derivatives
%\[
%   (\bs{a}\cdot\nabla_{\bs{r}})^n f(\bs{r}) \big|_{\bs{r} = \bs{b}+\bs{c}}
%    = (\bs{a}\cdot\nabla_{\bs{b}})^n f(\bs{b}+\bs{c})
%   \qquad \text{provided that} \qquad
%   \bs{a} \cdot \bs{c} = 0 .
%\]

In this way, we find that the momentum equation of motion is solved by
\begin{align}
   \bs{p}(t) = &\,
   {-}\bigg[\frac{G m_{\rm a}^2 m_{\rm b}^2 \gamma^2}{2 E_{\rm a} E_{\rm b}}
      \sum_\pm (1 \mp v)^2
      {\cal C}_{\rm a}^\pm(\nabla_{\bs{b}})
      {\cal C}_{\rm b}^\pm(\nabla_{\bs{b}})
      \bigg]_\text{in}\,
      \frac{ \frac{\bs{b}}{\bs{b}^2} t
           - \big[ \frac{ E_{\rm a} E_{\rm b} \bs{p} }
                        { (E_{\rm a} + E_{\rm b}) \bs{p}^2 }
             \big]_\text{in} }
           { \sqrt{ \bs{b}^2\!
           + \big[\frac{m_{\rm a} m_{\rm b} \gamma v}{E_{\rm a} E_{\rm b}}
             \big]^2_\text{in} t^2}}
    + \bs{C} + {\cal O}(G^2) \nn \\ &
      \xrightarrow[t \to {\color{black} \pm \infty}]{} {\color{black} \mp}
      \frac{G m_{\rm a} m_{\rm b} \gamma_\text{in}}{2 v_\text{in}}
      \sum_\pm (1 \mp v)_\text{in}^2
      {\cal C}_{{\rm a},\text{in}}^\pm(\nabla_{\bs{b}})
      {\cal C}_{{\rm b},\text{in}}^\pm(\nabla_{\bs{b}})
      \frac{\bs{b}}{\bs{b}^2\!}
    + \bs{C} + {\cal O}(G^2)
\label{MomLimits3d}
\end{align}
where the two time limits differ only by an overall sign.
The impulse $\Delta \bs{p}$ is the difference between these two limits.
It can thus be read off directly from \eqn{MomLimits3d}
even without specifying the constant of integration~$\bs{C}$,
which, incidentally, can be determined simply by enforcing
$\bs{p}(-\infty) = \bs{p}_\text{in}$.
Recalling the definition~\eqref{CoeffDiffOperator}
of the differential operators, we can express the net momentum change as
\[
\begin{aligned}
   \Delta \bs{p} = -G m_{\rm a} m_{\rm b} \Bigg[ \frac{\gamma}{v}
      \sum_\pm (1 \mp v)^2\!\!
      \sum_{n_1,n_2=0}^\infty\!\!\!
      \frac{C_{{\rm a}\:\!n_1} C_{{\rm b}\:\!n_2}\!}{n_1!n_2!} &
      \bigg( {\pm} \frac{1}{m_{\rm a}\!}
            [\hat{\bs{p}} \times \bs{S}_{\rm a}] \cdot \nabla_{\bs{b}}
      \bigg)^{\!n_1} \\ \times &
      \bigg( {\pm} \frac{1}{m_{\rm b}\!}
            [\hat{\bs{p}} \times \bs{S}_{\rm b}] \cdot \nabla_{\bs{b}}
      \bigg)^{\!n_2} \Bigg]_\text{in}
      \frac{\bs{b}}{\bs{b}^2\!} \,.
\label{ImpulseMomLOGrav3d}
\end{aligned}
\]
We can now easily recognize that this leading-order solution
to the Hamiltonian equations of motion is nothing but the COM-frame
version of the Lorentz-covariant impulse observable~\eqref{ImpulseMomLOGrav},
which was computed earlier using scattering amplitudes.

%%%%%%%%%%%%%%%%%%%%%%%%%%%%%%%%%%%%%%%%%%%%%%%%%%
\subsubsection{Spin kick from motion}
\label{sec:ImpulseSpin3d}
%%%%%%%%%%%%%%%%%%%%%%%%%%%%%%%%%%%%%%%%%%%%%%%%%%

Let us now find the leading solution to the angular-momentum
equation of motion
\[\!\!
   \dot{\bs{S}}_{\rm a} =-\bs{S}_{\rm a} \times
      \frac{\partial H}{\partial \bs{S}_{\rm a}}
    = \frac{G m_{\rm a}^2 m_{\rm b}^2 \gamma^2}{2 E_{\rm a} E_{\rm b}}
      \sum_\pm (1 \mp v)^2
      {\cal C}_{\rm b}^\pm(\nabla_{\bs{r}})
      \bigg[ \bs{S}_{\rm a}\!\times\!
      \frac{ \partial {\cal C}_{\rm a}^\pm(\nabla_{\bs{r}}) }
           { \partial \bs{S}_{\rm a} } \bigg]
      \frac{1}{|\bs{r}|} + {\cal O}(G^2) .\!\!
\]
The new differential operator
\[
   \frac{ \partial {\cal C}_{\rm a}^\pm(\nabla_{\bs{r}}) }
        { \partial \bs{S}_{\rm a} } 
    = \sum_{n=1}^\infty
      \frac{\mp C_{{\rm a}\:\!n}\!}{(n-1)!\:\!m_{\rm a}\!}
      \bigg( {\pm} \frac{1}{m_{\rm a}\!}
            [\hat{\bs{p}} \times \bs{S}_{\rm a}] \cdot \nabla_{\bs{r}}
      \bigg)^{\!n-1}
      [\hat{\bs{p}} \times \nabla_{\bs{r}}]
\label{CoeffDiffOperatorDiff}
\]
is clearly transverse as well.
We may therefore convert between derivatives
$\nabla_{\bs{r}}$ and $\nabla_{\bs{b}}$ as before
%\[
%\begin{aligned}
%   {\cal C}_{{\rm b},\text{in}}^\pm(\nabla_{\bs{r}})
%   \bigg[ \bs{S}_{\rm a}\!\times\!
%          \frac{ \partial {\cal C}_{\rm a}^\pm(\nabla_{\bs{r}}) }
%               { \partial \bs{S}_{\rm a} } & \bigg]_\text{in}
%      \frac{1}{|\bs{r}|}
%      \bigg|_{ \bs{r}\,=\,\bs{b}\,+
%               \big[\frac{E_{\rm a} + E_{\rm b}}{E_{\rm a} E_{\rm b}} \bs{p}
%               \big]_\text{in}\!t} \\
%    = {\cal C}_{{\rm b},\text{in}}^\pm(\nabla_{\bs{b}}) &
%      \bigg[ \bs{S}_{\rm a}\!\times\!
%          \frac{ \partial {\cal C}_{\rm a}^\pm(\nabla_{\bs{b}}) }
%               { \partial \bs{S}_{\rm a} } \bigg]_\text{in}
%      \bigg( \bs{b}^2\!
%           + \Big[\frac{m_{\rm a} m_{\rm b} \gamma v}{E_{\rm a} E_{\rm b}}
%             \Big]^2_\text{in} t^2\bigg)^{\!-1/2} .
%\label{r2b2}
%\end{aligned}
%\]
and integrate the spin evolution all the way to
\[
\begin{aligned}
   \bs{S}_{\rm a}(t)
    = \bigg[\frac{G m_{\rm a} m_{\rm b} \gamma}{4 v}
      \sum_\pm (1 \mp v)^2
      {\cal C}_{\rm b}^\pm(\nabla_{\bs{b}})
      \bs{S}_{\rm a}\!\times\!
      \frac{ \partial {\cal C}_{\rm a}^\pm(\nabla_{\bs{b}}) }
           { \partial \bs{S}_{\rm a} }
      \bigg]_\text{in} & \\ \times
      \log \frac{ \sqrt{ \bs{b}^2
                       + \big[\frac{m_{\rm a} m_{\rm b} \gamma v}
                                   {E_{\rm a} E_{\rm b}}
                         \big]^2_\text{in} t^2 }
                + \big[\frac{m_{\rm a} m_{\rm b} \gamma v}
                            {E_{\rm a} E_{\rm b}}
                  \big]_\text{in} t }
                { \sqrt{ \bs{b}^2
                       + \big[\frac{m_{\rm a} m_{\rm b} \gamma v}
                                   {E_{\rm a} E_{\rm b}}
                         \big]^2_\text{in} t^2 }
                - \big[\frac{m_{\rm a} m_{\rm b} \gamma v}
                            {E_{\rm a} E_{\rm b}}
                  \big]_\text{in} t } &
    + \bs{C} + {\cal O}(G^2) .
\end{aligned}
\]
In order to safely take $t \to \pm \infty$,
we need to evaluate the gradient that became
exposed in \eqn{CoeffDiffOperatorDiff}.
We find that its late/early time limits give simply
\[
   [\hat{\bs{p}}_\text{in}\!\times \nabla_{\bs{b}}]
%   \log \frac{ \sqrt{ \bs{b}^2 + \ldots } + \ldots }
%             { \sqrt{ \bs{b}^2 + \ldots } - \ldots }
      \log \frac{ \sqrt{ \bs{b}^2
                       + \big[\frac{m_{\rm a} m_{\rm b} \gamma v}
                                   {E_{\rm a} E_{\rm b}}
                         \big]^2_\text{in} t^2 }
                + \big[\frac{m_{\rm a} m_{\rm b} \gamma v}
                            {E_{\rm a} E_{\rm b}}
                  \big]_\text{in} t }
                { \sqrt{ \bs{b}^2
                       + \big[\frac{m_{\rm a} m_{\rm b} \gamma v}
                                   {E_{\rm a} E_{\rm b}}
                         \big]^2_\text{in} t^2 }
                - \big[\frac{m_{\rm a} m_{\rm b} \gamma v}
                            {E_{\rm a} E_{\rm b}}
                  \big]_\text{in} t }
   ~\xrightarrow[t \to {\color{black} \pm \infty}]{}~ {\color{black} \mp}
   \frac{2}{\bs{b}^2} [\hat{\bs{p}}_\text{in}\!\times \bs{b}] .
\]
Thus we arrive at the following expression
for the three-dimensional angular impulse:
\[
\begin{aligned}\!\!\!\!
   \tilde{\Delta} \bs{S}_{\rm a}
    = -G m_{\rm a} m_{\rm b} \Bigg[ \frac{\gamma}{v}
      \sum_\pm (1 \mp v)^2\!
      \sum_{n_1=1}^\infty
      \sum_{n_2=0}^\infty
      \frac{ \mp C_{{\rm a}\:\!n_1} C_{{\rm b}\:\!n_2}}
           { (n_1\!-\!1)!n_2! m_{\rm a}\!}
      \bigg( {\pm} \frac{1}{m_{\rm b}\!}
            [\hat{\bs{p}} \times \bs{S}_{\rm b}] \cdot \nabla_{\bs{b}}
      \bigg)^{\!n_2}\;\!\!\!\!\!\!&\!\! \\ \times
      \bigg( {\pm} \frac{1}{m_{\rm a}\!}
            [\hat{\bs{p}} \times \bs{S}_{\rm a}] \cdot \nabla_{\bs{b}}
      \bigg)^{\!(n_1-1)}
      \frac{1}{\bs{b}^2}
      \big[ (\bs{b} \cdot \bs{S}_{\rm a}) \hat{\bs{p}}
          - (\hat{\bs{p}} \cdot \bs{S}_{\rm a}) \bs{b} \big] &
      \Bigg]_\text{in}\!.\!\!\!\!\!\!
\label{ImpulseSpinLOGrav3dPart}
\end{aligned}
\]

%%%%%%%%%%%%%%%%%%%%%%%%%%%%%%%%%%%%%%%%%%%%%%%%%%
\subsubsection{Frame-choice subtlety}
\label{sec:FrameSubtlety}
%%%%%%%%%%%%%%%%%%%%%%%%%%%%%%%%%%%%%%%%%%%%%%%%%%

The reason why we have labeled the above expression as
$\tilde{\Delta} \bs{S}_{\rm a}$ is to discern it from
a similar result that follows from scattering amplitudes.
Namely, taking $\Delta \bs{a}_{\rm a}$ to be the three-dimensional part
of the Lorentz-covariant spin kick~\eqref{ImpulseSpinLOGrav}
in the COM frame,
we can convert it to its rest-frame version $\Delta \bs{S}_{\rm a}$
by appropriately perturbing the boost relationship~\eqref{SpinBoost}:
\[
   \Delta \bs{S}_{\rm a}
    = m_{\rm a} \Delta \bs{a}_{\rm a}
    - \frac{ m_{\rm a} (\bs{p} \cdot \Delta \bs{a}_{\rm a})
           + \Delta \bs{p} \cdot \bs{S}_{\rm a} }
           { E_{\rm a} (E_{\rm a} + m_{\rm a}) } \bs{p}
    - \frac{ \bs{p} \cdot \bs{S}_{\rm a} }
           { m_{\rm a} (E_{\rm a} + m_{\rm a}) } \Delta \bs{p}
    + {\cal O}(G^2) ,
\]
where we have used $\Delta E_{\rm a} = {\cal O}(G^2)$.
In this way, we find the full angular impulse to be
\[
   \Delta \bs{S}_{\rm a}
    = \tilde{\Delta} \bs{S}_{\rm a}
    + \frac{ (\Delta \bs{p} \cdot \bs{S}_{\rm a}) \bs{p}
           - (\bs{p} \cdot \bs{S}_{\rm a}) \Delta \bs{p} }
           { m_{\rm a} (E_{\rm a} + m_{\rm a}) } ,
\label{ImpulseSpinLOGrav3d}
\]
where the $\tilde{\Delta} \bs{S}_{\rm a}$ portion
is given precisely by the solution~\eqref{ImpulseSpinLOGrav3dPart}
to the equations of motion.
The rest of the terms are evidently due to $\Delta \bs{p}$,
which was computed in \eqn{ImpulseMomLOGrav3d}.
In fact, they can be seen to descend from the contribution
in the master spin-kick formula~\eqref{ImpulseSpinLOFinal}
that involves $\partial {\cal A}_4^{(0)}(b)/\partial b^\nu$,
and they are therefore structurally different
from $\tilde{\Delta} \bs{S}_{\rm a}$.
Moreover, these terms can be traced further back
to the boost difference~\eqref{SpinBoostLinear}
between the angular-momentum operators
associated with the incoming and outgoing momenta $p_{\rm a} \mp k/2$
in the scattering amplitude.

It should then not come as a surprise that this difference between
$\tilde{\Delta} \bs{S}_{\rm a}$ and $\Delta \bs{S}_{\rm a}$
is explained by the fact that
these two changes in the rest-frame angular momenta
are actually set up in different frames.
Indeed, $\tilde{\Delta} \bs{S}_{\rm a}$
has been derived from the equations of motion
\[
   \dot{\bs{S}}_{\rm a}
    = -\bs{S}_{\rm a} \times \frac{\partial H}{\partial \bs{S}_{\rm a}}
    = \{ \bs{S}_{\rm a}, H\} , \qquad \quad
   \{ S_{\rm a}^i, S_{\rm a}^j \} = \epsilon^{ijk} S_{\rm a}^k ,
\label{SpinEoM3d}
\]
which rely on the classical analogue of the rest-frame spin algebra~\eqref{AngularMomentumAlgebra}, which disregards changes in momenta.
Therefore, $\tilde{\Delta} \bs{S}_{\rm a}$ follows the rest frame
of the initial momentum $(E_{\rm a},\bs{p})_\text{in}$,
whereas $\Delta \bs{S}_{\rm a}$ refers to the rest frame of the outgoing
momentum $(E_{\rm a}^\text{in},\bs{p}_\text{in} + \Delta \bs{p})$.
It is perhaps more clearly summarized
by the following scattering diagram:
\[
\begin{aligned}
   (S + \tilde{\Delta} S)^i &
    = L^i{}_\mu\big((m,\bs{0}),(E,\bs{p})\big)
      m (a + \Delta a)^\mu \\
   \overset{
   \substack{\!\begin{rotate}{137}
             $\xrightarrow[~~~~]{}$
             \end{rotate}}}{S^i} &
    = L^i{}_\mu\big((m,\bs{0}),(E,\bs{p})\big)
%      \overset{\!\!\!\!
%      \substack{\begin{rotate}{90}
%                $\!\!\!\xrightarrow[~~]{}$
%                \end{rotate}\\~}}{)}\big)
      m
      \overset{\substack{~\\~\\\begin{rotate}{60}
                               $\:\!\!\xrightarrow[~~~\;]{}$
                               \end{rotate}}}{a^\mu} \\
   \overset{\substack{~\\~~\,\begin{rotate}{43}
                         $\!\xleftarrow[~~~~]{}$
                         \end{rotate}}}{(S + \Delta S)^i} &
    = L^i{}_\mu\big((m,\bs{0}),(E,
      \overset{\substack{\begin{rotate}{-65}
                         $\!\!\xrightarrow[~~~]{}$
                         \end{rotate}\\~\\~}}{\bs{p}} + \Delta \bs{p}
      \overset{\substack{\begin{rotate}{-27}
                         $\!\!\!\!\!\!\!\xrightarrow[~~~~~~~]{}$
                         \end{rotate}\\~}}{)}\big)
      m (a + \Delta a)^\mu ,
\label{SpinKickDifference}
\end{aligned}
\]
where the ``${\rm a}$'' and ``$\text{in}$'' subscripts
are omitted for brevity.
The standard minimal boosts are given by
\[
   L^\lambda{}_\mu(p_2,p_1) = \delta^\lambda_\mu
    + \frac{2}{p_1^2} p_2^\lambda p_{1\mu}
    - \frac{(p_1+p_2)^\lambda(p_1+p_2)_\mu}{p_1^2 + p_1 \cdot p_2} ,
   \qquad \quad p_1^2 = p_2^2 ,
\]
and the arrows in the diagram~\eqref{SpinKickDifference}
show the time evolution from past to future infinity.
Note that in both cases the Lorentz-covariant angular momentum
develops the same spin kick $m_{\rm a} \Delta a_{\rm a}^\mu$,
as given by \eqn{ImpulseSpinLOGrav},
and it is only at the level of the three-dimensional frame choice
that the discrepancy between
$\tilde{\Delta} \bs{S}_{\rm a}$ and $\Delta \bs{S}_{\rm a}$ appears.

The self-consistent choice is of course to define
$\bs{S}_{\rm a}^\text{out}$ to be in the rest frame of
$\bs{p}_\text{out} = \bs{p}_\text{in} + \Delta \bs{p}$,
so we tend to regard
$\Delta\bs{S}_{\rm a}=\bs{S}_{\rm a}^\text{out}-\bs{S}_{\rm a}^\text{in}$
as the true rest-frame angular impulse.
One should therefore be aware of this subtlety when dealing with solutions
of the three-dimensional equations of motion~\eqref{SpinEoM3d}.

%%%%%%%%%%%%%%%%%%%%%%%%%%%%%%%%%%%%%%%%%%%%%%%%%%
\section{Summary and outlook}
\label{sec:Outro}
%%%%%%%%%%%%%%%%%%%%%%%%%%%%%%%%%%%%%%%%%%%%%%%%%%

In this paper, we have extended the KMOC formalism
\cite{Kosower:2018adc,Maybee:2019jus,delaCruz:2020bbn,Cristofoli:2021vyo}
to describe general spinning bodies.
Their classical angular momenta are build up
as coherent superpositions of massive quantum states
with arbitrarily large quantum spin numbers.
In \sec{sec:LowerSpins}
we have also commented on how quantum states with finite lower spins
can still be use to model lower-multipole interactions,
as it was done in \rcite{Maybee:2019jus}
following earlier intuition
from \rcites{Holstein:2008sw,Holstein:2008sx,Vaidya:2014kza}.
In many ways, we find that our approach provides
a more solid justification for the earlier
treatments of classical scattering of spinning black holes
\cite{Guevara:2017csg,Guevara:2018wpp,Guevara:2019fsj}.
Note that coherent spin states were also invoked
in the EFT approach of \rcites{Bern:2020buy,Kosmopoulos:2021zoq}.

We have observed throughout \secs{sec:KMOC}{sec:ClassicalAmplitudes}
that the ${\rm SU}(2)$ spinors,
on which coherent spin states depend,
naturally saturate the little-group indices
that represent the spin quantum numbers
in scattering amplitudes considered
within the massive spinor-helicity formalism~\cite{Arkani-Hamed:2019ymq}.
In fact, such spinors can even be used
as a convenient bookkeeping device for the spin degrees of freedom,
as recently employed in \rcite{Chiodaroli:2021eug}.

Here we have concentrated on the three-point amplitudes
involving graviton emission,
as well as the classical limit of the elastic scattering amplitude
for two massive bodies.
Although the coherent-spin summation involves amplitudes
with arbitrary combinations of definite massive spins,
we could prove that all three-point amplitudes
with two massive particles of unequal spins (and one massless particle)
are naturally suppressed in the classical limit.
From the four-point coherent-spin amplitude,
we have computed the leading-order impulse~\eqref{ImpulseMomLOGrav},
spin kick~\eqref{ImpulseSpinLOGrav}
and an effective two-body Hamiltonian~\eqref{ElasticScatteringHamiltonian},
which can be used beyond the scattering setting.
Unlike the Hamiltonians of
\rcites{Chung:2020rrz,Bern:2020buy,Kosmopoulos:2021zoq},
which were obtained directly in the isotropic gauge,
our result is in a different gauge.

We have chosen to verify the validity of our Hamiltonian
by direct time integration of the corresponding equations of motion,
from which we could rederive the impulse observables
in the center-of-mass frame.
We found perfect agreement for the net momentum change $\Delta \bs{p}$.
As for the angular impulse,
the net change in the rest-frame angular momentum
$\tilde{\Delta} \bs{S}_{\rm a}$,
which we obtained from the Hamiltonian equations of motion,
was found to be missing certain terms
that are present in the answer $\Delta \bs{S}_{\rm a}$
derived from scattering amplitudes.
The same superficial discrepancy was earlier noticed
in \rcite{Aoude:2020ygw}.
In \sec{sec:FrameSubtlety},
we have found that these angular-impulse terms
depend on the linear impulse $\Delta \bs{p}$
and are simply due to the mismatch between the three-dimensional frames,
in which they are set up.
In other words, they can both be obtained
by considering slightly different Lorentz boosts
of the same Lorentz-covariant answer~$\Delta S_{\rm a}^\mu$,
which is unequivocally given by the KMOC formalism.

It will be interesting to apply our formalism
to gravitational Compton scattering~\cite{Falkowski:2020aso,Bautista:2021wfy,Chiodaroli:2021eug}.
The amplitudes for such a process are known to
suffer from a spurious pole at higher spins
\cite{Arkani-Hamed:2017jhn,Chung:2018kqs,Johansson:2019dnu},
but solutions to this problem have already started to take shape
\cite{Falkowski:2020aso,Chiodaroli:2021eug}.
In this setting, the KMOC formalism will naturally integrate
our coherent-state approach to spin
and a similar approach to classical radiation~\cite{Cristofoli:2021vyo}.

%%%%%%%%%%%%%%%%%%%%%%%%%%%%%%%%%%%%%%%%%%%%%%%%%%
\section*{Acknowledgements}

We are grateful to Alfredo Guevara, Ben Maybee, Donal O'Connell
and Justin Vines for many invaluable discussions that led to this project.
In particular, we thank Donal O'Connell for comments
on an earlier version of this draft. 
We would also like to thank Kays Haddad and Andreas Helset for enlightening discussions.
AO is grateful to the Higgs Centre for Theoretical Physics for hospitality.
We also thank the Galileo Galilei Institute for Theoretical Physics (GGI) for hosting a workshop, conference and training week on ``Gravitational scattering, inspiral, and radiation'' which informed and enriched our work.
AO's research is funded by the STFC grant ST/T000864/1.
RA's research is funded by the F.R.S.-FNRS with the EOS - be.h project n. 30820817.

\appendix
%%%%%%%%%%%%%%%%%%%%%%%%%%%%%%%%%%%%%%%%%%%%%%%%%%
\section{Non-minimal spin multipoles}
\label{app:Nonminimal}
%%%%%%%%%%%%%%%%%%%%%%%%%%%%%%%%%%%%%%%%%%%%%%%%%%

Here we outline a connection between the Wilson coefficients
for the spin-induced multipole couplings
in the worldline effective action~\eqref{LSInteractions}
and the corresponding ``classical'' three-point amplitudes.
We start by expanding the curvature tensor in terms of the linear
gravitational perturbation $h_{\mu\nu} = g_{\mu\nu} - \eta_{\mu\nu}$,
\[
   R_{\lambda\mu\,\nu\rho} = \frac{1}{2}
      \big( \partial_\mu \partial_\nu h_{\lambda\rho}
          - \partial_\mu \partial_\rho h_{\lambda\nu}
          - \partial_\lambda \partial_\nu h_{\mu\rho}
          + \partial_\lambda \partial_\rho h_{\mu\nu}
      \big) + {\cal O}(h^2) ,
\]
and plugging this into the worldline effective action.
We get
\begin{align}
\label{LSInteractionsLinear}
   S_\text{Int} = -\frac{m}{2}\!\int\!d\tau
   \bigg[ & \sum_{n=0}^\infty \frac{(-1)^n}{(2n)!} C_{{\rm ES}^{2n}}
          (a \cdot \partial)^{2n} u^\mu u^\nu h_{\mu\nu} \\ &
        + \sum_{n=0}^\infty \frac{(-1)^n}{(2n+1)!} C_{{\rm BS}^{2n+1}}
          (a \cdot \partial)^{2n}\,
          u^\mu \epsilon^{\nu\rho\sigma\tau} u_\rho a_\sigma \partial_\tau
          h_{\mu\nu}
   \bigg]_{x = r(\tau)}\!+ {\cal O}(h^2) . \nn
\end{align}
In producing the above result, we were allowed to neglect
the time derivatives of $u^\mu$ and $a^\mu$, since they are ${\cal O}(h)$
and thus increase the order of approximation.
Moreover, we have taken care to introduce the $n=0$ terms,
which are hardwired into the worldline kinetic terms
\[
   S_\text{Kin} = \int\!d\tau
   \bigg[ {-m}\sqrt{u^2} - \frac{1}{2} S_{\mu\nu} \Omega^{\mu\nu} \bigg]
\]
along with their Wilson coefficients $C_{{\rm ES}^0} = -C_{{\rm BS}^1} = 1$,
see \eg \rcites{Chung:2018kqs,Guevara:2020xjx}.

Let us reinterpret the linearized action~\eqref{LSInteractionsLinear}
as the interaction
\[
   S_\text{Int}
    = -\frac{1}{2}\!\int\!d^4x\;\!h_{\mu\nu}(x) T^{\mu\nu}_\text{gen}(x)
    = -\frac{1}{2}\!\int\!\!\frac{d^4\bar{k}}{(2\pi)^4}
       h_{\mu\nu}(\bar{k}) T^{\mu\nu}_\text{gen}(-\bar{k}) ,
\label{InteractionsLinear}
\]
where the effective stress-energy tensor
can be read off \eqn{LSInteractionsLinear} as
\[
   T^{\mu\nu}_\text{gen}(\bar{k})
    = m\!\int\!d\tau\;\!e^{i\bar{k} \cdot r(\tau)}
      \sum_{n=0}^\infty (\bar{k} \cdot a)^{2n}
      \bigg[ \frac{C_{{\rm ES}^{2n}}}{(2n)!} u^\mu u^\nu 
        + \frac{C_{{\rm BS}^{2n+1}}}{(2n+1)!}
          iu^{(\mu} \epsilon^{\nu)\rho\sigma\tau}
          u_\rho a_\sigma \bar{k}_\tau
      \bigg] .
\label{LSEnergyTensor}
\]
Note that in the case where $C_{{\rm ES}^{2n}} = -C_{{\rm BS}^{2n+1}} = 1$
the stress-energy tensor~\eqref{LSEnergyTensor}
may be shown to be equivalent to the Kerr source
given in \eqn{KerrEnergyTensor}.

A recipe to obtain a scattering amplitude
from the worldline action~\eqref{InteractionsLinear}
is to consider a straight particle trajectory
and couple it to an on-shell graviton:
\[
   h^{\mu\nu}(\bar{k}) ~\to~ \kappa\;\!2\pi \delta(\bar{k}^2)
      \varepsilon_k^\mu \varepsilon_k^\nu , \qquad \quad
   r^\mu(\tau) = \frac{p^\mu}{m} \tau \qquad \Rightarrow \qquad
   u^\mu(\tau) = \frac{p^\mu}{m} .
\]
Then the interaction term becomes
\[
   S_\text{Int} = \int\!\!\frac{d^4\bar{k}}{(2\pi)^2}
      \delta(\bar{k}^2) \delta(2p\cdot\bar{k})
      {\cal A}_\text{gen}(p,k) ,
\label{Action2Amplitude}
\]
in terms of the ``classical amplitude''
\[
   {\cal A}^\pm_\text{gen}(p,k)
    = -\kappa (p\cdot\varepsilon_k^\pm)^2
      \bigg[ \sum_{n=0}^\infty \frac{C_{{\rm ES}^{2n}}}{(2n)!}
             (\bar{k} \cdot a)^{2n}
             \pm \sum_{n=0}^\infty \frac{C_{{\rm BS}^{2n+1}}}{(2n+1)!}
             (\bar{k} \cdot a)^{2n+1}
      \bigg] ,
\label{LSAmplitude3}
\]
where we have used
\[
   i\epsilon^{\mu\nu\rho\sigma}
      \varepsilon_{k\:\!\mu}^\pm \bar{k}_\nu p_\rho a_\sigma
    = \mp (p\cdot\varepsilon_k^\pm) (\bar{k} \cdot a) .
\label{Even2Odd3pt}
\]
This identity holds on the support of the delta functions
in \eqn{Action2Amplitude}, which is, strictly speaking,
incompatible with real momenta.
However, we can understand the above equations in the sense
of analytic continuation to complex kinematics,
which is precisely the context of \sec{sec:ClassicalAmplitudes}.
Alternatively, one might consider an analytic continuation
to ``split'' signature $(+,+,-,-)$,
as \eg in \rcites{Monteiro:2020plf,Guevara:2020xjx}.

An easy cross-check of \eqn{LSAmplitude3} is to observe that for
$C_{{\rm ES}^{2n}} = -C_{{\rm BS}^{2n+1}} = 1$
it reduces to the expected Kerr result~\eqref{A3ptMinClassical}:
\[
   {\cal A}^\pm_\text{min}(p,k)
    = -\kappa (p\cdot\varepsilon_k^\pm)^2
      \big[ \cosh(\bar{k} \cdot a) \mp \sinh(\bar{k} \cdot a) \big]
    = -\frac{\kappa}{2} m^2 x^{\pm 2} e^{\mp \bar{k} \cdot a} .
\]

%%%%%%%%%%%%%%%%%%%%%%%%%%%%%%%%%%%%%%%%%%%%%%%%%%
\bibliographystyle{JHEP}
\bibliography{references}
\end{document}